\documentclass[apjl,twocolumn,numberedappendix,appendixfloats]{openjournal}
\usepackage{natbib} 
\usepackage[dvipsnames]{xcolor}
\usepackage{aas_macros} 
\usepackage{amssymb}
\usepackage{amsmath}
\usepackage{threeparttable}
\usepackage{rotating}
\usepackage{booktabs}
\usepackage{multirow}
\usepackage[title]{appendix}
\usepackage{hyperref}
\hypersetup{colorlinks=true,linkcolor=blue,citecolor=blue,filecolor=blue,urlcolor=blue}
\usepackage[caption=false]{subfig}
\usepackage{soul}

\newcommand{\cii}{[C\,{\sc ii}]}

\newcommand{\oiiinowave}{[O\,{\sc iii}]}

\newcommand{\oi}{[O\,{\sc i}]}

\newcommand{\nii}{[N\,{\sc ii}]}

\newcommand{\oiiil}{[O\,{\sc iii}] 88\,$\mu{\rm m}$}
\newcommand{\ciil}{[C\,{\sc ii}] 158\,$\mu{\rm m}$}

\begin{document}

\title[Multi-band dust properties of REBELS galaxies]{The ALMA REBELS Survey: multi-band dust continuum observations reveal heavily obscured starbursts among UV-bright galaxies at $\MakeLowercase{z}\sim7$
\vspace{-15mm}}

\author{Hiddo S. B. Algera$^{1\,*}$}
\author{Lucy Astles$^{2}$}
\author{Hanae Inami$^{3}$}
\author{Manuel Aravena$^{4,5}$}
\author{Tom Bakx$^{6}$}
\author{Rychard Bouwens$^{7}$}
\author{Rebecca Bowler$^{8}$}
\author{Karin Cescon$^{7}$}
\author{Chian-Chou Chen$^{1}$}
\author{Elisabete da Cunha$^{9}$}
\author{Claudia Di Cesare$^{10}$}
\author{Andreas Faisst$^{11}$}
\author{Andrea Ferrara$^{12}$}
\author{Rebecca Fisher$^{8}$}
\author{Yoshinobu Fudamoto$^{13}$}
\author{Luca Graziani$^{14,15,16}$}
\author{Ikki Mitsuhashi$^{17}$}
\author{Shoichiro Mizukoshi$^{1}$}
\author{Themiya Nanayakkara$^{18}$}
\author{Katherine Ormerod$^{2}$}
\author{Lucie Rowland$^{19,7}$}
\author{Raffaella Schneider$^{14,15,16,20}$}
\author{Sander Schouws$^{21}$}
\author{Renske Smit$^{2}$}
\author{Laura Sommovigo$^{22,23}$}
\author{Mauro Stefanon$^{21,24}$}
\author{Aswin P. Vijayan$^{25}$}
\author{Paul van der Werf$^{7}$}
\author{Joris Witstok$^{26,27}$}
\thanks{$^*$E-mail: \href{mailto:hsbalgera@asiaa.sinica.edu.tw}{hsbalgera@asiaa.sinica.edu.tw}}

\affiliation{$^{1}$Institute of Astronomy and Astrophysics, Academia Sinica, 11F of Astronomy-Mathematics Building, No.1, Sec. 4, Roosevelt Rd, Taipei 106319, Taiwan, R.O.C.}
\affiliation{$^{2}$Astrophysics Research Institute, Liverpool John Moores University, Liverpool, UK}
\affiliation{$^{3}$Hiroshima Astrophysical Science Center, Hiroshima University, Hiroshima 739-8526, Japan}
\affiliation{$^{4}$Instituto de Estudios Astrofísicos, Facultad de Ingeniería y Ciencias, Universidad Diego Portales, Av. Ejército 441, Santiago, Chile}
\affiliation{$^{5}$Millennium Nucleus for Galaxies (MINGAL)}
\affiliation{$^{6}$Department of Space, Earth and Environment, Chalmers University of Technology, SE-412 96 Gothenburg, Sweden}
\affiliation{$^{7}$Leiden Observatory, Leiden University, P.O. Box 9513, 2300 RA Leiden, The Netherlands}
\affiliation{$^{8}$Jodrell Bank Centre for Astrophysics, Department of Physics and Astronomy, University of Manchester, Manchester, UK}
\affiliation{$^{9}$International Centre for Radio Astronomy Research, University of Western Australia, ICRAR M468, 35 Stirling Hwy, Crawley 6009, Western Australia, Australia}
\affiliation{$^{10}$Institute of Science and Technology Austria (ISTA), Am Campus 1, 3400 Klosterneuburg, Austria}
\affiliation{$^{11}$IPAC, California Institute of Technology, 1200 E. California Blvd., Pasadena, CA 91125, USA}
\affiliation{$^{12}$Scuola Normale Superiore, Piazza dei Cavalieri 7, 56126 Pisa, Italy}
\affiliation{$^{13}$Center for Frontier Science, Chiba University, 1-33 Yayoi-cho, Inage-ku, Chiba 263-8522, Japan}
\affiliation{$^{14}$INAF/Osservatorio Astronomico di Roma, Via Frascati 33, 00078 Monte Porzio Catone, Italy}
\affiliation{$^{15}$Dipartimento di Fisica, ``Sapienza'' Università di Roma, Piazzale Aldo Moro 2, 00185 Roma, Italy}
\affiliation{$^{16}$INFN, Sezione Roma1, Dipartimento di Fisica, ``Sapienza'' Università di Roma, Piazzale Aldo Moro 2, 00185, Roma, Italy}
\affiliation{$^{17}$Department for Astrophysical \& Planetary Science, University of Colorado, Boulder, CO 80309, USA}
\affiliation{$^{18}$Sydney Institute for Astronomy, School of Physics, The University of Sydney, Sydney, NSW 2006, Australia}
\affiliation{$^{19}$European Southern Observatory, Karl-Schwarzschild-Straße 2, D-85748, Garching bei M\"{u}nchen, Germany}
\affiliation{$^{20}$Sapienza School for Advanced Studies, Viale Regina Elena 291, 00161 Roma, Italy}
\affiliation{$^{21}$Departament d'Astronomia i Astrofísica, Universitat de València, C. Dr Moliner 50, E-46100 Burjassot, València, Spain}
\affiliation{$^{22}$Department of Astronomy, Columbia University, New York, NY 10027, USA}
\affiliation{$^{23}$Center for Computational Astrophysics, Flatiron Institute, 162 Fifth Avenue, New York, NY 10010, USA}
\affiliation{$^{24}$Unidad Asociada CSIC `Grupo de Astrofísica Extragaláctica y Cosmología' (Instituto de Física de Cantabria -- Universitat de València), Spain}
\affiliation{$^{25}$Astronomy Centre, University of Sussex, Falmer, Brighton BN1 9QH, UK}
\affiliation{$^{26}$Cosmic Dawn Center (DAWN)}
\affiliation{$^{27}$Niels Bohr Institute, University of Copenhagen, Jagtvej 128, DK-2200, Copenhagen, Denmark}

\begin{abstract}
Detailed studies with the Atacama Large Millimeter/submillimeter Array (ALMA) have shown that dust is widespread even in UV-luminous galaxies at $z\gtrsim6.5$. Multi-band continuum observations are critical to fully characterize these early dust reservoirs. We present a homogeneous analysis of multi-band ALMA observations from the REBELS Large Program and seven follow-up programs, targeting thirteen massive galaxies at $6.5\lesssim z\lesssim7.7$ in at least two bands, including Band 8. Through optically thin modified blackbody fitting, we infer the dust masses, temperatures, and infrared luminosities of our targets. Their dust temperatures span $T_\mathrm{dust}\approx30-70\,\mathrm{K}$, and a stack reveals an average $T_\mathrm{dust}=49_{-8}^{+11}\,\mathrm{K}$. Three sources nominally qualify as ultra-luminous infrared galaxies (ULIRGs; $L_\mathrm{IR}>10^{12}\,L_\odot$), with total star formation rates of $\mathrm{SFR}_\mathrm{UV+IR}\approx160-300\,M_\odot\,\mathrm{yr}^{-1}$, the highest of which is found in our most distant target, REBELS-18 at $z=7.67$. Their star formation is heavily obscured, with $\gtrsim90\%$ emerging in the infrared, compared to an already large average obscured fraction of $f_\mathrm{obs}=0.75\pm0.11$ for the stack. These SFRs are consistent with the reconstructed growth histories of some massive quiescent galaxies at $z\approx4-5$, suggesting that dusty yet UV-bright galaxies at $z\sim7$ are plausible progenitors of such systems. The sample moreover displays large dust-to-stellar mass ratios ($M_\mathrm{dust}/M_\star = 0.4_{-0.2}^{+0.3}\%$), best reproduced by models in which dust grows efficiently in the interstellar medium. This demonstrates that a dust-rich, IR-luminous, and heavily obscured galaxy population is already in place well into the Epoch of Reionization.
\end{abstract}

\section{Introduction}
\label{sec:introduction}

One of the key aims in extragalactic astrophysics is to understand the growth of the galaxy population over time. One particularly powerful means of doing so is by tracing the typical star formation rates (SFRs) of galaxies across redshift \citep[e.g.,][]{madau2014}. Generally, this is done through observations in the rest-frame ultraviolet (UV), which directly probes the light emitted by young stars \citep[e.g.,][]{lilly1996,mclure2013,bouwens2015,bouwens2023,finkelstein2015,finkelstein2024,oesch2018,harikane2023,donnan2024}. However, UV-based SFRs are prone to underestimating the true star formation rates of galaxies in the presence of dust, which irrevocably alters their observable spectral energy distributions (SEDs) by absorbing UV emission and re-emitting it in the far-infrared \citep[see][for reviews]{galliano2018,calura2026}.

Far-infrared and (sub)millimeter surveys have demonstrated that dust-obscured star formation dominates the cosmic star formation rate density out to at least $z\approx3$ \citep[e.g.,][]{gruppioni2013,gruppioni2020,magnelli2013,casey2018,zavala2021,traina2024}. The most luminous dusty sources, the submillimeter galaxies (SMGs), peak in number density around Cosmic Noon ($z\approx1-4$; e.g., \citealt{simpson2014,dudzeviciute2020,chen2022_as2cospec}). Beyond $z\approx5$, however, the number density of SMGs rapidly drops, and only a handful of dusty star-forming galaxies (DSFGs) selected directly at far-infrared or (sub)millimeter wavelengths are known in the Epoch of Reionization (here loosely defined as $z\gtrsim6$; e.g., \citealt{riechers2013,strandet2017,zavala2018,bing2025}). 

Instead, most of our understanding of dust at these early epochs comes from dedicated follow-up observations of galaxies initially selected at shorter wavelengths. The Atacama Large Millimeter/submillimeter Array (ALMA) has played a particularly important role in these follow-up efforts, first revealing dust in a UV-selected, gravitationally lensed Lyman break galaxy (LBG) at $z > 7$ \citep{watson2015} and not long after in a UV-luminous system at $z > 8$ \citep{tamura2019}. Subsequent follow-up observations of larger samples of UV-selected galaxies gradually increased the number of known dusty galaxies in the first billion years of cosmic time \citep[e.g.,][see also the recent review by \citealt{smit_bowler2026}]{laporte2017,bowler2018,hashimoto2019,schouws2022}. 

Significant progress was made with the \textit{Reionization Era Bright Emission Line Survey} \citep[REBELS;][]{bouwens2022}, which newly identified dust continuum emission from a dozen galaxies spectroscopically confirmed at $z > 6.5$ through the \ciil{} line, tripling the then-known number of dusty galaxies at this epoch \citep[][]{inami2022}. In total, there are now 26 dust-detected, spectroscopically confirmed LBGs at $z > 6.5$ \citep{watson2015,laporte2017,bowler2018,bowler2022,tamura2019,fudamoto2021,inami2022,schouws2022,hashimoto2023_protocluster}, with MACS0416-Y1 at $z=8.31$ remaining the most distant dusty galaxy since its discovery by \citet[][see also \citealt{bakx2020,bakx2025,tamura2023}]{tamura2019}. Additional dusty galaxies in the $z > 6.5$ Universe have furthermore been found serendipitously around reionization-era quasars \citep[e.g.,][]{decarli2017,venemans2020} -- which themselves are also often dusty \citep[e.g.,][]{bertoldi2003,wang2008,tripodi2024} -- as well as around other massive galaxies \citep[][]{fudamoto2021,vanleeuwen2024}. 

For dust-detected galaxies at these epochs, the main quantities of interest are generally the dust mass ($M_\mathrm{dust}$) and infrared luminosity ($L_\mathrm{IR}$), with the latter providing a direct measure of the obscured SFR. Both quantities depend sensitively on the shape of the underlying dust SED, which is governed mostly by the dust temperature ($T_\mathrm{dust}$), and to a lesser extent by the dust emissivity index ($\beta_\mathrm{IR}$). With only a single continuum measurement, the dust SED must therefore be assumed, either by adopting a fixed dust temperature or through physically motivated models that link it to other observables \citep[e.g.,][]{inoue2020,sommovigo2021,ferrara2022}. To directly constrain the dust SED, continuum observations at multiple frequencies are required, ideally with at least one near the peak \citep[e.g.,][]{casey2014,hodge2020,dacunha2021,sommovigo_algera2025}. Consequently, dust temperatures of high-redshift galaxies have garnered significant attention in the last decade, both from an observational \citep[e.g.,][]{magnelli2014,faisst2017,faisst2020,schreiber2018,bakx2020,bakx2021,witstok2022,witstok2023_beta,viero2022,algera2024,mitsuhashi2024_serenade,mitsuhashi2026_serenade,rajulal2026} and theoretical perspective \citep[e.g.,][]{behrens2018,liang2019,sommovigo2020,sommovigo2021,sommovigo2022,vijayan2022,lower2024,parente2026}. 

Recent multi-band dust continuum observations have begun to provide precisely these constraints for galaxies in the Epoch of Reionization. Initial measurements using two or three bands already suggested a large scatter in the dust temperatures of high-redshift systems \citep[from $T_\mathrm{dust} \sim 30\,\mathrm{K}$ to $>80\,\mathrm{K}$; e.g.,][]{hashimoto2019,bakx2020,faisst2020,harikane2020,witstok2022,algera2024,mitsuhashi2024_serenade}, which has been confirmed for a smaller subset of galaxies with additional continuum datapoints \citep{bakx2021,bakx2025,akins2022,villanueva2024,algera2024b,algera2025_hz10,rajulal2026}. On average, however, and in line with theoretical expectations, dust in $z\gtrsim6$ galaxies is found to be warmer than at later times, which is likely mainly driven by their high star formation rate surface densities and low overall dust masses, enabling efficient heating \citep[e.g.,][]{sommovigo2022,parente2026}. 

Given the steep scaling of infrared luminosity with dust temperature -- approximately $L_\mathrm{IR} \propto M_\mathrm{dust} T_\mathrm{dust}^{6}$ -- warm galaxies may be particularly IR-luminous at fixed dust mass. Recently, \citet{bakx2025} measured a hot dust temperature of $T_\mathrm{dust}\sim90\,\mathrm{K}$ for MACS0416-Y1 \citep[see also][]{harshan2024,takechi2026}, while \citet{rajulal2026} measured a similarly warm $T_\mathrm{dust}\sim80\,\mathrm{K}$ for the Big Three Dragons, a merging pair of LBGs at $z=7.15$ \citep[see also][]{bowler2018,hashimoto2019,sugahara2021,sugahara2025}. Both MACS0416-Y1 and the Big Three Dragons are classified as ultra-luminous infrared galaxies (ULIRGs; defined as having $L_\mathrm{IR} > 10^{12}\,L_\odot$) and have $>90\%$ of their star formation obscured by dust despite their initial selection in the UV. While statistics still remain limited, especially at $z > 7$, if such galaxies with hot dust turn out to be common, this could imply that the obscured star formation rate density at this epoch is currently significantly underestimated \citep{sommovigo2020,viero2022,algera2023,rajulal2026}. 

Such heavily obscured, highly star-forming galaxies are particularly intriguing in the context of the massive quiescent galaxy population that the \textit{James Webb Space Telescope} (\textit{JWST}) is now uncovering at $z\gtrsim3$ in large numbers \citep[e.g.,][]{carnall2023,glazebrook2024,degraaff2025}. The reconstructed star formation histories of these systems generally require intense star formation episodes at $z\gtrsim6$, with typical SFRs of up to several hundred solar masses per year \citep[e.g.,][]{carnall2023,nanayakkara2025,xiao2026}. Dusty, UV-selected galaxies at $z\sim7$ may thus represent (a subset of) the progenitors of these massive quiescent systems, and robustly constraining the obscured star formation rates of the former provides a first important step toward testing this connection.

While warm dust boosts the infrared luminosity, cold dust instead implies large dust masses. Multi-band ALMA observations have now revealed a number of $z\sim7$ LBGs with massive, cold dust reservoirs \citep[e.g.,][]{witstok2022,algera2024,algera2024b}, challenging our understanding of how rapidly dust can build up in the early Universe. Reproducing these large dust masses in simulations generally requires a combination of efficient supernova dust production and rapid ISM dust growth \citep[e.g.,][]{mancini2015,popping2017,vijayan2019,vijayan2026,graziani2020,dayal2022,dicesare2023,choban2025,osman2025}. However, the typical supernova dust yields and dust growth timescales remain highly uncertain even in the local Universe \citep[e.g.,][]{devis2019,delooze2020,galliano2021,schneider2024}. In this regard, robustly measuring the dust masses of $z\gtrsim6$ galaxies provides a complementary avenue to constrain the dominant dust build-up pathways and their efficiencies.

The constraints on early dust build-up become even more stringent at $z\gtrsim8$, where the number of spectroscopically confirmed galaxies continues to rapidly increase with the advent of \textit{JWST} \citep[e.g.,][]{arrabal-haro2023,curtis-lake2023,roberts-borsani2025,tang2025}. However, dust continuum searches at this epoch have mostly come up empty-handed \citep[e.g.,][]{ciesla2025,algera2026b,bakx2026,mitsuhashi2026}, hinting at a rapid change in galaxy dust contents or properties \citep[e.g.,][]{ferrara2023,burgarella2025,burgarella2026,markov2025,narayanan2025,shivaei2025,sommovigo2026}. Despite this, \textit{JWST} has uncovered a number of galaxies with red UV continua at $z\gtrsim10$ \citep[e.g.,][]{donnan2025,rodighiero2026}. Whether these red colors are due to dust, however, remains unclear, as they are degenerate with contributions from nebular continuum emission and AGN activity \citep[e.g.,][]{napolitano2025,mitsuhashi2026_red}. Even if dust is responsible, the inferred UV attenuation depends strongly on the star-dust geometry and the assumed dust properties, and therefore provides only weak constraints on the underlying dust masses \citep[e.g.,][]{ferrara2023,bakx2026,sommovigo2026}. Far-infrared continuum observations instead provide a more direct handle on dust masses and obscured star formation. With direct dust detections at $z \gtrsim 8$ remaining scarce, robust multi-band measurements for a statistical sample of galaxies at $z\sim7$ are therefore essential to establish a baseline for the dust properties of reionization-era galaxies against which any evolution toward earlier epochs can be assessed.

In this work, we take a step toward constructing such a statistical sample by presenting multi-band ALMA observations of 13 galaxies at $6.5 \lesssim z \lesssim 7.7$. We constrain the dust SEDs of the 12 that are detected in dust continuum emission, including nine sources newly detected in ALMA Band 8, which more than doubles the number of $z > 6.5$ LBGs detected in this band. These observations provide new constraints on the dust temperatures, masses, and infrared luminosities of reionization-era galaxies, allowing us to investigate both their obscured star formation and their dust build-up. We focus on galaxies drawn from the Cycle 7 REBELS Large Program \citep[LP;][]{bouwens2022,inami2022} that have received ALMA follow-up observations in the years since the LP was conducted. We describe our targets in Section \ref{sec:targets}, and the ALMA data and their reduction in Section \ref{sec:observations}. We measure the multi-band flux densities of our sample in Section \ref{sec:methods} and infer their dust properties through modified blackbody fitting in Section \ref{sec:results}. We discuss our results in the context of high-redshift obscured star formation and dust production in Section \ref{sec:discussion} before summarizing our findings in Section \ref{sec:conclusions}. 

Throughout this work, we adopt a standard flat $\Lambda$CDM cosmology with $H_0 = 70\,\mathrm{km\,s}^{-1}\,\mathrm{Mpc}^{-1}$ and $\Omega_m = 0.30$. We assume a \citet{chabrier2003} Initial Mass Function (IMF) whereby unobscured SFRs are determined as $\mathrm{SFR}_\mathrm{UV}/(M_\odot\,\mathrm{yr}^{-1}) = 7.1 \times 10^{-29}\,L_\nu / (\mathrm{erg\,s}^{-1}\mathrm{Hz}^{-1})$ following \citet{bouwens2022}, with $L_\nu$ the specific luminosity at $1600\,$\AA. Equivalently, $\mathrm{SFR}_\mathrm{UV}/(M_\odot\,\mathrm{yr}^{-1}) = 1.45 \times 10^{-10}\,(L_\mathrm{UV} / L_\odot)$ where $L_\mathrm{UV} = \nu_{1600}L_\nu$. Obscured SFRs are computed assuming the same IMF, with $\mathrm{SFR}_\mathrm{IR} / (M_\odot\,\mathrm{yr}^{-1}) = 1.2 \times 10^{-10}\,(L_\mathrm{IR}/L_\odot)$ following \citet{inami2022}. Here the IR luminosity is defined across a rest-frame wavelength range of $8-1000\,\mu\mathrm{m}$.

\section{Targets and observations}
\label{sec:targets}

\subsection{REBELS}
\label{sec:targets_rebels}

The galaxies studied in this work are drawn from the Cycle 7 ALMA Large Program REBELS \citep{bouwens2022}. REBELS carried out spectral scans for the \ciil{} or \oiiil{} lines in, respectively, 36 and 4 galaxies with photometric redshifts $z > 6.5$ using ALMA Bands 5, 6, and/or 7. In total, 22 galaxies were detected in \cii{} (Schouws et al.\ in preparation; see also \citealt{herard-demanche2026}), while none were detected in \oiiinowave{} \citep{vanleeuwen2025}. REBELS simultaneously delivered sensitive continuum observations, resulting in 16 dust-detected galaxies, 15 of which were also detected in \cii{} \citep{inami2022}.\footnote{The exact redshift of the remaining source, REBELS-04, remains elusive, but is likely to be $z>8$ \citep{vanleeuwen2025}. \textit{JWST}/NIRSpec observations are planned to verify this (GO 10141, PI Rowland).} 

This widespread detection of dust in REBELS galaxies motivated ALMA follow-up observations to constrain the shape of their far-infrared SEDs, and better determine their dust masses and temperatures. \citet{algera2024} presented ALMA Band 8 observations toward three REBELS galaxies previously detected in Band 6, detecting two, and revealing relatively cold dust reservoirs in both ($T_\mathrm{dust} \approx 35\,\mathrm{K}$), under the assumption of optically thin dust and a fixed emissivity index $\beta_\mathrm{IR} = 2.0$. The REBELS galaxy brightest in Band 6 continuum, REBELS-25, was subsequently followed up in an additional four ALMA bands (3, 4, 5, and 9), further refining the constraints on its dust temperature and mass, as well as enabling a direct measurement of the emissivity index \citep[$\beta_\mathrm{IR} = 2.5 \pm 0.4$;][]{algera2024b}. 

Since then, a larger number of REBELS galaxies have been targeted in additional ALMA bands, generally with the aim of studying emission lines such as \oiiil{} \citep{algera2025_oiii}, \nii{} $205\,\mu\mathrm{m}$ and \oi{} $145\,\mu\mathrm{m}$ \citep{fudamoto2025}, and CO \citep{cescon2026}. Moreover, several REBELS galaxies have been targeted with deeper and higher-resolution \cii{} -- and in one case \oiiil{} -- observations to study their gas morphologies and kinematics \citep[][Phillips et al.\ in preparation, Astles et al.\ in preparation]{rowland2024,rowland2026_r25}. Naturally, these observations also probe the underlying continuum emission, although the main focus of these previous and ongoing works is generally the emission lines rather than the dust continuum (but see Fisher et al.\ in preparation, who use high-resolution Band 6 dust continuum data to measure spatial offsets between the rest-frame UV and far-infrared emission). The aim of this work, then, is to provide a comprehensive overview of the multi-band dust continuum observations available for the REBELS sample.

\subsection{Multi-band targets}
\label{sec:targets_multiband}

\begin{figure}
    \centering
    \includegraphics[width=0.5\textwidth]{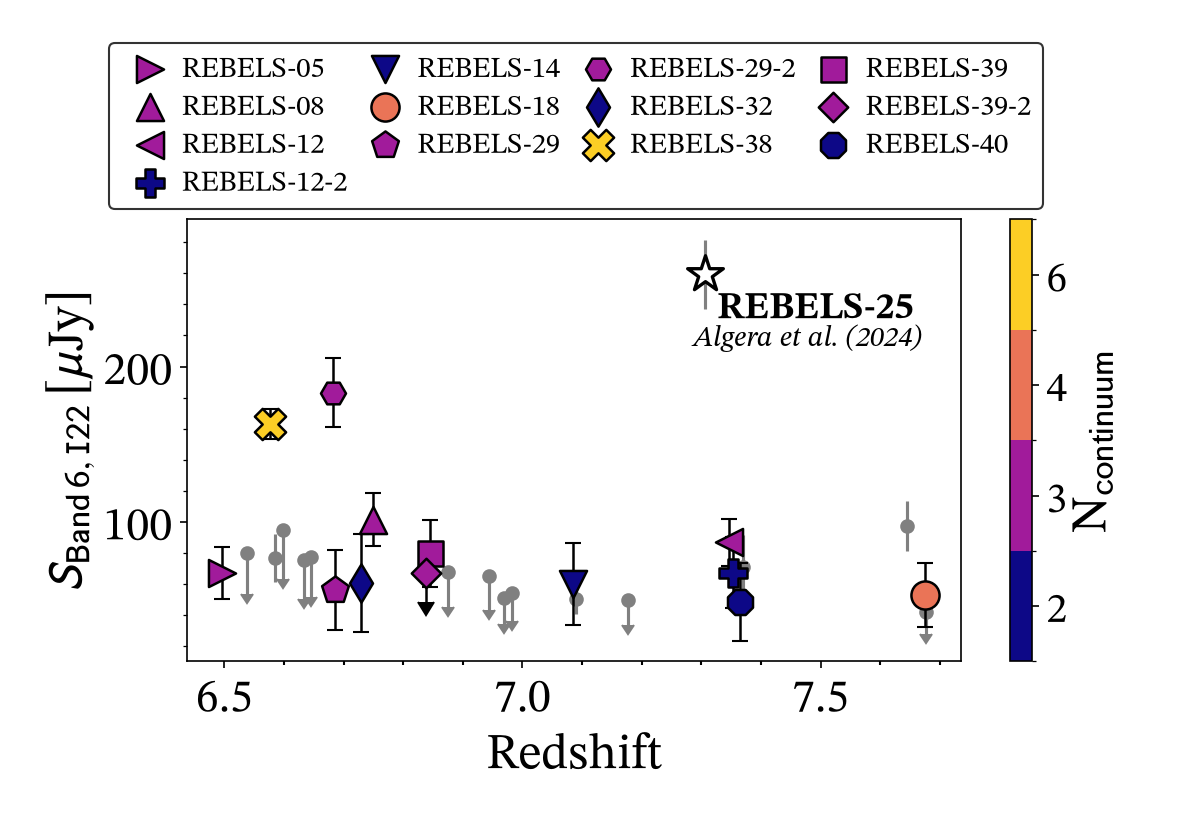}
    \caption{\textbf{The sample of 13 $z=6.5 - 7.7$ galaxies with multi-band ALMA observations studied in this work.} We show the original REBELS Band 6 continuum flux density from \citet{inami2022} versus spectroscopic \cii{}-based redshift. Our targets are shown through the larger symbols, and are colored by the number of unique ALMA projects in which they have been targeted (Table \ref{tab:observations}). Other \cii{}-detected REBELS galaxies are shown as the grey symbols. REBELS-25, highlighted through the black star, is not re-analyzed in this work.}
    \label{fig:sampleDefinition}
\end{figure}

\begin{table*}
    \def\arraystretch{1.4}
    \centering
    \caption{ALMA observations and naturally weighted continuum image properties of the REBELS galaxies.}
    \label{tab:observations}
    \begin{threeparttable}
    \begin{tabular}{lccccccl}
    
    \hline\hline
    ID & Band & $\nu_\mathrm{cont}$ (GHz) & $\sigma_\mathrm{cont}$ ($\mu$Jy beam$^{-1}$) & $\theta_M$ (arcsec) & $\theta_m$ (arcsec) & PA (deg) & Project ID \\
    \hline 
    
REBELS-05 & 6 & 245.63 & 6.9 & 0.36 & 0.29 & 38.0 & 2022.1.01131.S (Smit) \\
 & 8 & 405.03 & 35.2 & 0.77 & 0.67 & -86.9 & 2024.1.00406.S (Algera) \\
REBELS-08 & 6 & 237.25 & 6.5 & 0.37 & 0.32 & 49.6 & 2022.1.01131.S (Smit) \\
 & 8 & 405.01 & 39.1 & 0.77 & 0.67 & -75.1 & 2024.1.00406.S (Algera) \\
REBELS-12 & 8 & 401.37 & 64.9 & 0.61 & 0.45 & 58.5 & 2021.1.01297.S (Fudamoto) \\
 & 8 & 401.55 & 42.0 & 0.72 & 0.56 & -67.2 & 2024.1.00406.S (Algera) \\
REBELS-12-2 & 8 & 401.37 & 70.0 & 0.61 & 0.45 & 58.6 & 2021.1.01297.S (Fudamoto) \\
REBELS-14 & 8 & 412.77 & 33.7 & 0.90 & 0.68 & -69.9 & 2024.1.00406.S (Algera) \\
REBELS-18 & 6 & 229.57 & 6.2 & 1.35 & 1.18 & -77.5 & 2022.1.00446.S (Fudamoto) \\
 & 6 & 226.08 & 7.7 & 0.37 & 0.36 & 20.2 & 2022.1.01131.S (Smit) \\
 & 8 & 398.19 & 37.6 & 0.67 & 0.56 & 41.7 & 2024.1.00406.S (Algera) \\
REBELS-29 & 6 & 239.29 & 4.1 & 0.28 & 0.27 & 8.6 & 2022.1.01131.S (Smit) \\
 & 8 & 404.97 & 26.4 & 0.83 & 0.75 & -75.5 & 2024.1.00406.S (Algera) \\
REBELS-29-2 & 6 & 239.29 & 4.1 & 0.29 & 0.27 & 5.9 & 2022.1.01131.S (Smit) \\
 & 8 & 404.97 & 52.8 & 0.83 & 0.75 & -75.5 & 2024.1.00406.S (Algera) \\
REBELS-32 & 8 & 404.97 & 31.0 & 0.85 & 0.74 & 76.5 & 2024.1.00406.S (Algera) \\
REBELS-38 & 3 & 99.08 & 4.2 & 1.69 & 1.34 & -54.4 & 2021.1.01389.S (Gonzalez) \\
 & 5 & 199.87 & 10.6 & 1.14 & 0.91 & -67.3 & 2022.1.01384.S (Fudamoto) \\
 & 6 & 263.98 & 10.5 & 1.33 & 1.04 & 72.4 & 2022.1.00446.S (Fudamoto) \\
 & 6 & 243.13 & 12.6 & 0.61 & 0.52 & 45.8 & 2022.1.01131.S (Smit) \\
 & 8 & 404.98 & 42.4 & 0.61 & 0.48 & 56.8 & 2021.1.00318.S (Inami) \\
REBELS-39 & 8 & 425.65 & 49.3 & 0.53 & 0.40 & -85.5 & 2021.1.01297.S (Fudamoto) \\
 & 8 & 425.58 & 42.7 & 0.77 & 0.70 & -88.4 & 2024.1.00406.S (Algera) \\
REBELS-39-2 & 8 & 425.65 & 52.4 & 0.53 & 0.40 & -85.5 & 2021.1.01297.S (Fudamoto) \\
 & 8 & 425.58 & 45.6 & 0.77 & 0.70 & -88.4 & 2024.1.00406.S (Algera) \\
REBELS-40 & 8 & 400.54 & 65.5 & 0.82 & 0.74 & 85.4 & 2021.1.00318.S (Inami) \\

\hline\hline 
\end{tabular}
    \textit{Notes}: By construction, all REBELS galaxies have also been observed as part of the Large Program (2019.1.01634.L; PI Bouwens). We here list only follow-up observations, obtained in later ALMA cycles, while the continuum properties of the original REBELS Band 6 data can be found in \citet{inami2022}. Details of the follow-up observations of REBELS-25 -- the fourteenth REBELS galaxy with multi-band data -- can be found in \citet{algera2024b}, \citet{cescon2026}, and \citet{rowland2026_r25}. For sources outside the pointing center, the quoted RMS is corrected for primary beam attenuation and thus represents the local noise.
\end{threeparttable}
\end{table*}

As the main goal of this work is to study the dust properties of REBELS galaxies through multi-band continuum data, we here focus only on those with observations in at least two distinct ALMA bands, which is the case for 14 REBELS galaxies in total.\footnote{We note that REBELS-18 was observed across a contiguous frequency range in both Bands 5 and 6 as part of the original REBELS \cii{} spectral scans \citep{bouwens2022}. Following \citet{inami2022}, these observations have been concatenated into a single continuum map at $\nu_\mathrm{obs} \approx 200\,\mathrm{GHz}$. This nominally falls into the Band 5 receiver range; however, we will collectively refer to the LP data as `Band 6' for simplicity, and count these observations as being conducted in a single band. The exact central frequencies of the REBELS LP data can be found in \citet{inami2022}, while those of all follow-up observations are given in Table \ref{tab:observations}.\label{footnote:rebels18}} All of these have a robust spectroscopic redshift in the range $z=6.5 - 7.7$. Eleven of these are `primary' REBELS targets, being UV-luminous sources that were explicitly targeted in the LP, while the remaining three sources (REBELS-12-2, REBELS-29-2 and REBELS-39-2) were serendipitously found in the LP data through a blind detection of their \cii{} lines \citep{fudamoto2021,vanleeuwen2024}. 

Ten of the eleven primary targets -- that is, all except for REBELS-40 -- have moreover been targeted by \textit{JWST}/NIRSpec in Integral Field Unit (IFU) mode as part of the REBELS-IFU program \citep[][]{fisher2025,algera2026,rowland2026}. This program delivered low-resolution ($R\sim100$) prism spectroscopy of twelve REBELS galaxies in total, ten of which are detected in dust continuum emission. Stellar masses and metallicities for the REBELS-IFU sample have been derived in \citet[][see also Stefanon et al.\ in preparation]{rowland2026}. For REBELS-40, we rely on the pre-\textit{JWST} stellar mass from \citet{topping2022}, while the stellar masses of the serendipitous sample are derived in Van Leeuwen et al.\ (in preparation) based on NIRCam photometry obtained as part of GO 6480 (PI Schouws). 

One of the 14 REBELS sources with multi-band observations is the $z=7.31$ galaxy REBELS-25, whose dust SED has already been studied across six ALMA bands by \citet{algera2024b}, on spatially resolved scales by \citet{rowland2026_r25}, and jointly with its gas content by \citet{cescon2026}. We therefore do not re-analyze REBELS-25 in this work, but include its published dust properties in our comparison sample of high-redshift galaxies (Section~\ref{sec:discussion}). While dual-band ALMA observations of REBELS-12 and REBELS-38 have previously been presented by \citet{algera2024}, new data have since been obtained for both, and we therefore re-analyze their dust SEDs here.

In total, we thus analyze multi-band ALMA observations of 13 REBELS galaxies in this work, although one of the serendipitous \cii{}-emitters (REBELS-39-2) remains undetected in dust continuum in all bands (Section \ref{sec:methods_source_detection}), leaving 12 galaxies for which dust properties can be inferred. The list of ALMA programs from which these observations were drawn is provided in Table \ref{tab:observations}. We note that we include only ALMA projects conducted prior to Cycle 12.

All 13 galaxies have observations in ALMA Band 8, drawn from three separate projects (2021.1.00318.S, 2021.1.01297.S and 2024.1.00406.S; PIs Inami, Fudamoto, and Algera, respectively). Six also have Band 6 observations from the high-resolution project 2022.1.01131.S (PI Smit), which provide deeper continuum imaging than the original REBELS LP data. Two galaxies, REBELS-18 and REBELS-38, have additional ALMA Band 6 observations from 2022.1.00446.S (PI Fudamoto), targeting the \oi{} $145\,\mu\mathrm{m}$ line. REBELS-38 moreover has been targeted in \nii{} $205\,\mu\mathrm{m}$ and underlying continuum in ALMA Band 5 (2022.1.01384.S, PI Fudamoto), as well as in CO(7--6), CO(6--5) and $\mathrm{[C\,I]}(2$--$1)$ in ALMA Band 3 (2021.1.01389.S, PI Gonzalez).

We show the multi-band REBELS sample in Figure \ref{fig:sampleDefinition}, in terms of its redshift range ($z=6.5 - 7.7$) and Band 6 flux density from \citet{inami2022}. The galaxies are color-coded by the number of unique ALMA projects in which they have been targeted -- including the LP data itself -- which ranges from two (four sources) to six (REBELS-38).

\section{Calibration and imaging}
\label{sec:observations}

\begin{figure}
    \centering
    \includegraphics[width=0.5\textwidth]{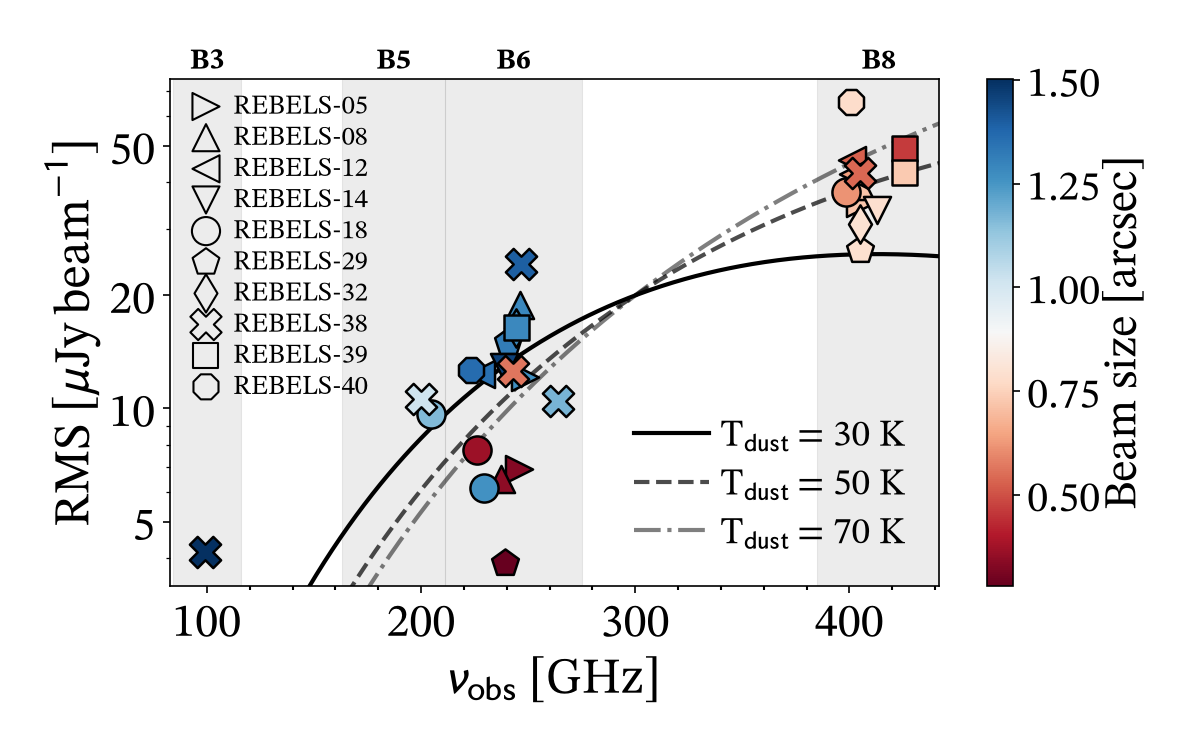}
    \caption{\textbf{The depth, frequency coverage and native angular resolution of the multi-band data presented in this work.} We show the continuum RMS ($1\sigma$) at the phase center versus observing frequency, colored by naturally weighted angular resolution. Overplotted are three $z=7$ modified blackbody models with $T_\mathrm{dust} = 30, 50$ and $70\,\mathrm{K}$ and $\beta_\mathrm{IR} = 2.0$, arbitrarily normalized at $\nu_\mathrm{obs} = 300\,\mathrm{GHz}$ and corrected for the cosmic microwave background. The grey shading shows the nominal frequency ranges of the ALMA bands used in this work.}
    \label{fig:multiBandDepths}
\end{figure}

In this work, we image and analyze the ALMA observations of the multi-band REBELS sample in a homogeneous manner. For the observations taken after the LP, we retrieve the visibility data from the ALMA archive, and restore the pipeline calibrations using the standard {\texttt{scriptForPI.py}} in the appropriate version of {\sc{CASA}} \citep{casa2022}. For the LP data, we start from the pipeline-calibrated measurement sets created by Schouws et al.\ (in preparation), which have been used throughout the series of observational REBELS papers \citep[e.g.,][]{fudamoto2022,heintz2022,inami2022,algera2023,barrufet2023,hygate2023,aravena2024,bowler2024,herard-demanche2026}. 

\subsection{Excluding emission line-contaminated channels}
\label{sec:observations_line_exclusion}

\begin{figure*}
    \centering
    \includegraphics[width=0.92\textwidth]{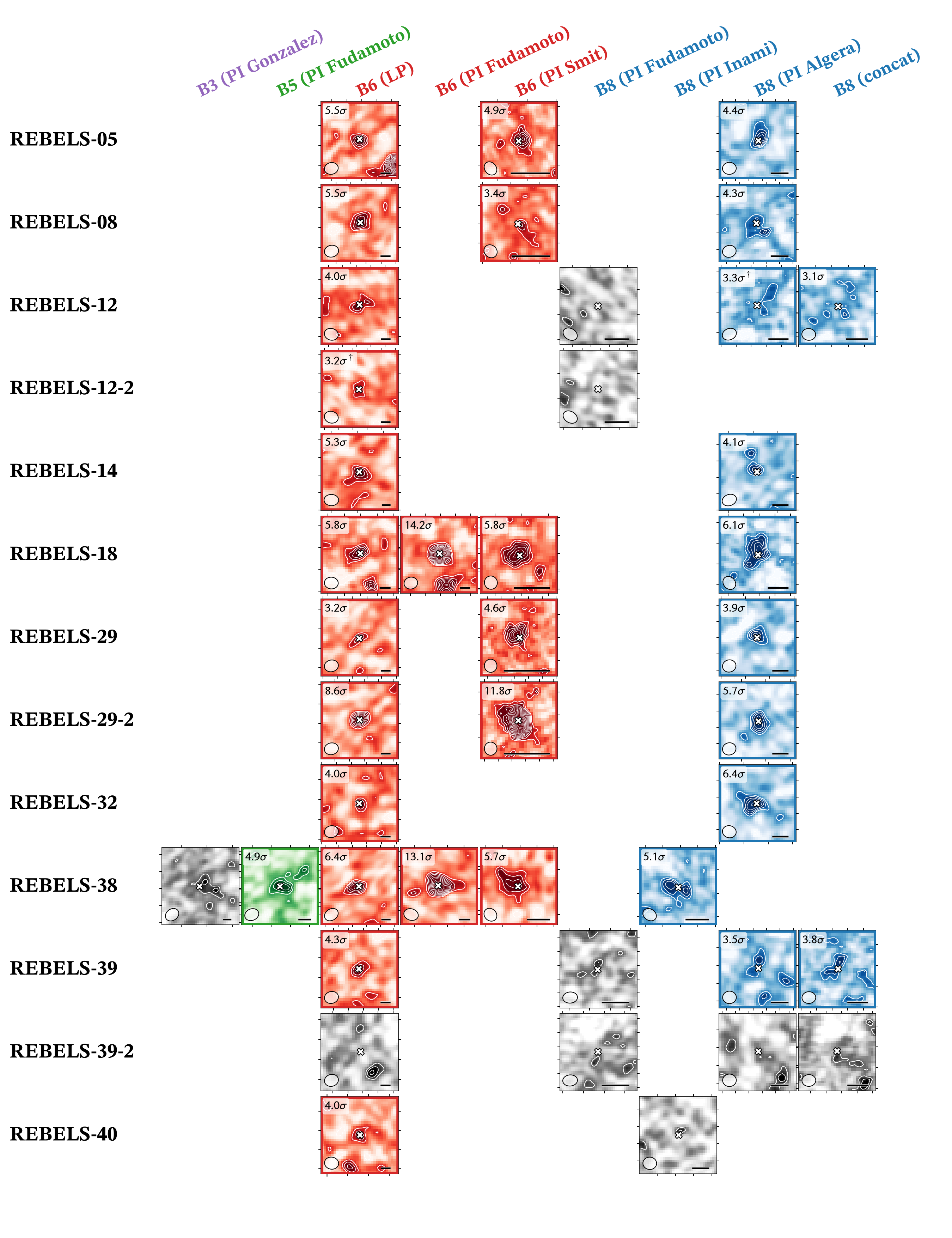}
    \vspace*{-1cm}\caption{\textbf{Compilation of the naturally weighted ALMA dust continuum observations for the multi-band REBELS sample presented in this work.} Rows represent different targets ($N=13$), while the columns represent different projects (except the final column, which is the concatenation of the Band 8 projects from PI Fudamoto and PI Algera; Section \ref{sec:methods_concat}). Cutouts are colored by ALMA band, or are shown in grey if no dust emission is detected. Numbers in the top-left corner indicate the peak-pixel S/N of each detection, measured in the $1.5''$-tapered images; daggers mark sources that are not detected at $>3\sigma$ in the naturally weighted images. Contours are shown at the $2,3,4,\ldots\sigma$ level, where $\sigma$ is the continuum RMS, and the color scale ranges from $-1.5\sigma$ (white) to $3.5\sigma$ (color). The white cross highlights the fiducial dust continuum centroid of our targets, as determined in Section \ref{sec:centroids}. Given the spread in native beam sizes of the data, the angular scale is different across the cutouts; the black scalebar in the lower right corner represents $1''$ in each panel.}
    \label{fig:allCutouts}
\end{figure*}

Several of the datasets presented in this work simultaneously target line and continuum emission. For accurate continuum measurements, it is therefore necessary to exclude the channels `contaminated' by bright emission lines. We do so by flagging all channels within one line full width at half maximum (FWHM) of the line center, i.e., a window spanning twice the FWHM.\footnote{For REBELS-39 and REBELS-39-2 we are slightly more conservative: since these two galaxies are separated by just $\sim2''$, we simultaneously exclude the channels spanning both emission lines to prevent residual line emission from one galaxy blending with the other's continuum.} For a Gaussian profile, this removes $\sim98\%$ of the total line flux, thus leaving essentially only the continuum.

By construction, all of the REBELS LP data used in this work contain the \cii{} line, and we adopt the \cii{} redshifts and FWHMs obtained by Schouws et al.\ (in preparation) to exclude the emission lines. We adopt the same set of parameters for the high-resolution \cii{} observations conducted as part of 2022.1.01131.S (PI Smit), which are available for six of our targets. A subset of seven REBELS sources observed as part of the various ALMA Band 8 projects (2021.1.00318.S, 2021.1.01297.S, and 2024.1.00406.S) were furthermore targeted in the \oiiil{} line.\footnote{This subset consists of REBELS-12, REBELS-12-2, REBELS-14, REBELS-18, REBELS-39, REBELS-39-2 and REBELS-40, all of which were detected in the \oiiil{} line.} Their \oiiinowave{}-based redshifts and FWHMs were presented by \citet{algera2025_oiii} and these values are adopted here to remove the line-contaminated channels. 

As part of 2022.1.00446.S (PI Fudamoto), REBELS-18 and REBELS-38 were targeted in the \oi{} $145\,\mu\mathrm{m}$ line. As described in \citet{fudamoto2025}, both galaxies were detected in \oi{}, and we adopt their measured FWHMs to exclude the line emission from the data. REBELS-38 was moreover targeted in the \nii{} $205\,\mu\mathrm{m}$ line in ALMA Band 5, but as discussed in \citet{fudamoto2025}, the line was not detected, and as such we do not exclude any channels prior to continuum imaging.

Finally, REBELS-38 has been targeted in the CO(7--6), CO(6--5) and $\mathrm{[C\,I]}(2$--$1)$ lines in ALMA Band 3 as part of 2021.1.01389.S (PI Gonzalez). These data have not previously been presented, and we therefore first assess whether any emission lines are detected. We create a datacube using the {\sc{tclean}} task in {\sc{CASA}}, adopting a velocity resolution of $30\,\mathrm{km\,s^{-1}}$ and Briggs weighting with a robust parameter of $2.0$, which is close to natural weighting \citep{briggs1995}. We search for the CO and $\mathrm{[C\,I]}$ line emission by creating moment-0 maps around the expected line frequencies, as well as by extracting aperture spectra centered on the \cii{} position of REBELS-38, but find no clear line emission above the $3\sigma$ level. We therefore do not exclude any channels when imaging the Band 3 continuum emission.

\subsection{Continuum imaging}
\label{sec:observations_continuum_imaging}

Armed with line-free measurement sets, we proceed by creating continuum images of our targets. We make use of the {\sc{tclean}} task, adopting the standard H\"{o}gbom deconvolver, and use Briggs weighting with a robust parameter equal to 2.0 to maximize sensitivity. This is akin to natural weighting, and henceforth we will therefore refer to these as the `naturally weighted' images. Cleaning is performed down to $2.0\sigma$, where $\sigma$ is the root-mean-square (RMS) noise obtained from the dirty image based on $3.5\sigma$ clipping. We utilize automasking with the recommended parameters for ALMA data, following \citet{kepley2020}, and thus clean any sufficiently bright sources in the field of view.

\begin{figure*}[!t]
    \centering
    \includegraphics[width=0.95\textwidth]{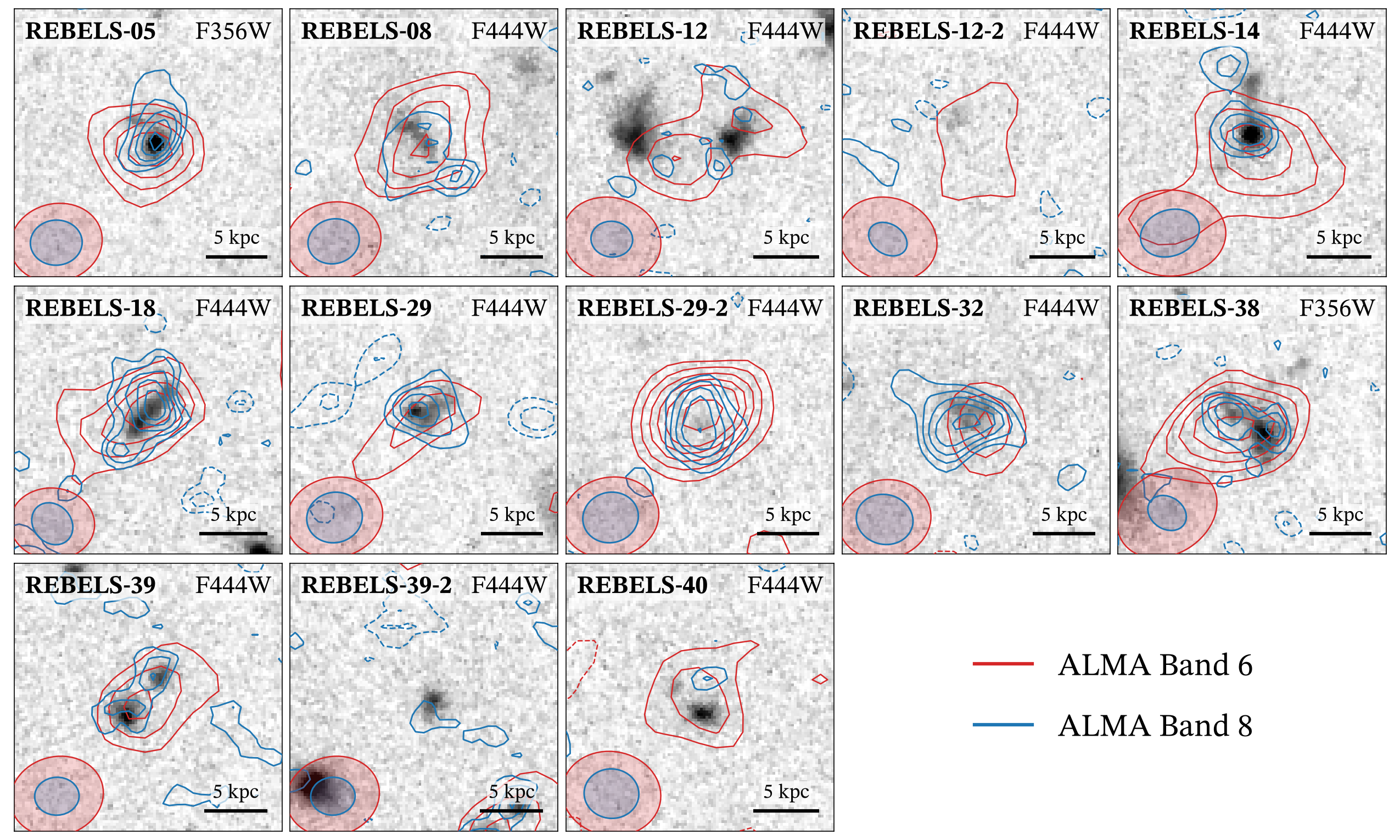}
    \caption{\textbf{Naturally weighted ALMA Band 6 (red) and Band 8 (blue) contours on NIRCam F356W or F444W imaging.} The Band 6 data correspond to those from the REBELS Large Program \citep{bouwens2022,inami2022}, while the Band 8 continua are newly presented in this work (except for REBELS-38, where the same data as in \citealt{algera2024} are shown). Cutouts span $4''\times4''$, and contour levels represent $2, 3, 4, 5, 6, 8, 10\sigma$. The beam sizes in both ALMA bands are shown in the lower left corner, and a scalebar of $5\,\mathrm{kpc}$ is indicated on the lower right. A beam-matched variant, using $1.5''$-tapered images, is shown in Figure \ref{fig:NIRCamALMAoverlaysTapered} (Appendix \ref{app:tapered_imaging}).}
    \label{fig:NIRCamALMAoverlays}
\end{figure*}

We detail the image properties, such as their continuum frequency, beam size, and RMS in Table \ref{tab:observations}. We moreover show the continuum RMS as a function of frequency and angular resolution in Figure \ref{fig:multiBandDepths}, and present the naturally weighted images themselves in Figure \ref{fig:allCutouts}. We also show the ALMA Band 6 and 8 contours on top of NIRCam F356W or F444W imaging from GO 6480 (PI Schouws) in Figure \ref{fig:NIRCamALMAoverlays}. Throughout this work, we use the naturally weighted images to present the data, while the tapered maps described below serve primarily as a beam-matched tool for source detection and for determining the dust continuum centroids (Section \ref{sec:methods_source_detection}). Our fiducial flux densities are measured directly from the visibilities (Section \ref{sec:methods_fluxdensity_measurements}), for which the naturally weighted images provide the closest image-plane equivalent in terms of noise properties and angular resolution.

As is clear from Figure \ref{fig:multiBandDepths}, the naturally weighted data are inhomogeneous in terms of RMS and beam size, as expected given that the observations are drawn from eight separate ALMA projects. The median RMS at the phase center of the Band 6 images equals $\sigma_\mathrm{nat} = 12.4\,\mu\mathrm{Jy\,beam}^{-1}$, with a range of $\sigma_\mathrm{nat} = 3.9 - 24.2\,\mu\mathrm{Jy\,beam}^{-1}$, while for the Band 8 data this is $\sigma_\mathrm{nat} = 40.6\,\mu\mathrm{Jy\,beam}^{-1}$ with a range of $\sigma_\mathrm{nat} = 26.4 - 65.5\,\mu\mathrm{Jy\,beam}^{-1}$. The resolution of the Band 6 continuum data varies considerably, from $>1''$ for the REBELS LP data and \oi{} follow-up to $\lesssim 0.6''$ for the high-resolution \cii{} follow-up. The resolution of the Band 8 observations is more homogeneous, with a median beam size of $0.77'' \times 0.67''$, and a range of $\approx 0.4'' - 0.9''$. 

To homogenize the resolution between the various datasets presented in this work -- which in turn facilitates uniform flux density measurements across multiple bands (Section \ref{sec:methods_fluxdensity_measurements} and Appendix \ref{app:flux_measurement_comparison}) -- we also create tapered continuum maps by specifying the {\sc{uvtaper}} parameter in {\sc{tclean}}. However, by down-weighting the contribution of long baselines, tapering reduces the overall continuum sensitivity -- in particular for the high-resolution data, where long baselines make up a larger fraction of the total visibilities. 

We create a range of tapered images with target resolutions between the naturally weighted beam size and $2.5''$, in steps of $0.1''$. In what follows, we will focus specifically on images tapered to a resolution of $1.5''$, which is approximately the lowest common resolution across the various datasets.\footnote{The naturally weighted beams of REBELS-14 (Band 6 LP data; \citealt{inami2022}) and REBELS-38 (Band 3; Table \ref{tab:observations}) are already $\sim1.5''$, albeit asymmetric. For these two datasets, we use the naturally weighted images in place of the $1.5''$-tapered maps.} We show the $1.5''$-tapered maps in Figure \ref{fig:allCutoutsTapered}, and overlay the Band 6 and 8 contours on NIRCam imaging in Figure \ref{fig:NIRCamALMAoverlaysTapered}. The median RMS at the phase center of the tapered Band 6 maps equals $\sigma_{1.5''} = 15.6\,\mu\mathrm{Jy\,beam}^{-1}$ with a range of $6.7 - 25.1\,\mu\mathrm{Jy\,beam}^{-1}$. For the Band 8 data, the median tapered RMS equals $\sigma_{1.5''} = 68.3\,\mu\mathrm{Jy\,beam}^{-1}$, with a range of $40.2 - 109\,\mu\mathrm{Jy\,beam}^{-1}$. The median beam size of the tapered maps equals $1.53'' \times 1.47''$, approximately equal to the requested $1.5''$.

\section{Methods}
\label{sec:methods}

\subsection{Source detection}
\label{sec:methods_source_detection}

\begin{figure}
    \centering
    \includegraphics[width=0.5\textwidth]{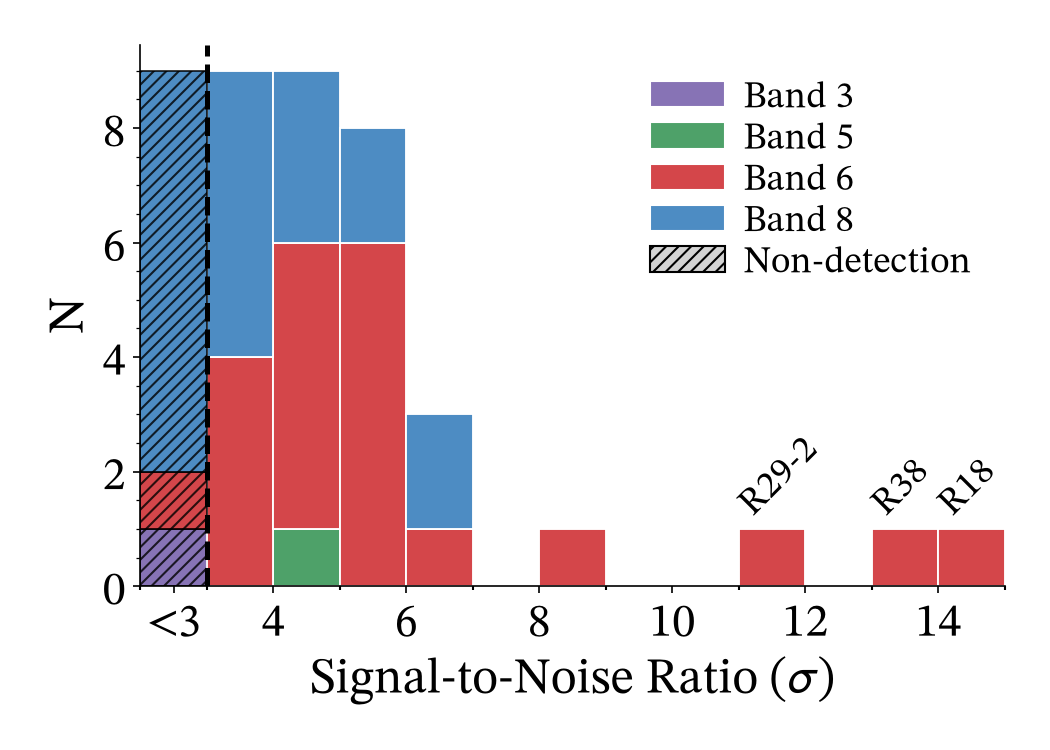}
    \vspace*{-0.6cm}
    \caption{\textbf{Our targets span a range of signal-to-noise ratios across the various ALMA bands.} The median S/N, measured as the peak flux density in $1.5''$-tapered imaging divided by the RMS, equals $4.4\sigma$ and increases to $4.9\sigma$ upon excluding the non-detections. The highest-significance detections are all in Band 6.}
    \vspace*{0.2cm}
    \label{fig:snr_hist}
\end{figure}

We first aim to establish which of our targets are detected in the multi-band imaging, recalling that all primary REBELS galaxies analyzed in this work are continuum-detected in the original (naturally weighted) LP data \citep{inami2022}. This is not by construction, as our sample consists of any REBELS galaxies with multi-band ALMA data; however, it turns out that so far only dust-detected REBELS sources have been followed up in additional bands, and this selection bias should be borne in mind throughout this work.

We use a similar approach to that adopted in the original REBELS continuum detection paper \citep{inami2022}, and run the Python Blob Detector and Source Finder \citep[{\sc{PyBDSF}};][]{mohanrafferty2015} on the full set of multi-band ALMA images. {\sc{PyBDSF}} searches for islands of contiguous emission above some user-specified threshold, and fits detected sources with one or multiple Gaussian components. We adopt a peak and island threshold of {\sc{thresh\_pix}}~$=3.0$ and {\sc{thresh\_isl}}~$=2.0$, respectively, the former being slightly lower than the $3.3\sigma$ peak threshold used by \citet{inami2022}. This is motivated by the fact that \citeauthor{inami2022} performed a blind search for dust emission within the central $1.5''$ of all REBELS ALMA maps, whereas we search for multi-band emission from galaxies already known to be dusty. We list a source as detected if {\sc{PyBDSF}} finds a continuum counterpart within $1.0''$ of the nominal \cii{} position of the galaxy, though in practice all counterparts are found within $0.5''$ (Section \ref{sec:centroids}). 

\medskip

Given the wide range of native angular resolutions of the ALMA data, we run {\sc{PyBDSF}} on both the naturally weighted and the $1.5''$-tapered images. The former are most sensitive to compact emission, as tapering increases the noise, while the latter can recover faint, extended emission in high-resolution data that may otherwise remain below the detection threshold if diluted across multiple beams. We consider a source detected if it is identified in either set of images. In practice, all sources detected in the naturally weighted images are also recovered in the tapered maps, such that adopting only the tapered imaging would yield identical results.

In the naturally weighted maps, we recover all continuum sources identified in the original REBELS Band 6 imaging. The only two sources undetected in the LP data are the serendipitous neighbors REBELS-12-2 and REBELS-39-2. While neither of the two was previously reported to show dust emission at the $>3\sigma$ level, \citet{fudamoto2021} identified a tentative, possibly extended $2.8\sigma$ signal at the position of REBELS-12-2, which we recover at $3.2\sigma$ in the $1.5''$-tapered LP data (Figures \ref{fig:allCutouts} and \ref{fig:NIRCamALMAoverlays}).

In the follow-up observations, we identify counterparts in all naturally weighted images except for seven Band 8 datasets, covering REBELS-12 ($2\times$), REBELS-12-2, REBELS-39, REBELS-39-2 ($2\times$) and REBELS-40. Tapering to $1.5''$ yields one additional detection: REBELS-12 at $3.3\sigma$ in the Band 8 data from PI Algera. REBELS-39, while undetected in the shallower data from PI Fudamoto, is detected in the observations from PI Algera and more significantly so in their concatenation (Section \ref{sec:methods_concat}). Finally, REBELS-38 is additionally detected in Bands 5 and 3, although the latter detection is likely spurious (Section \ref{sec:centroids}), and we therefore treat it as an upper limit in what follows.

\medskip

In summary, all 10 primary REBELS targets are continuum-detected in at least one -- and up to three -- Band 6 projects. All but REBELS-40 are also detected in precisely one Band 8 project. REBELS-38 is moreover detected in Band 5. Among the three serendipitous neighbors, REBELS-29-2 is detected at high significance in both Bands 6 and 8, REBELS-12-2 only in Band 6, and REBELS-39-2 not at all.

We show a histogram of the detection significance -- defined as the peak flux density in the $1.5''$-tapered imaging divided by the RMS -- across all projects in Figure~\ref{fig:snr_hist}. The median signal-to-noise ratio (S/N) equals $4.4$, or $4.9$ when non-detections are excluded. An S/N above 10 is reached only in the Band 6 observations targeting \oi{} $145\,\mu\mathrm{m}$ in REBELS-18 and REBELS-38 (PI Fudamoto), and in the high-resolution \cii{} data of REBELS-29-2 (PI Smit). In Band 8, the ten continuum-detected sources reach $\mathrm{S/N} = 3.3-6.4$. 

Excluding REBELS-38, which was previously detected in Band 8 by \citet{algera2024}, this work thus presents Band 8 continuum detections of eight $z > 6.5$ LBGs -- in addition to the serendipitous \cii{} emitter REBELS-29-2 -- for the first time, increasing the number of such galaxies with Band 8 continuum detections from six to 14 \citep[cf.][see also Section \ref{sec:discussion}]{hashimoto2019,inoue2020,witstok2022,algera2024,harshan2024}.

\subsubsection{Concatenating similar observations}
\label{sec:methods_concat}

Given that some of our targets have been observed in the same band by multiple ALMA projects, these datasets can in principle be concatenated to obtain deeper continuum constraints. However, collapsing observations at different frequencies into a single continuum image discards information on the spectral slope, even within the same band. Moreover, separate datasets at similar frequencies -- in particular when taken at different native resolutions -- are useful for assessing the accuracy of the flux density measurements, for example by revealing offsets due to resolved-out flux or residual calibration errors (Appendix~\ref{app:fluxdensity_measurements_B6consistency}).

We therefore opt to concatenate data only if two conditions are met: 1) the central frequencies of the observations differ by less than $5\%$, and 2) the native beam sizes of the data differ by less than a factor of two.\footnote{Specifically, we require the frequency offset between two observations $(i,j)$ to be $|\nu_{i} - \nu_{j}| / \max(\nu_{i}, \nu_{j}) < 0.05$.} This is only the case for the Band 8 observations targeting REBELS-12, REBELS-39 and REBELS-39-2. The Cycle 8 (PI Fudamoto) and Cycle 11 (PI Algera) observations share a near-identical frequency setup covering the \oiiil{} line and underlying continuum, though the latter are $1.55\times$ and $1.15\times$ deeper at the positions of REBELS-12 and REBELS-39, respectively. We note that REBELS-12-2 is not covered by the Cycle 11 data, whereas the Cycle 8 data adopted a pointing center in between REBELS-12 and its serendipitous neighbor \citep[for details, see][]{algera2025_oiii}.

We first run the {\sc{CASA}} task {\sc{statwt}} on the line-free datasets to reweight the visibilities according to their scatter, prior to combining them via {\sc{concat}}. We subsequently re-image the data following the approach outlined in Section \ref{sec:observations_continuum_imaging}. The concatenated continuum images are shown in the last column of Figures \ref{fig:allCutouts} (naturally weighted) and \ref{fig:allCutoutsTapered} (tapered). The concatenated images are also overplotted on NIRCam imaging in Figures \ref{fig:NIRCamALMAoverlays} and \ref{fig:NIRCamALMAoverlaysTapered}. 

Compared to the Cycle 11 Band 8 data, the $1.5''$-tapered, concatenated images are $\sim1.1-1.2\times$ deeper at the position of our targets. While REBELS-39-2 remains undetected in the concatenated dataset, the detection significance of REBELS-39 increases from $3.5\sigma$ in the Cycle 11 data to $3.8\sigma$ in the combined observations. For REBELS-12, on the other hand, the detection significance slightly decreases from $3.3\sigma$ in the Cycle 11 data to $3.1\sigma$ in the concatenated data. This could suggest that the flux in the Cycle 11 data is boosted by noise, or that the signal is spurious altogether. However, given that the Band 6 and 8 continua share similar morphologies once tapered to a common $1.5''$ resolution (this is most readily seen in Figure~\ref{fig:NIRCamALMAoverlaysTapered}), we consider the Band 8 emission likely to be genuine.

\subsubsection{Dust continuum centroids}
\label{sec:centroids}

In Section \ref{sec:methods_fluxdensity_measurements} and Appendix \ref{app:flux_measurement_comparison} we perform homogeneous flux density measurements across the full set of multi-band observations, including through aperture photometry and $(u,v)$-plane fitting. These techniques require a consistent definition of the center of the galaxy, and we thus first define one global dust continuum centroid for each target. 

We compile the centroids obtained by {\sc{PyBDSF}} -- which employs 2D Gaussian fitting -- for each band, utilizing the $1.5''$-tapered images. We choose to taper to a consistent, coarse resolution to ensure our targets indeed show smooth emission profiles that resemble Gaussians (Figure \ref{fig:allCutoutsTapered}). We then adopt the inverse-variance-weighted position across the available multi-band detections as the fiducial dust continuum centroid. For our two targets detected in only a single band (REBELS-12-2 and REBELS-40), we simply adopt the Band 6 centroid, while for REBELS-39-2, which is not continuum-detected at all, we adopt the \cii{} centroid instead.

We show the per-band offsets with respect to the fiducial centroid for each of our targets in Figure \ref{fig:positionalOffsets} in Appendix \ref{app:spatialOffsets}. Overall, all multi-band continuum positions are consistent with this centroid within $0.5''$, except for the Band 3 emission from REBELS-38 which is offset by $\sim0.8''$. Indeed, this emission has a peculiar, extended morphology that is not seen at shorter wavelengths (Figure \ref{fig:allCutouts}). Taken at face value, the inferred flux density (Section \ref{sec:methods_fluxdensity_measurements}) moreover exceeds that expected from a standard modified blackbody (MBB), which further suggests that the Band 3 emission is boosted by, or entirely due to, a noise peak close to REBELS-38. We therefore exclude it from the centroid determination, and treat it as an upper limit in what follows.

Across the full set of observations, the mean separation with respect to the inverse-variance-weighted centroid is $0.17'' \pm 0.10''$, where the uncertainty represents the standard deviation. This thus confirms that the positions of our targets are consistent between the different bands, and supports the reality of their multi-band counterparts.

\subsection{Flux density measurements}
\label{sec:methods_fluxdensity_measurements}

There are multiple ways in which flux densities can be measured from radio interferometric data, each with their own advantages and disadvantages. We discuss these in detail in Appendix \ref{app:flux_measurement_comparison}, while we here adopt flux densities from $(u,v)$-plane model fitting as our fiducial approach. On average, these correspond to $0.81\times$ and $0.89\times$ the integrated flux densities from 2D Gaussian fitting to the naturally weighted and $1.5''$-tapered images, respectively, while being slightly larger than aperture flux densities measured from the same images (by $1.04\times$ and $1.10\times$). Finally, the $(u,v)$-plane flux densities are $1.19\times$ larger than peak flux densities, which treat the galaxies as spatially unresolved (Figures~\ref{fig:flux_comparison_matrix} and \ref{fig:flux_comparison_B6_B8}).

In practice, we fit an exponential disk model in the $(u,v)$-plane, following Astles et al.\ (in preparation), who constrain the dust and \cii{} sizes and morphologies of the subset of REBELS galaxies with high-resolution Band 6 observations. We here adopt this approach for the full REBELS sample with multi-band data, though solely to measure their continuum flux densities; a discussion of multi-band dust continuum sizes is left for a separate work.

To facilitate this $(u,v)$-plane analysis, it is necessary to first remove any bright non-target sources from the visibility data. We identify these sources by re-running {\sc{PyBDSF}} on the naturally weighted images with conservative detection thresholds ({\sc{thresh\_pix}} = 5.0 and {\sc{thresh\_isl}} = 3.0), which are more robust against picking up off-center noise peaks and thus identify only sufficiently bright sources that could potentially contaminate the flux measurements of our targets (see also Nozawa et al.\ in preparation for a discussion of the serendipitous continuum sources in the REBELS LP Band 6 pointings). We write out a binary segmentation map where {\sc{PyBDSF}} has identified significant emission, and ensure that a circular region with radius $1.5''$ around the sky position of our target is unmasked. We then utilize this as a clean mask for imaging in {\sc{tclean}}, following the approach in Section \ref{sec:observations_continuum_imaging}. Essentially, this thus ensures that bright, non-target sources are cleaned, while the main target is not. Serendipitous companions are treated as non-target sources when fitting the primary target, and vice versa.

We subsequently subtract the cleaned non-target sources from the visibility data using the task {\sc{uvsub}} in {\sc{CASA}} and shift the phase center to the continuum centroid measured in Section \ref{sec:centroids} using the task {\sc{phaseshift}}. We next reduce the measurement set to a single frequency channel using the task {\sc{split}}, and also apply $30\,\mathrm{s}$ time-averaging to reduce the data volume. The visibility data are subsequently exported to an ASCII file via the {\sc{uvplot}} library \citep{tazzari2017}. We fit the azimuthally averaged real visibilities, denoted $\mathrm{Re}(V)$ and binned in steps of $1\,\mathrm{k}\lambda$, with an exponential disk profile following the Bayesian approach from Astles et al.\ (in preparation) based on the nested-sampling code {\sc{PyMultiNest}} \citep{feroz2009,buchner2014}. 

In the fitting, we adopt a prior on the half-light radius that is uniform in $\log R$ between $0.1$ and $5.0\,\mathrm{kpc}$, and require the amplitude to be non-negative. Adopting a prior that is uniform in $R$ places stronger weight on larger sizes and would yield fluxes that are, on average, $\sim14\%$ larger. The flux density of each target is given by the amplitude of the exponential disk model at zero $(u,v)$-distance, marginalized over the source size, and -- where necessary -- corrected for primary beam attenuation. 

We find that the native visibility weights do not fully capture the noise in our data: even after re-computing the weights with the {\sc{CASA}} task {\sc{statwt}}, the RMS noise of the naturally weighted images exceeds that expected from the visibility weights by up to a factor of two. As the image-plane RMS agrees with the noise predicted from the on-source time and precipitable water vapor (PWV), we attribute this discrepancy to the visibility weights, and rescale them accordingly. In practice, we fit each dataset twice: after an initial fit, we rescale the weights by a single factor such that the reduced $\chi^2$ of the best-fitting model is unity, and repeat the fit with these rescaled weights. This rescaling inflates the flux density uncertainties, bringing them in line with those of the alternative flux density measurements (Appendix \ref{app:flux_measurement_comparison}), but leaves the best-fitting flux densities unchanged. We present the fiducial $(u,v)$-plane flux densities in Table \ref{tab:flux_measurements} in Appendix \ref{app:fluxdensity_table} and demonstrate that our Band 6 LP flux measurements agree with those by \citet{inami2022} in Appendix \ref{app:comparison_inami2022}.

We note that this $(u,v)$-plane fitting approach yields a flux measurement even for galaxies not formally detected in the image plane. We confirm that this $(u,v)$-plane measurement is consistent with the $3\sigma$ image-plane upper limits in all cases (Figure \ref{fig:flux_comparison_B6_B8}). Since the fitted fluxes retain more information than an upper limit, we adopt these flux densities and their (large) uncertainties directly in the MBB fits in Section~\ref{sec:results_MBB_fits}, with the exception of the Band 3 data of REBELS-38 which we deem unreliable (Section~\ref{sec:centroids}).

As a further check on our flux measurements, we compare the flux densities of the six targets with multiple independent observations in Bands 5 and 6 in Appendix~\ref{app:fluxdensity_measurements_B6consistency}, where we show them as a function of observed frequency together with MBB models spanning a range of plausible dust temperatures (Figure~\ref{fig:fluxComparisonB6}). In all cases, the measurements are consistent with each other given the expected spectral slope. The largest difference is found for REBELS-08, whose two Band 6 flux densities differ by a factor of $\sim2$, but remain consistent at the $\sim1.5\sigma$ level given their sizable uncertainties.

\subsection{Modified blackbody fitting}
\label{sec:methods_mbb_fitting}

To determine the dust properties of our targets, we fit their multi-band flux densities with an optically thin modified blackbody \citep{hildebrand1983}, as is common in the high-redshift literature \citep[e.g.,][]{bakx2021,mitsuhashi2024_serenade}. In doing so, we correct for heating by and contrast against the warm cosmic microwave background (CMB; $T_{\mathrm{CMB},z} \approx 22\,\mathrm{K}$ at $z=7$) following \citet{dacunha2013}. We adopt the Bayesian fitting framework from \citet{algera2024}, which relies on the {\sc{emcee}} library \citep{foreman-mackey2013}. The precise functional form being fitted is reproduced from \citet{algera2024}:

{\footnotesize
\begin{align}
    S_\nu = \left(\frac{1+z}{d_L^2}\right) M_\mathrm{dust} \kappa_0 \left(\frac{\nu}{\nu_0}\right)^{\beta_\mathrm{IR}} \left[B_\nu(T_{\mathrm{dust},z}) - B_\nu(T_{\mathrm{CMB},z}) \right] \ .
    \label{eq:mbb_fit}
\end{align}
}

\noindent The free parameters are the dust mass ($M_\mathrm{dust}$), the CMB-heated dust temperature ($T_{\mathrm{dust},z}$), and the dust emissivity index $\beta_\mathrm{IR}$. In what follows, we will quote dust temperatures corrected for CMB heating following \citet{dacunha2013}, and denote these simply as $T_\mathrm{dust}$. In Equation \ref{eq:mbb_fit}, $d_L$ furthermore denotes the luminosity distance to redshift $z$, and $B_\nu$ the Planck function. We adopt a dust opacity coefficient of $\kappa_0 = 10.41\,\mathrm{cm}^2\,\mathrm{g}^{-1}$ at $\nu_0 = 1900\,\mathrm{GHz}$, based on \citet{weingartner2001}, noting that the inferred dust masses scale inversely with $\kappa_0$ and thus depend on the assumed grain composition and size distribution \citep[e.g.,][]{fanciullo2020,fanciullo2026,sommovigo_algera2025}. We furthermore discuss the optically thin assumption in Section \ref{sec:discussion_Tdust_optically_thick}, where we show that it is appropriate for our sample.

The Bayesian fitting framework allows for physically motivated priors on the free parameters, and we here adopt a simple flat prior on the logarithm of the dust mass, $\log(M_\mathrm{dust}/M_\odot) \in [4, 12]$. For the dust emissivity index, we consider two scenarios. Given that the number of distinct ALMA bands is limited to two for all but one of our sources, our fiducial approach is to adopt a fixed $\beta_\mathrm{IR} = 2.0$, motivated by empirical studies of the dust emissivity index at slightly lower redshifts \citep[e.g.,][]{dacunha2021,bendo2023,bendo2025,witstok2023_beta,liao2024}. However, we also present fits assuming a Gaussian prior on $\beta_\mathrm{IR}$, with a mean and standard deviation of $1.8$ and $0.5$, respectively (Appendix \ref{app:mbb_fits_gaussian_prior}), following \citet{algera2024} and \citet{mitsuhashi2024_serenade,mitsuhashi2026_serenade}. In principle, marginalizing over $\beta_\mathrm{IR}$ yields more realistic uncertainties on the dust masses and temperatures, but owing to the degeneracy between $\beta_\mathrm{IR}$ and $T_\mathrm{dust}$, fixing the emissivity index is standard practice when only two or three bands are available.

The dust temperature prior requires more care, given the generally modest S/N of our data, and a lack of high-frequency coverage beyond ALMA Band 8. A flat prior on $T_{\mathrm{dust},z}$ is frequently adopted in the literature \citep[e.g.,][]{algera2024,mitsuhashi2024_serenade}, although we show in Appendix~\ref{app:mock_mbb_fitting} that this tends to overestimate dust temperatures for low-S/N dual-band data. The primary reason is that the dust temperature depends on the measured flux ratio (generally $S_\mathrm{B8} / S_\mathrm{B6}$ in our case); at low dust temperatures, the flux ratio changes quickly with increasing temperature, whereas at high temperatures it becomes nearly insensitive to $T_\mathrm{dust}$, and eventually converges to a fixed value when both bands probe the Rayleigh-Jeans tail of the dust emission. As a result, a flat prior assigns substantial weight to hot dust solutions that the data cannot distinguish, shifting the posterior median to higher temperatures.

We therefore adopt a dust temperature prior that is uniform in inverse temperature ($1 / T_{\mathrm{dust},z}$, i.e., with a probability density of $p(T_{\mathrm{dust},z}) \propto T_{\mathrm{dust},z}^{-2}$), bounded from below by the CMB temperature at the specific redshift of each target. Although the prior probability converges to zero at high temperatures and an upper limit is therefore not formally needed in the fitting, we adopt a conservative upper bound of $300\,\mathrm{K}$, validated by extensive testing and well above the dust temperatures measured for $z \gtrsim 5$ galaxies in the literature (Section \ref{sec:discussion}). 

Using MBB fits to mock sources, we show in Appendix~\ref{app:mock_mbb_fitting} that this inverse-temperature prior recovers the input dust properties with considerably smaller biases at limited S/N. This can also be understood physically, as discussed in detail in that appendix. Briefly, the dust temperature enters the Planck function through the quantity $h\nu / (k_\mathrm{B} T_{\mathrm{dust},z})$, making the inverse temperature the natural variable. For a fixed dust emissivity index, the peak wavelength of the dust SED moreover scales as $1 / T_{\mathrm{dust},z}$ (Wien's displacement law), such that a prior uniform in inverse temperature can be thought of as being flat in the wavelength where the dust SED peaks \citep[see also][]{drew2022}. In addition, it can be shown analytically that a flux density ratio constrains the inverse temperature about equally well for cold and for hot dust (Equations \ref{eq:fluxRatioRayleighJeansLimit} and \ref{eq:fluxRatioWienLimit} and Figure \ref{fig:Tdust_prior}), whereas the corresponding uncertainty on the temperature itself grows rapidly toward high temperatures. A prior uniform in inverse temperature therefore assigns equal prior weight to temperatures that are equally well distinguished by the data.

\subsection{Dual-band stacking}
\label{sec:methods_stacking}

\begin{figure}
    \centering
    \includegraphics[width=0.45\textwidth]{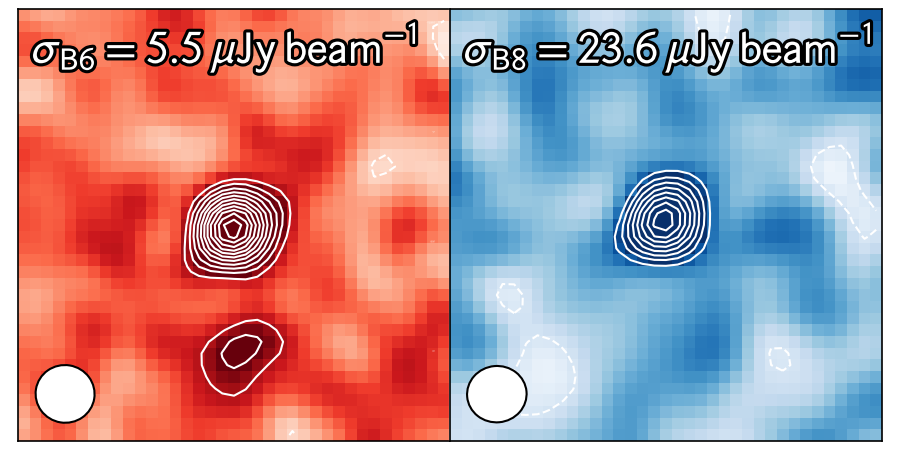}
    \caption{\textbf{Image-plane stacks of our ten primary targets in Bands 6 (\textit{left}) and 8 (\textit{right}).} Contours are $\pm2,3,\ldots\sigma$ where $\sigma$ is the RMS annotated on the panels. The cutouts span $12''\times12''$ and the (tapered) beam is shown in the lower left corner. The stacks are detected at a peak S/N of 13 (Band 6) and 10 (Band 8).}
    \label{fig:dualBandStacks}
    \vspace*{0.2cm}
\end{figure}

\begin{figure*}
    \centering
    \includegraphics[width=0.90\textwidth]{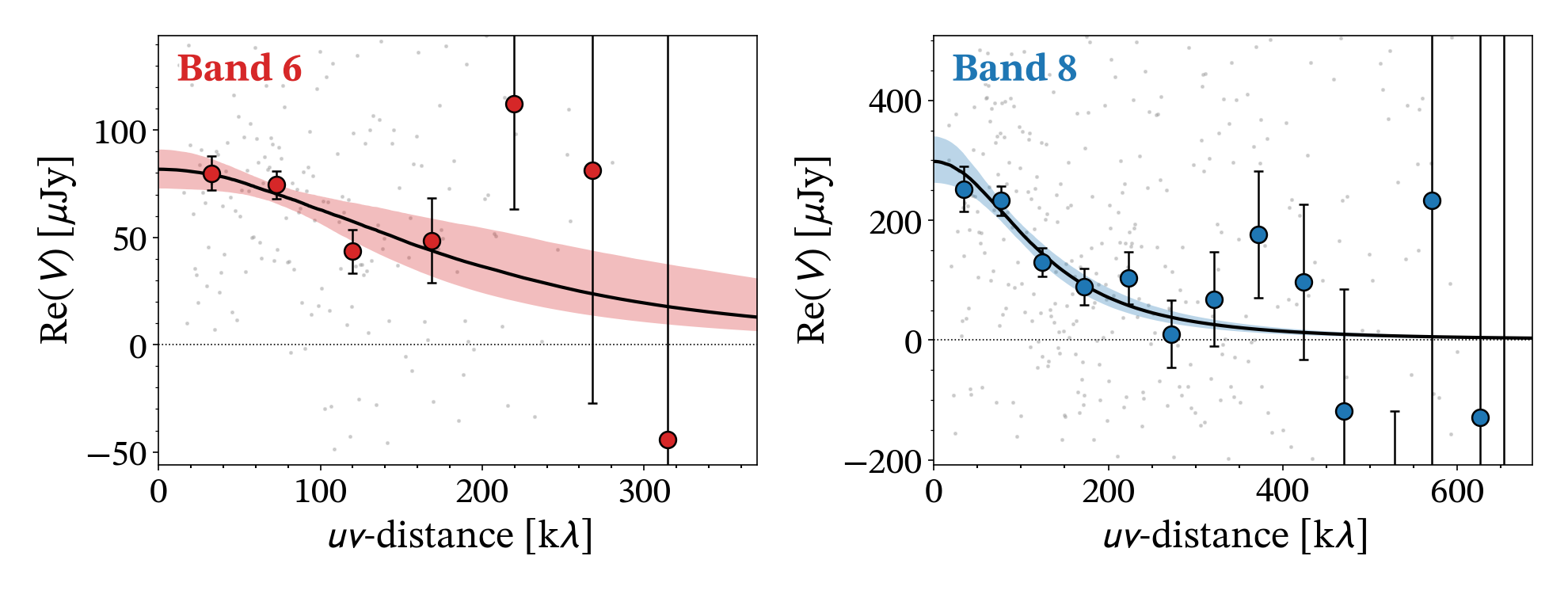}
    \vspace*{-0.15cm}
    \caption{\textbf{(u,v)-plane stacks of the combined visibility data of our ten primary targets in Bands 6 (\textit{left}) and 8 (\textit{right}).} The black line and shaded regions represent the best fit and $1\sigma$ confidence interval, and are fitted to the visibility data binned in $1\,\mathrm{k}\lambda$ bins (small grey points). The larger markers correspond to $50\,\mathrm{k}\lambda$ bins and are shown for visualization purposes.}
    \label{fig:dualBandUVStacks}
    \vspace*{0.2cm}
\end{figure*}

Given the modest S/N at which the bulk of our targets is detected (Figure \ref{fig:snr_hist}), we employ stacking to gain further insight into their average dust properties. We focus on the Band 6 and 8 data, which are available for all targets, and stack only the ten primary REBELS sources, thereby excluding the serendipitous \cii{}-emitters REBELS-12-2, REBELS-29-2 and REBELS-39-2. As such, the stacks represent the average dust properties of UV-bright, dust-continuum-detected galaxies at $z\approx7$. Our fiducial stacks are conducted in the $(u,v)$-plane, although we additionally perform image-plane stacking using the beam-matched, $1.5''$-tapered images. For the latter, we subtract bright non-target sources via 2D Gaussian fitting using {\sc{PyBDSF}}, while for the former these are subtracted from the visibility data as outlined in Section \ref{sec:methods_fluxdensity_measurements}. 

We stack the original REBELS LP Band 6 data for all targets except REBELS-18, for which we instead use the Band 6 map from PI Fudamoto. The central frequency of this image is closer to that of the Band 6 data used for the rest of the sample (see footnote \ref{footnote:rebels18}), which makes it better suited for stacking. In Band 8, we adopt the data from PI Inami for REBELS-38 and REBELS-40, and those from PI Algera for the remaining eight targets to similarly construct a homogeneous dataset.

As we stack in two bands simultaneously to determine the average dust properties of our targets, each galaxy must carry the same relative weight in the Band 6 and Band 8 stacks; otherwise, the two stacked fluxes do not describe the same average galaxy. Inverse-variance weighting applied separately in each band does not guarantee this. For instance, the Band 8 integration times in program 2024.1.00406.S (PI Algera) were deliberately shorter for galaxies that are bright in Band 6 continuum. Per-band inverse-variance weighting would therefore down-weight these Band 6-bright galaxies in the Band 8 stack only, and consequently bias the inferred dust properties.

To circumvent this bias, we follow \citet{bakx2024} and assign each galaxy $j$ a single weight $w_j$, which we apply in both bands. We first compute the inverse-variance weight $w_{i,j}$ of each galaxy in each band $i \in \{\mathrm{B6}, \mathrm{B8}\}$. For the image-plane stack, $w_{i,j} \propto 1/\sigma^2_{i,j}$, where $\sigma_{i,j}$ is the RMS of the $1.5''$-tapered image. For the $(u,v)$-plane stack, we take $w_{i,j}$ to be the sum of the visibility weights (rescaled as in Section \ref{sec:methods_fluxdensity_measurements}). In both cases, we normalize the weights to sum to unity within each band, i.e., $\sum_j w_{i,j} = 1$. Each galaxy then receives the smaller of its two normalized weights, $w_j = \min\left(w_{\mathrm{B6},j}, w_{\mathrm{B8},j}\right)$, such that its weight is set by the band in which its data are noisiest relative to the rest of the sample. Finally, we re-normalize the $w_j$ to sum to unity.

In the image plane, the stack in each band is then the weighted mean of the individual images centered at the fiducial continuum positions. In the $(u,v)$-plane, we multiply all visibility weights of a given galaxy by the same factor, chosen such that they sum to $w_j$. This sets the contribution of each galaxy to the stack, without changing the relative weights of its individual visibilities. We then combine the visibilities of all galaxies and, as for the individual targets, bin them in $(u,v)$-distance and fit them with an exponential profile.

Due to the different redshifts and spectral setups of our targets, the individual datasets probe slightly different rest-frame wavelengths, spanning $\lambda_\mathrm{B6} = 150.6 - 162.4\,\mu\mathrm{m}$ and $\lambda_\mathrm{B8} = 86.8 - 98.8\,\mu\mathrm{m}$, with weighted standard deviations of $3.7\,\mu\mathrm{m}$ and $4.3\,\mu\mathrm{m}$, respectively. We fit the stacks at their weighted mean observed frequencies of $238.6\,\mathrm{GHz}$ and $406.3\,\mathrm{GHz}$, at the weighted mean redshift of the sample ($z=6.99 \pm 0.14$). Given the only minor variation in the rest-frame wavelengths probed, we stack the data without rescaling them to a common rest-frame wavelength.

We show the dual-band image-plane and $(u,v)$-plane stacks in Figures \ref{fig:dualBandStacks} and \ref{fig:dualBandUVStacks}, respectively. In the image plane, emission is detected at a peak S/N of $13$ and $10$ in Bands 6 and 8, respectively. For the $(u,v)$-plane stacks, which we adopt as fiducial, we infer flux densities of $S_\mathrm{B6} = 81.6 \pm 9.2\,\mu\mathrm{Jy}$ and $S_\mathrm{B8} = 298.4 \pm 38.3\,\mu\mathrm{Jy}$ from an exponential fit. We fit the stack, in addition to the full REBELS multi-band sample, with an MBB in the following section.

\section{Results}
\label{sec:results}

\subsection{Modified blackbody fits}
\label{sec:results_MBB_fits}

\begin{figure*}
    \centering\includegraphics[width=1.0\textwidth]{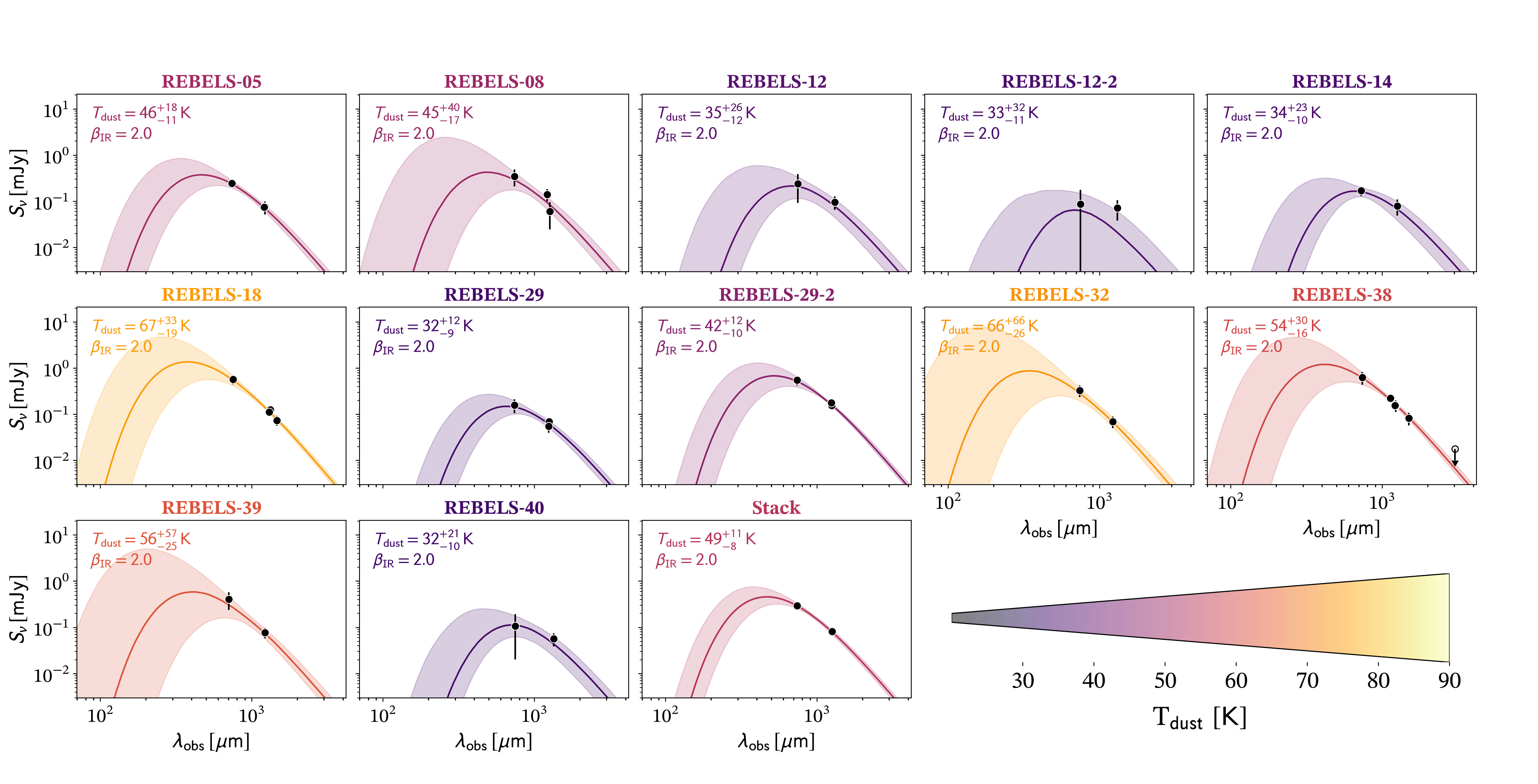}
    \caption{\textbf{Modified blackbody fits to our 12 targets with multi-band data and a dual-band stack of the ten primary REBELS sources (final panel).} A fixed emissivity index of $\beta_\mathrm{IR}=2.0$ is assumed, and galaxies are color-coded by their median $T_\mathrm{dust}$, from cool (purple) to hot (yellow). The dual-band stack reveals a typical $T_\mathrm{dust}\approx50\,\mathrm{K}$.}
    \label{fig:MBBfitting}
\end{figure*}

\begin{table*}
\centering
\begin{threeparttable}
\caption{Dust properties from MBB fitting with a fixed $\beta_\mathrm{IR}=2.0$, plus ancillary and derived parameters.}
\label{tab:dust_properties}
\begin{tabular}{lccccccccc}
\toprule
Galaxy & $z$ & $\log M_{\mathrm{dust}}$ & $T_{\mathrm{dust}}$ & $\log L_{\mathrm{IR}}$ & $\log M_\star$ & $\log(M_\mathrm{dust}/M_\star)$ & $\log\,\mathrm{SFR}_{\mathrm{UV+IR}}$ & $\log\,\mathrm{IRX}$ & $f_{\mathrm{obs}}$ \\
 &  & [$M_\odot$] & [K] & [$L_\odot$] & [$M_\odot$] &  & [$M_\odot\,\mathrm{yr}^{-1}$] &  &  \\
\midrule
REBELS-05 & 6.4963 & $7.11_{-0.37}^{+0.36}$ & $46_{-11}^{+18}$ & $11.53_{-0.34}^{+0.50}$ & $9.42 \pm 0.10$ & $-2.33_{-0.37}^{+0.38}$ & $1.73_{-0.24}^{+0.42}$ & $0.61_{-0.35}^{+0.50}$ & $0.77_{-0.17}^{+0.14}$ \\
REBELS-08 & 6.7495 & $7.24_{-0.66}^{+0.70}$ & $45_{-17}^{+40}$ & $11.61_{-0.56}^{+1.02}$ & $9.33_{-0.06}^{+0.07}$ & $-2.11_{-0.65}^{+0.71}$ & $1.82_{-0.33}^{+0.90}$ & $0.52_{-0.57}^{+1.02}$ & $0.73_{-0.31}^{+0.23}$ \\
REBELS-12 & 7.3459 & $7.51_{-0.81}^{+0.77}$ & $35_{-12}^{+26}$ & $11.29_{-0.34}^{+0.70}$ & $9.54_{-0.04}^{+0.03}$ & $-2.05_{-0.80}^{+0.78}$ & $1.71_{-0.13}^{+0.46}$ & $0.00_{-0.35}^{+0.71}$ & $0.45_{-0.18}^{+0.36}$ \\
REBELS-12-2$^\dagger$ & 7.3524 & $6.80_{-1.62}^{+1.18}$ & $33_{-11}^{+32}$ & $10.80_{-1.45}^{+0.64}$ & $9.30_{-0.50}^{+0.40}$ & $-2.57_{-1.60}^{+1.34}$ & $0.88_{-1.45}^{+0.66}$ & $0.49_{-1.45}^{+0.66}$ & $0.72_{-0.63}^{+0.20}$ \\
REBELS-14 & 7.0842 & $7.50_{-0.87}^{+0.74}$ & $34_{-10}^{+23}$ & $11.12_{-0.19}^{+0.52}$ & $9.24_{-0.06}^{+0.07}$ & $-1.77_{-0.85}^{+0.77}$ & $1.62_{-0.06}^{+0.27}$ & $-0.13_{-0.19}^{+0.52}$ & $0.38_{-0.09}^{+0.29}$ \\
REBELS-18 & 7.6750 & $6.98_{-0.32}^{+0.33}$ & $67_{-19}^{+33}$ & $12.37_{-0.53}^{+0.72}$ & $9.71_{-0.04}^{+0.06}$ & $-2.75_{-0.32}^{+0.34}$ & $2.47_{-0.45}^{+0.70}$ & $1.19_{-0.52}^{+0.73}$ & $0.93_{-0.13}^{+0.06}$ \\
REBELS-29 & 6.6847 & $7.50_{-0.46}^{+0.57}$ & $32_{-9}^{+12}$ & $11.02_{-0.26}^{+0.39}$ & $9.69_{-0.05}^{+0.07}$ & $-2.21_{-0.45}^{+0.57}$ & $1.51_{-0.08}^{+0.19}$ & $-0.12_{-0.26}^{+0.39}$ & $0.39_{-0.13}^{+0.22}$ \\
REBELS-29-2$^\dagger$ & 6.6818 & $7.58_{-0.27}^{+0.35}$ & $42_{-10}^{+12}$ & $11.77_{-0.35}^{+0.39}$ & $9.50_{-0.50}^{+0.40}$ & $-1.89_{-0.52}^{+0.58}$ & $1.86_{-0.36}^{+0.38}$ & $1.24_{-0.36}^{+0.38}$ & $0.94_{-0.07}^{+0.04}$ \\
REBELS-32 & 6.7290 & $6.74_{-0.57}^{+0.57}$ & $66_{-26}^{+66}$ & $12.09_{-0.76}^{+1.25}$ & $9.57_{-0.08}^{+0.10}$ & $-2.86_{-0.56}^{+0.59}$ & $2.22_{-0.67}^{+1.23}$ & $1.31_{-0.78}^{+1.26}$ & $0.94_{-0.20}^{+0.05}$ \\
REBELS-38 & 6.5770 & $7.31_{-0.39}^{+0.41}$ & $54_{-16}^{+30}$ & $12.13_{-0.54}^{+0.79}$ & $9.75_{-0.06}^{+0.09}$ & $-2.46_{-0.39}^{+0.42}$ & $2.27_{-0.45}^{+0.74}$ & $1.02_{-0.55}^{+0.78}$ & $0.90_{-0.19}^{+0.09}$ \\
REBELS-39 & 6.8449 & $6.91_{-0.59}^{+0.72}$ & $56_{-25}^{+57}$ & $11.85_{-0.81}^{+1.25}$ & $9.35_{-0.08}^{+0.09}$ & $-2.46_{-0.59}^{+0.73}$ & $2.05_{-0.44}^{+1.15}$ & $0.56_{-0.81}^{+1.26}$ & $0.75_{-0.43}^{+0.23}$ \\
REBELS-40 & 7.3650 & $7.44_{-0.84}^{+0.77}$ & $32_{-10}^{+21}$ & $10.96_{-0.29}^{+0.56}$ & $9.82_{-0.50}^{+0.30}$ & $-2.35_{-0.89}^{+0.88}$ & $1.44_{-0.10}^{+0.31}$ & $-0.10_{-0.30}^{+0.56}$ & $0.40_{-0.15}^{+0.31}$ \\
Stack & 6.9909 & $7.10_{-0.24}^{+0.23}$ & $49_{-8}^{+11}$ & $11.70_{-0.23}^{+0.30}$ & $9.53 \pm 0.06$ & $-2.44_{-0.24}^{+0.24}$ & $1.91_{-0.16}^{+0.24}$ & $0.55_{-0.24}^{+0.31}$ & $0.75 \pm 0.11$ \\
\bottomrule
\end{tabular}
\begin{tablenotes}
        \footnotesize
        \item[$\dagger$] Only an upper limit on $\mathrm{SFR}_\mathrm{UV}$ is available for REBELS-12-2 and REBELS-29-2. We therefore list $\mathrm{SFR}_\mathrm{IR}$ in place of $\mathrm{SFR}_\mathrm{UV+IR}$, and the corresponding IRX and $f_\mathrm{obs}$ are thus lower limits.
\end{tablenotes}
\end{threeparttable}
\end{table*}

We fit optically thin MBBs -- adopting a dust temperature prior that is uniform in $1/T_\mathrm{dust}$ (Section~\ref{sec:methods_mbb_fitting}) -- to the multi-frequency flux densities of our targets, as well as to the dual-band stack of the ten primary REBELS sources (Section~\ref{sec:methods_stacking}). As detailed in Section~\ref{sec:methods_fluxdensity_measurements}, all flux densities are measured in the $(u,v)$-plane through an exponential disk fit. The MBB fits for our fiducial fixed $\beta_\mathrm{IR} = 2.0$ are shown in Figure \ref{fig:MBBfitting} and the inferred dust properties are presented in Table \ref{tab:dust_properties}. Fits adopting a Gaussian prior on $\beta_\mathrm{IR}$ can be found in Figure \ref{fig:MBBfitting_gaussianPrior} in Appendix \ref{app:mbb_fits_gaussian_prior}, while MBB fits to alternative flux density measurements (Appendix \ref{app:flux_measurement_comparison}) are discussed in Appendix \ref{app:mbb_properties_across_flux_densities}. In what follows, we discuss the dust properties inferred from the fiducial fixed-$\beta_\mathrm{IR}$ fits.

\subsubsection{Dust temperatures}

Our MBB fits reveal a range of dust temperatures between $T_\mathrm{dust} \approx 30 - 70\,\mathrm{K}$, with a median of $T_\mathrm{dust} \approx 44\,\mathrm{K}$. The warmest dust is found for REBELS-18 and REBELS-32, with $T_\mathrm{dust} \approx 65 - 70\,\mathrm{K}$, while five sources are found to have relatively cold dust temperatures of $T_\mathrm{dust} \approx 32 - 35\,\mathrm{K}$. However, the data do not constrain the dust temperatures of all targets equally well. For six of them (REBELS-08, REBELS-12, REBELS-12-2, REBELS-14, REBELS-39, and REBELS-40), the posterior is shaped substantially by the adopted inverse-temperature prior, and their individual dust temperatures are correspondingly less certain (see Figure~\ref{fig:Tdust_posteriors} and the discussion in Appendix~\ref{app:dust_temperature_posteriors}). This subsample includes four sources with relatively cold fitted dust temperatures, two of which are not detected in Band 8 ($<3\sigma$; REBELS-12-2 and REBELS-40). The warm dust temperature of REBELS-32 is moreover only loosely constrained, given the modest S/N of both its Band 6 and Band 8 data, while REBELS-18, in contrast, is well detected in several Band 6 projects as well as in Band 8, such that its elevated dust temperature ($T_\mathrm{dust} = 67_{-19}^{+33}\,\mathrm{K}$) is robust. Further follow-up in Band 9 will be needed to more accurately measure its dust temperature and infrared luminosity \citep[as in, e.g.,][]{bakx2021,algera2024b,mitsuhashi2026_serenade,rajulal2026}. 

For REBELS-38, where we can leverage a Band 5 detection (and a weakly constraining Band 3 upper limit), we infer $\beta_\mathrm{IR} = 1.9_{-0.4}^{+0.5}$ and $T_\mathrm{dust} = 56_{-20}^{+51}\,\mathrm{K}$ when fitted with a Gaussian prior on $\beta_\mathrm{IR}$ (Figure \ref{fig:MBBfitting_gaussianPrior}). However, we caution that the dust emissivity index is poorly constrained given the lack of detections deep on the Rayleigh-Jeans tail, such that its posterior largely reflects the prior. The fit with a free dust emissivity index in any case yields a similar dust temperature to that with $\beta_\mathrm{IR} = 2.0$, which we adopt as our fiducial result for consistency with the rest of the sample.

We note that the fitted dust temperature for REBELS-38 is higher than what was inferred by \citet{algera2024} based on just the REBELS LP Band 6 observations and the Band 8 data re-analyzed in this work ($T_\mathrm{dust,A24} = 32_{-5}^{+8}\,\mathrm{K}$ versus the fiducial $T_\mathrm{dust} = 54_{-16}^{+30}\,\mathrm{K}$ derived here). This is primarily due to our $(u,v)$-plane flux measurement in Band 8 ($S_\mathrm{B8} = 629 \pm 191\,\mu\mathrm{Jy}$) exceeding the peak flux density inferred by \citet{algera2024} based on a $1.2''$-tapered image ($S_\mathrm{B8,A24} = 334 \pm 63\,\mu\mathrm{Jy\,beam}^{-1}$). We confirm that we infer a consistent peak flux density from our $1.5''$-tapered map of $S_\mathrm{B8} = 397 \pm 78\,\mu\mathrm{Jy\,beam}^{-1}$, and had we adopted peak flux measurements for all bands, we would infer a lower dust temperature of $T_\mathrm{dust} = 43_{-7}^{+10}\,\mathrm{K}$ for REBELS-38, consistent with the measurement from \citet{algera2024}. MBB fits to the aperture and integrated flux densities, measured on the $1.5''$-tapered images, both yield dust temperatures of $T_\mathrm{dust} \sim 45-47\,\mathrm{K}$, consistent with both our fiducial value and the peak flux measurement (Figure \ref{fig:MBB_fits_across_flux_measurements}). Overall, the fitted dust temperature of REBELS-38 is thus higher than, but still consistent with, that of \citet{algera2024}.

A representative dust temperature for the REBELS sample can be obtained from the dual-band stack of our ten primary targets (Section \ref{sec:methods_stacking}). An MBB fit to the stack yields a well-constrained $T_\mathrm{dust} = 49_{-8}^{+11}\,\mathrm{K}$ (assuming $\beta_\mathrm{IR} = 2.0$), which is consistent with the median across the full sample of 12 galaxies. This therefore suggests that dust-detected, UV-bright Lyman break galaxies at $z\approx7$ have a typical dust temperature of $T_\mathrm{dust} \approx 50\,\mathrm{K}$. We compare this to empirical and theoretical trends for the cosmic dust temperature evolution in Section \ref{sec:discussion_Tdust_evolution}. Had we instead adopted a Gaussian prior on $\beta_\mathrm{IR}$, we would infer $T_\mathrm{dust} = 55_{-15}^{+34}\,\mathrm{K}$ for the stack, which is consistent with -- albeit more uncertain than -- our fiducial measurement with $\beta_\mathrm{IR} = 2.0$. Our typical dust temperature is moreover in good agreement with the sample-averaged values inferred from the single-band REBELS LP data by \citet{ferrara2022} and \citet{sommovigo2022}, although the agreement varies for individual galaxies, as we discuss in Appendix~\ref{app:singleband_predictions}.

\subsubsection{Dust masses}

Again adopting a fiducial $\beta_\mathrm{IR} = 2.0$, our targets span a range of $\log(M_\mathrm{dust}/M_\odot) = 6.7 - 7.6$, with a median of $\log(M_\mathrm{dust}/M_\odot) \approx 7.3$. The lowest dust mass is inferred for REBELS-32, $\log(M_\mathrm{dust}/M_\odot) = 6.7 \pm 0.6$, owing to its likely warm dust temperature. The highest dust mass is obtained for REBELS-29-2, $\log(M_\mathrm{dust}/M_\odot) = 7.6 \pm 0.3$, reinforcing the highly obscured nature of this serendipitously identified galaxy.

The inferred dust masses are consistent with what has previously been adopted for the REBELS sample, based on either a fixed dust temperature of $T_\mathrm{dust} = 45 \pm 15\,\mathrm{K}$ by \citet{algera2026}, or temperatures inferred using the \cii{}-based method from \citet{sommovigo2021,sommovigo2022}, which yields a similar average value of $T_\mathrm{dust} \approx 45 - 50\,\mathrm{K}$ \citep[see also][]{ferrara2022}. Quantitatively, for the nine galaxies in common between this work and \citet{algera2026}, we infer dust masses that are on average $0.03\,\mathrm{dex}$ lower with a standard deviation of $0.33\,\mathrm{dex}$ and a range of $-0.46\,\mathrm{dex}$ to $+0.59\,\mathrm{dex}$. This is well within the uncertainties of the individual fitted dust masses of these nine sources, which range from $\sim0.3 - 0.9\,\mathrm{dex}$.

The dual-band stack yields an average dust mass of $\log(M_\mathrm{dust}/M_\odot) = 7.1 \pm 0.2$. With a Gaussian prior on $\beta_\mathrm{IR}$, the stacked dust mass is similar albeit more uncertain, corresponding to $\log(M_\mathrm{dust}/M_\odot) = 7.0 \pm 0.4$.

\subsubsection{Infrared luminosities}

We determine the infrared luminosities of our targets by integrating a representative subset of the MBB-fitting posterior samples across rest-frame $8$--$1000\,\mu\mathrm{m}$. We infer -- again for a fiducial $\beta_\mathrm{IR} = 2.0$ -- a range of $\log(L_\mathrm{IR}/L_\odot) = 10.8 - 12.4$, with REBELS-12-2 and REBELS-18 being the least and most IR-luminous galaxies among the sample, respectively. The median IR luminosity is $\log(L_\mathrm{IR}/L_\odot) \approx 11.6$, which corresponds to an obscured star formation rate of $\mathrm{SFR}_\mathrm{IR} \approx 45\,M_\odot\,\mathrm{yr}^{-1}$. 

Three of our targets -- REBELS-18, REBELS-32, and REBELS-38 -- have infrared luminosities that nominally exceed $10^{12}\,L_\odot$, classifying them as ULIRGs. If their far-infrared emission is powered fully through star formation, this implies high obscured star formation rates of $\mathrm{SFR}_\mathrm{IR} \sim 150 - 280\,M_\odot\,\mathrm{yr}^{-1}$. In contrast, single-band predictions suggested that only one or two of the sixteen dust-detected REBELS galaxies would exceed $L_\mathrm{IR} = 10^{12}\,L_\odot$ \citep{ferrara2022,inami2022,sommovigo2022}. The apparently high incidence of ULIRGs among our sample is nevertheless consistent with recent ALMA Band 9 observations of MACS0416-Y1 \citep{bakx2025} and the Big Three Dragons \citep{rajulal2026}, which revealed these UV-bright galaxies to also be highly infrared-luminous owing to their warm dust. However, we caution that the uncertainties on the infrared luminosities of our targets remain substantial; we therefore classify the ULIRG fraction of our sample probabilistically in Section \ref{sec:discussion_obscured_SF}.

The IR luminosity inferred from an MBB fit to the dual-band stack equals $\log(L_\mathrm{IR}/L_\odot) = 11.7_{-0.2}^{+0.3}$, which corresponds to an average obscured star formation rate of $\mathrm{SFR}_\mathrm{IR} = 60_{-25}^{+61}\,M_\odot\,\mathrm{yr}^{-1}$, consistent with the sample median. With a Gaussian prior on the dust emissivity index in the fit, we infer a consistent $\log(L_\mathrm{IR}/L_\odot) = 11.8_{-0.3}^{+0.5}$ for the stack, which corresponds to an obscured star formation rate of $\mathrm{SFR}_\mathrm{IR} = 72_{-36}^{+172}\,M_\odot\,\mathrm{yr}^{-1}$, reinforcing the importance of obscured star formation even in galaxies at $z\approx7$.

\section{Discussion}
\label{sec:discussion}

In this section, we place our targets in the context of earlier observational and theoretical studies of high-redshift dusty galaxies. In particular, we discuss our sample in the context of the $T_\mathrm{dust} - z$ and $T_\mathrm{dust} - L_\mathrm{IR}$ relations in Section \ref{sec:discussion_Tdust}, present their obscured star-formation properties -- their infrared luminosities, the $f_\mathrm{obs}-M_\star$ and IRX-$\beta_\mathrm{UV}$ relations, and a possible connection to high-redshift quiescent galaxies -- in Section \ref{sec:discussion_obscured_SF}, and discuss their likely dust production pathways in Section \ref{sec:discussion_dust_production}.

Throughout these sections, we compare to high-redshift galaxies from the literature with dust properties inferred from MBB fits to multi-band continuum data, focusing primarily on Lyman break galaxies for a fair comparison to our UV-selected targets. This comparison sample consists of 25 galaxies at $z=5.3 - 8.3$ (20 LBGs and 5 DSFGs), drawn from the following studies:

\begin{itemize}
\setlength{\itemsep}{0.05cm}

    \item Four $z\approx5.3 - 5.7$ LBGs from \citet{faisst2020}, of which one -- HZ10 at $z=5.65$ -- has now been followed up in six ALMA bands, including Bands 9 and 10 \citep{villanueva2024,algera2025_hz10}. For the remaining three, we adopt the dust temperatures from the re-analysis by \citet{mitsuhashi2024_serenade}, who -- similar to our work -- adopt optically thin dust.
    
    \item The merging pair of $z=5.85$ DSFGs MAMBO-9-A/B \citep{casey2019,akins2025}.
    
    \item Nine UV-luminous LBGs at $z=5.8 - 6.2$ from SERENADE \citep[][]{harikane2020,mitsuhashi2024_serenade,mitsuhashi2026_serenade}.
    
    \item The IR-luminous DSFG HFLS3 at $z=6.34$ \citep{riechers2013,witstok2023_beta}.
    
    \item The lensed pair of $z=6.9$ DSFGs SPT0311-58 \citep{strandet2017,marrone2018,jarugula2021}.
    
    \item Three UV-bright $z=6.7 - 7.1$ LBGs in COSMOS that served as pilots for the REBELS program \citep{schouws2022,witstok2022}.
    
    \item The lensed $z=7.13$ galaxy A1689-zD1, which was the first dust detection beyond $z = 7$ \citep{watson2015,knudsen2017,bakx2021,akins2022,killi2023,heintz2025_zD1}.
    
    \item The $z=7.15$ LBG B14-65666, also known as the Big Three Dragons \citep{bowler2018,hashimoto2019,sugahara2021,rajulal2026}.
    
    \item The $z=7.31$ LBG REBELS-25 \citep{hygate2023,algera2024,algera2024b,rowland2024,rowland2026_r25}. As REBELS-25 is part of the REBELS sample, we show it with a distinct marker in the following figures, adopting its published dust properties from \citet{algera2024b}.
    
    \item The currently most distant-known dusty galaxy, MACS0416-Y1 at $z=8.31$ \citep{tamura2019,bakx2020,bakx2025}. 
    
\end{itemize}

\begin{figure*}
    \centering
    \includegraphics[width=0.95\textwidth]{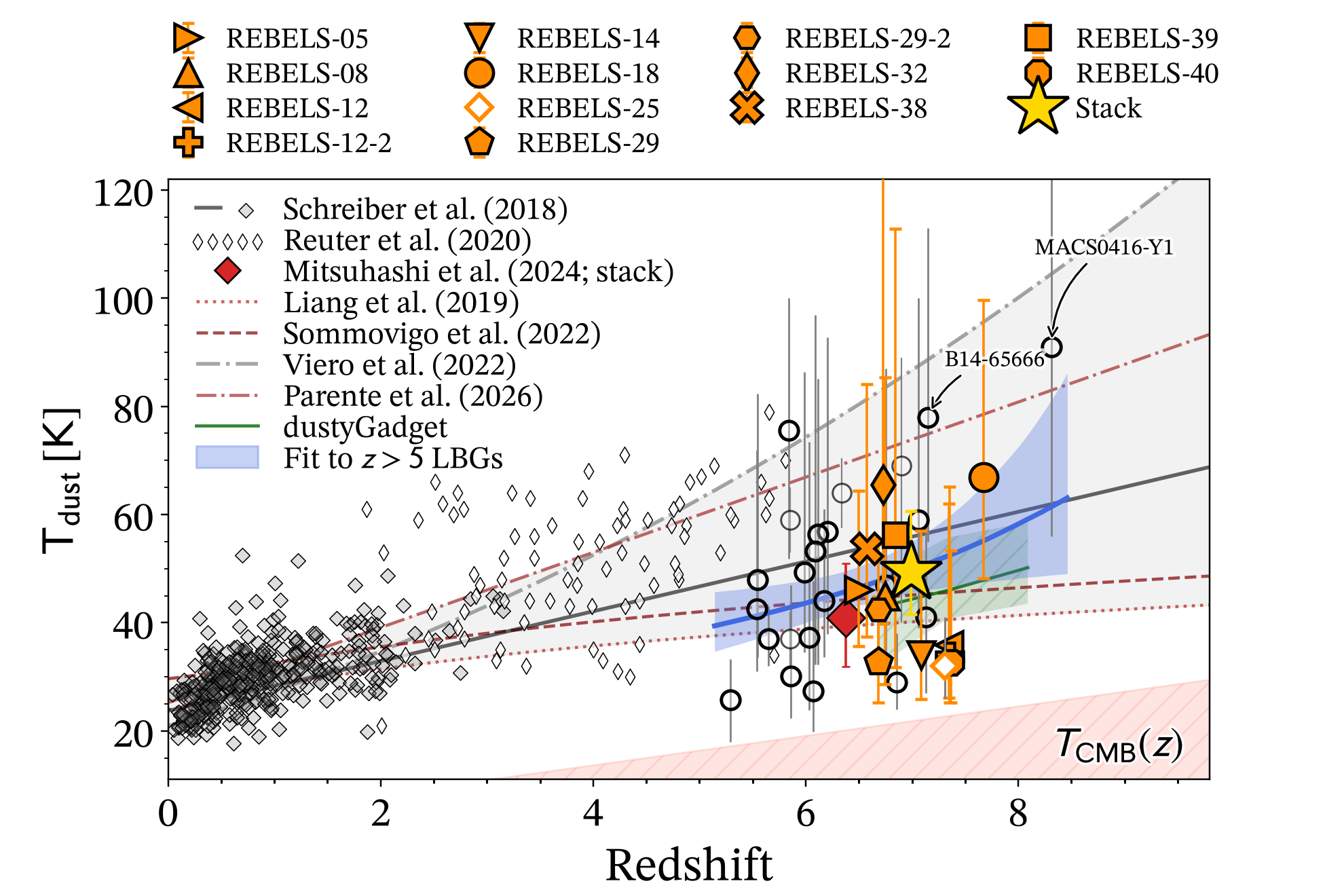}
    \caption{\textbf{The cosmic evolution of dust temperatures from $\mathbf{z=0 - 8.3}$.} Our $z\sim7$ targets (orange) show a variety of dust temperatures spanning $T_\mathrm{dust} \approx 30 - 70\,\mathrm{K}$ given an assumed $\beta_\mathrm{IR}=2.0$, with an average (stacked) $T_\mathrm{dust}\approx50\,\mathrm{K}$ (yellow star). These temperatures are similar to those inferred for LBGs at $z\sim5 - 8$ (open circles with thick black outline; DSFGs at $z\gtrsim6$ are shown with thin grey outlines). A linear fit in $1 / T_\mathrm{dust}$ to all $z\gtrsim5$ LBGs is shown in blue, but does not reveal clear evidence for an increase in their typical dust temperatures with redshift. Also overplotted are lower-redshift star-forming galaxies \citep{schreiber2018} and SPT-selected DSFGs \citep{reuter2020}, though we caveat that the latter are generally more IR-luminous than our targets. We moreover show several empirical \citep{schreiber2018,viero2022} and theoretical \citep[][{\sc{dustyGadget}}]{liang2019,sommovigo2022,parente2026} relations, as well as an LBG stack from \citet[][red diamond]{mitsuhashi2024_serenade}. Two $z > 7$ literature LBGs with particularly well-constrained dust SEDs suggesting hot dust are highlighted \citep[][]{bakx2025,rajulal2026}.}
    \label{fig:Tdust_vs_redshift}
\end{figure*}

We note that the literature dust temperatures are taken as published, and are therefore based on a variety of priors and fitting approaches. As shown in Appendix~\ref{app:mock_mbb_fitting}, the commonly adopted flat temperature prior tends to overestimate dust temperatures at low S/N, although this effect is expected to be smaller for the typically brighter literature sources with multi-band coverage.

Where relevant throughout the following sections, we will perform simple linear fits (in linear- or log-space, as appropriate) to the combined REBELS and literature LBG samples to investigate any potential evolutionary trends within this population (e.g., in $T_\mathrm{dust} - z$). We do so with the Bayesian {\sc{linmix}} package \citep{kelly2007}, which can incorporate uncertainties in both variables as well as an intrinsic scatter term. We fit only the LBGs, thus excluding both the DSFGs and the serendipitous REBELS companions, which were selected through their \cii{} emission rather than in the rest-frame UV. We stress, however, that by focusing on LBGs with multi-band ALMA observations, our analysis is inevitably biased toward a particularly dusty subset of the high-redshift LBG population. In a follow-up work (Algera et al.\ in preparation), we will explore this in more detail by also incorporating galaxies with single-band detections or continuum upper limits, focusing in particular on the pathways of dust build-up in the $z\gtrsim6$ Universe. Here, we restrict ourselves to comparing our REBELS targets with other high-redshift galaxies that have multi-band dust measurements.

\subsection{Dust temperatures}
\label{sec:discussion_Tdust}

\subsubsection{Cosmic dust temperature evolution}
\label{sec:discussion_Tdust_evolution}

We show the MBB-fitted dust temperatures of our targets as a function of redshift in Figure~\ref{fig:Tdust_vs_redshift}, together with the $z\gtrsim5$ literature sample. To investigate whether $z\gtrsim5$ LBGs show evidence for redshift evolution in their typical dust temperature, we fit a linear relation to this combined sample. Given the large and asymmetric uncertainties on $T_\mathrm{dust}$, particularly for warmer sources, a fit linear in $T_\mathrm{dust}$ would be dominated by the colder galaxies with tighter constraints. As demonstrated in Appendix~\ref{app:mock_mbb_fitting}, the inverse dust temperature is instead constrained with approximately fixed precision and symmetric uncertainties at a given S/N, irrespective of $T_\mathrm{dust}$ itself. We therefore use {\sc{linmix}} to fit a relation of the form

\begin{align}
    \frac{10^{3}\,\mathrm{K}}{T_\mathrm{dust}} = A \times (z - 7) + b
\label{eq:Tdust_evolution}
\end{align}

\noindent The best-fit parameters are $A = -2.9_{-2.1}^{+2.2}$ and $b = 20.1 \pm 1.9$, corresponding to $T_\mathrm{dust} = 50_{-4}^{+5}\,\mathrm{K}$ at $z=7$. This is consistent with our dual-band stack -- which is not included in the fit -- as well as with the dual-band stack of 22 $z\approx6-7$ LBGs from \citet[][note that this includes REBELS-12 and REBELS-38]{mitsuhashi2024_serenade}, who find $T_\mathrm{dust} = 41_{-9}^{+10}\,\mathrm{K}$ at an average $z\approx6.4$. The slope of our fit differs from zero at only the $\sim1.3\sigma$ level, such that there is little evidence for redshift evolution in the typical dust temperature of $z > 5$ LBGs alone.

Our dual-band stack and the fit to the REBELS plus literature LBG sample thus both indicate a typical $T_\mathrm{dust}\approx50\,\mathrm{K}$ at $z\approx7$. We therefore recommend adopting $T_\mathrm{dust} = 50\,\mathrm{K}$ (for optically thin dust with $\beta_\mathrm{IR} = 2.0$) when inferring dust masses and infrared luminosities from single-band continuum observations of UV-bright galaxies at $z\approx7$.

\medskip

To place these results in the context of lower-redshift galaxies, Figure~\ref{fig:Tdust_vs_redshift} additionally shows $z\lesssim3$ star-forming galaxies from \citet[][as compiled by \citealt{sommovigo2022}]{schreiber2018} and dusty star-forming galaxies from the SPT-SZ survey \citep{reuter2020}, as well as the empirical relations from \citet{schreiber2018} and \citet{viero2022}, which constrain the evolution of dust temperatures through stacking. Most evolutionary trends, both empirical and theoretical, predict dust temperatures to increase with redshift. Extrapolated to $z\approx7$, the empirical relations of \citet{schreiber2018} and, in particular, \citet{viero2022} predict warmer dust than our best-fit relation.

We further compare to four sets of theoretical predictions: those from the FIRE simulations \citep{liang2019}, the analytical model by \citet{sommovigo2022}, the semi-analytical model from \citet{parente2026}, and those from the {\sc{dustyGadget}} cosmological simulations \citep{graziani2020,dicesare2023}, whose dust temperature predictions will be presented in Schneider et al.\ (in preparation). At $z\approx7$, our results agree well with the model from \citet{sommovigo2022}, shown here for their fiducial parameter set. Their predicted dust temperature additionally depends on metallicity and gas column density, such that in practice the model spans a range of temperatures at fixed redshift. The {\sc{dustyGadget}} simulations predict dust temperatures of $T_\mathrm{dust}\approx43-46\,\mathrm{K}$ at $z\approx6.7-7.3$, similarly in good agreement with our measurements \citep[see also][]{algera2024}.

Our typical dust temperature slightly exceeds that predicted by \citet{liang2019}, although a subset of our targets is consistent with, or even colder than, their prediction. Conversely, none of our sample reaches the high dust temperatures predicted by \citet{parente2026}, although our warmest sources (e.g., REBELS-18 and REBELS-32) are consistent with their predictions within the uncertainties. A few galaxies in the literature, however, do reach dust temperatures comparable to the \citet{parente2026} predictions at $z\approx6-8$ \citep{bakx2025,mitsuhashi2026_serenade,rajulal2026}.

\subsubsection{Optically thick dust}
\label{sec:discussion_Tdust_optically_thick}

\begin{figure}
    \centering
    \includegraphics[width=0.50\textwidth]{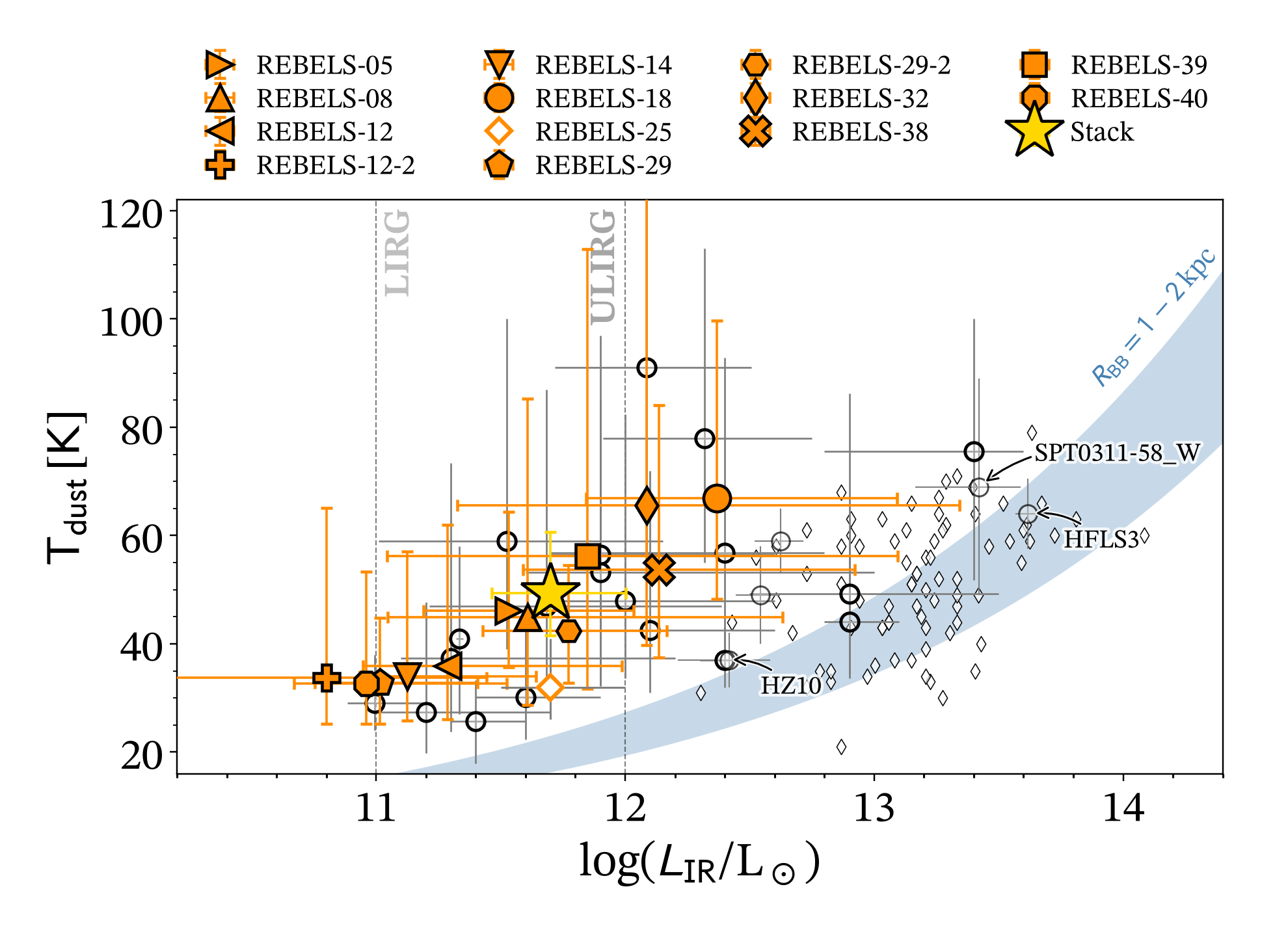}
    \caption{\textbf{Our targets are consistent with optically thin dust.} Dust temperature vs.\ infrared luminosity for our targets (orange), their stack (yellow star), $z\gtrsim5$ LBGs (open markers with thick, black outline), and DSFGs (open markers with thin, grey outline). Small open diamonds show the SPT-selected DSFGs at $z\approx2-6$ from \citet{reuter2020}, whose dust temperatures are based on MBB fits with a non-zero optical depth. The blue shaded region represents the dust temperature of a blackbody with radius $R=1 - 2\,\mathrm{kpc}$; sources below this region either have larger sizes, or are more optically thick than assumed, in which case their dust temperatures are likely underestimated. The REBELS targets consistently fall above the blackbody region, suggesting their dust is unlikely to be globally optically thick.}
\label{fig:Tdust_vs_Lir}
\end{figure}

We show the dust temperatures of our targets, the dual-band stack, and the $z \gtrsim5$ literature sample against their infrared luminosities in Figure \ref{fig:Tdust_vs_Lir}. For comparison, we plot the $T_\mathrm{dust} - L_\mathrm{IR}$ relation for a blackbody with a characteristic size of $R_\mathrm{BB} = 1$--$2\,\mathrm{kpc}$, expressed as $L_\mathrm{IR} = 4\pi R_\mathrm{BB}^2\sigma_\mathrm{SB} T_\mathrm{dust}^4$ with $\sigma_\mathrm{SB}$ being the Stefan-Boltzmann constant. As a galaxy cannot be more luminous than a blackbody of the same size and temperature, those that fall below this curve are either more spatially extended, or have their dust temperatures underestimated. The latter scenario is likely to occur if the galaxies are more optically thick than assumed, as MBB fits with non-zero optical depth yield higher dust temperatures than optically thin models \citep[e.g.,][]{dacunha2021,lower2024,algera2024,costa2026}. Given the anti-correlation between dust temperature and dust mass, an optically thin fit consequently yields a higher $M_\mathrm{dust}$ than would be inferred from an optically thick one. Infrared luminosities, on the other hand, do not strongly depend on the assumed optical depth.

The REBELS galaxies and their stack fall well above the blackbody relation, and therefore do not require globally optically thick dust. For a few literature galaxies, in particular luminous DSFGs such as HFLS3 \citep{riechers2013} and several of the \citet{reuter2020} sources, the dust may instead be more optically thick than assumed.

We further verify the optically thin assumption a posteriori, following \citet[][see also \citealt{algera2024}]{witstok2022} by estimating the wavelength $\lambda_\mathrm{thick}$ at which the dust becomes optically thick ($\tau_{\lambda_\mathrm{thick}} = 1$). For a dust mass surface density $\Sigma_\mathrm{dust}$, this wavelength follows from $\lambda_\mathrm{thick} = \lambda_0 \left( \kappa_0 \Sigma_\mathrm{dust} \right)^{1/\beta_\mathrm{IR}}$. We adopt $\Sigma_\mathrm{dust} = 0.5\,M_\mathrm{dust} / (\pi R_{1/2}^2)$, where $R_{1/2}$ is the dust continuum half-light radius obtained from our $(u,v)$-plane fits and the factor $0.5$ assumes that half of the total dust mass is contained within this radius. For 9/12 of our continuum-detected targets, we find that the dust size can be well constrained in at least one band, while for the remaining three we adopt the typical size of the rest of the sample ($R_{1/2} = 1.9_{-0.8}^{+1.3}\,\mathrm{kpc}$), which we note is furthermore consistent with the size of the dual-band $(u,v)$-plane stack. A detailed analysis of the dust and \cii{} sizes of the REBELS sample will be presented in Astles et al.\ (in preparation).

We infer $\lambda_\mathrm{thick} \approx 3 - 20\,\mu\mathrm{m}$ for the 12 continuum-detected targets, and $\lambda_\mathrm{thick} \approx 6\,\mu\mathrm{m}$ for the stack, well shortward of the rest-frame wavelengths probed by our data ($\lambda_\mathrm{rest} \gtrsim 87\,\mu\mathrm{m}$). For the dust to become optically thick at these wavelengths, its surface density would need to be at least $\sim20$ times larger, and typically more than two orders of magnitude larger. As the optically thin dust masses are likely overestimated if the dust were in fact optically thick, these estimates of $\lambda_\mathrm{thick}$ are moreover conservative. We therefore conclude that the dust in our targets is unlikely to be optically thick on galaxy-wide scales, although we cannot rule out compact, optically thick regions on smaller scales than those probed by our data.

\subsection{Obscured star formation}
\label{sec:discussion_obscured_SF}

\begin{figure}
    \centering
    \includegraphics[width=0.50\textwidth]{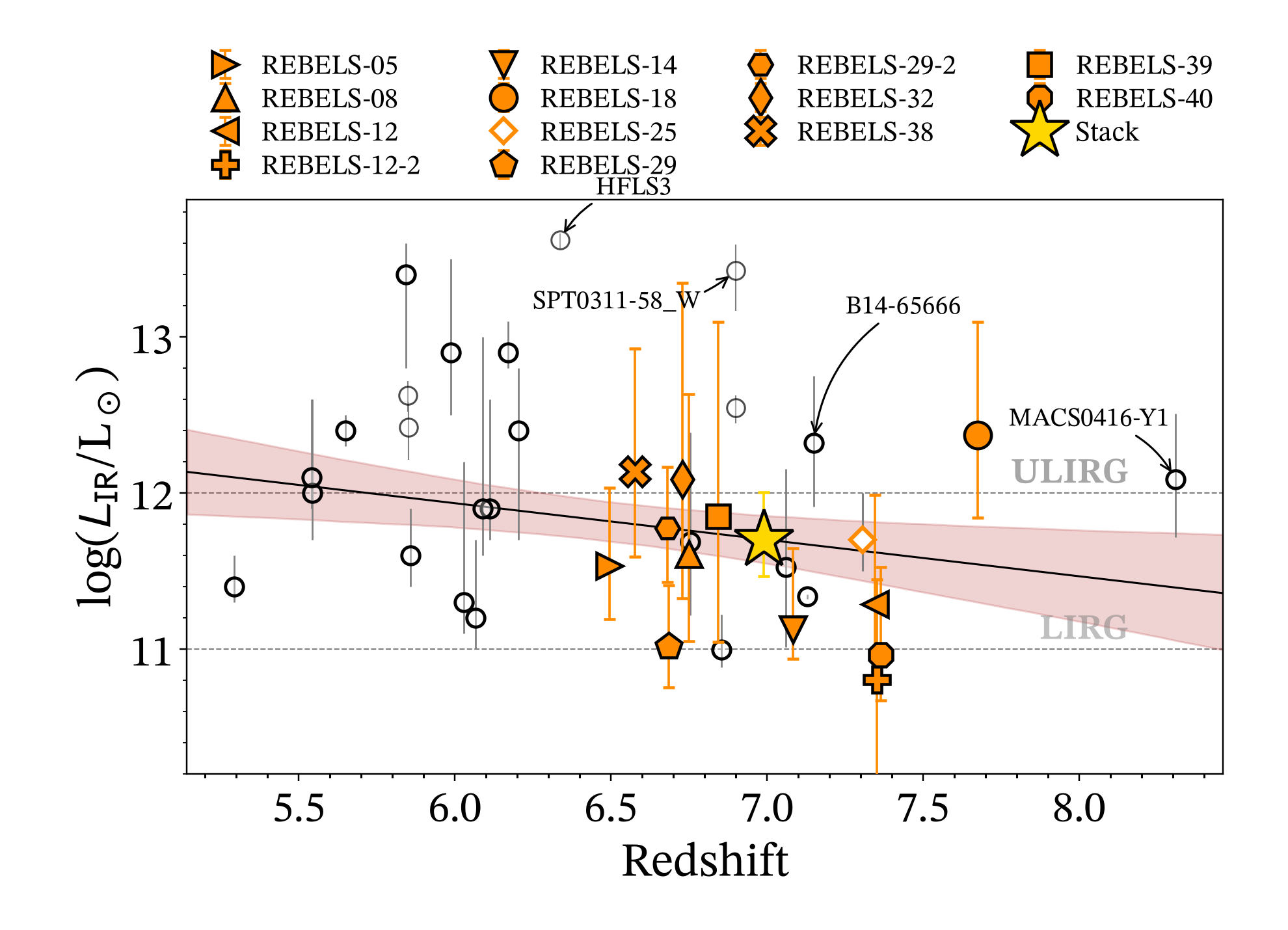}
    \caption{\textbf{Even at $z\gtrsim6.5$, roughly one in three UV-luminous, dust-detected LBGs are classified as ULIRGs ($L_\mathrm{IR} > 10^{12}\,L_\odot$).} The IR luminosities of our targets (orange), their stack (yellow star), and other $z\gtrsim5$ galaxies with multi-band dust continuum detections as a function of redshift (open markers; LBGs are shown with a thicker outline than DSFGs). A linear fit to the LBGs reveals no significant evolution in their typical IR luminosity across $z\approx5 - 8$, with a slope of $-0.23 \pm 0.18\,\mathrm{dex}$.}
\label{fig:Lir_vs_redshift}
\end{figure}

We show the IR luminosities of our sample in Figure~\ref{fig:Lir_vs_redshift}, and compare these against literature $z > 5$ LBGs and DSFGs. Considering only the former population, there is little evidence of evolution in the typical IR luminosities of $z\gtrsim5$ Lyman break galaxies, with the slope of the linear fit, $\alpha = -0.23 \pm 0.18\,\mathrm{dex}$ per unit redshift, differing from zero at only the $\sim1.3\sigma$ level. Individually, however, several of our targets are remarkably IR-luminous: three of them -- REBELS-18, REBELS-32, and REBELS-38 -- have median infrared luminosities exceeding $10^{12}\,L_\odot$, despite being UV-selected sources. For comparison, only a handful of bona fide SMGs are known at this epoch, most notably the lensed SPT0311-58 system at $z=6.9$ \citep{marrone2018}.

At $z>7$, the Big Three Dragons \citep[$z=7.15$;][]{rajulal2026} and the most distant known dusty galaxy, MACS0416-Y1 \citep[$z=8.31$;][]{bakx2025}, were also recently shown to be ULIRGs, and are now joined by REBELS-18 ($z=7.67$). While it remains unclear precisely how common such galaxies are, these detections suggest that IR-luminous, heavily obscured galaxies are already in place within the first $\sim700\,\mathrm{Myr}$ of cosmic history. 

We can make a simple estimate of the ULIRG fraction of our sample using the MCMC posterior chains for our ten primary targets, denoting for each source the probability of being a ULIRG as the fraction of posterior samples with $L_\mathrm{IR} > 10^{12}\,L_\odot$. This yields a ULIRG fraction of $f_\mathrm{ULIRG} = 0.32_{-0.17}^{+0.19}$, where the uncertainty accounts for both the individual $L_\mathrm{IR}$ posteriors and the finite sample size. 

With the new observations presented in this work, there are now 17 LBGs at $z \gtrsim 6.5$ with multi-band ALMA data.\footnote{We include REBELS-05 at a \cii{}-based redshift of $z=6.496$ in this statistic and, as before, exclude DSFGs and quasars.} While we do not have posterior samples for the seven literature LBGs, we can extend the above calculation assuming their uncertainties on $L_\mathrm{IR}$ are represented by a two-piece Gaussian (representing the generally asymmetric lower and upper error). This yields a ULIRG fraction that is consistent with that of our targets, $f_\mathrm{ULIRG} = 0.31_{-0.13}^{+0.15}$. This suggests that nearly one in three dust-detected, UV-bright LBGs at $z\gtrsim6.5$ are, in fact, highly IR-luminous galaxies with ULIRG-like luminosities. As galaxies without dust continuum detections, and hence without multi-band follow-up, are absent from this sample, this fraction should be regarded as an upper limit for the general $z \gtrsim 6.5$ LBG population.

\subsubsection{Obscured fraction of star formation}

\begin{figure}
    \centering
    \includegraphics[width=0.5\textwidth]{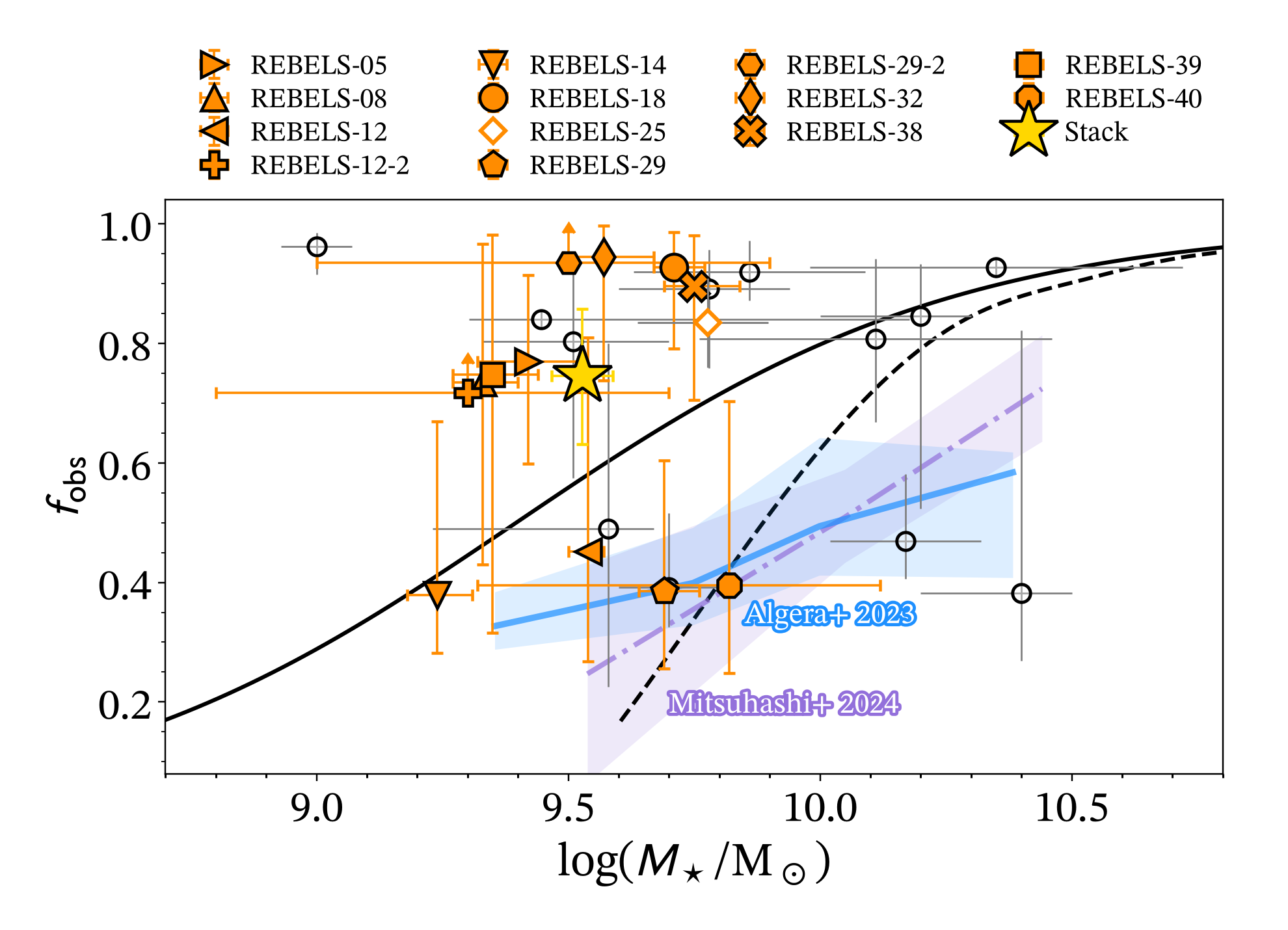}
    \caption{\textbf{Obscured fraction ($f_\mathrm{obs} = \mathrm{SFR}_\mathrm{IR} / \mathrm{SFR}_\mathrm{UV+IR}$) against stellar mass.} The bulk of our multi-band REBELS sample (orange) shows high obscured fractions compared to the $z\lesssim2.5$ relations from \citet[][solid and dashed lines, assuming different IR templates; see their Appendix A]{whitaker2017} and the high-redshift relations from \citet[][$z\sim7$; blue]{algera2023} and \citet[][$z\sim5$; purple]{mitsuhashi2024_cristal}, despite modest stellar masses of $\log(M_\star / M_\odot) \sim 9.5$. The dual-band stack (yellow star) suggests a typical obscured fraction of $f_\mathrm{obs} = 0.75 \pm 0.11$. Literature $z \gtrsim 5$ LBGs with multi-band observations are shown as open black markers.}
    \label{fig:fobs_Mstar}
\end{figure}

We quantify the obscured fractions of our sample, defined as $f_\mathrm{obs} = \mathrm{SFR}_\mathrm{IR} / \mathrm{SFR}_\mathrm{UV + IR}$, as a function of stellar mass in Figure \ref{fig:fobs_Mstar}. The UV SFRs of our targets with \textit{JWST}/NIRSpec IFU observations are taken from \citet{fisher2025b}, while for REBELS-40 we adopt the value from \citet{bouwens2022}. For the serendipitous neighbors, we adopt the UV SFRs from Van Leeuwen et al.\ (in preparation), based on \textit{JWST}/NIRCam observations from GO 6480 (PI Schouws). We adopt stellar masses from \citet{rowland2026} for the REBELS-IFU targets, from \citet{topping2022} for REBELS-40, and from Van Leeuwen et al.\ (in preparation) for the neighbors.

Altogether, our sample spans a range of obscured fractions from $f_\mathrm{obs} \approx 0.4 - 0.95$ (Table \ref{tab:dust_properties}). Particularly obscured is the serendipitous neighbor REBELS-29-2, for which only a lower limit of $f_\mathrm{obs} \gtrsim 0.94$ is available given a non-detection in the rest-UV (Van Leeuwen et al.\ in prep; see also \citealt{fudamoto2021,vanleeuwen2024}). For REBELS-12-2, we can similarly only infer a lower limit on the obscured fraction of $f_\mathrm{obs} \gtrsim 0.72$ given the lack of a UV counterpart.

Eight of our twelve targets, including the two serendipitous sources with lower limits, show obscured fractions $f_\mathrm{obs} \gtrsim 0.7$, while the stack of just the 10 primary targets suggests an average $f_\mathrm{obs} = 0.75 \pm 0.11$ at a stellar mass of $\log(M_\star/M_\odot) = 9.53 \pm 0.06$. This is consistent with earlier analyses of the full sample of dust-detected REBELS galaxies by \citet{inami2022} and the REBELS-IFU subset by \citet{fisher2025b}, who also find high obscured fractions, although under the assumption of a fixed dust temperature.

The obscured fractions measured here are higher than those inferred by \citet{algera2023} of $f_\mathrm{obs} \approx 0.3 - 0.6$ for the entire $z\sim7$ REBELS sample, based on a stacking analysis including also galaxies not detected in dust continuum. Both their work and a similar stacking approach carried out on the $z\sim5$ CRISTAL sample by \citet[][see also \citealt{herrera-camus2025}]{mitsuhashi2024_cristal} suggest a typical obscured fraction of $f_\mathrm{obs} \sim 0.5$ at a stellar mass of $\log(M_\star/M_\odot) \sim 10$.\footnote{We note that the stellar masses of the REBELS-IFU sample have decreased slightly compared to the pre-\textit{JWST} values from \citet{topping2022} that were used in \citet{algera2023}. For our 10 primary targets, the median difference equals $-0.23\,\mathrm{dex}$.} By virtue of also including continuum non-detections, these studies investigate a less biased subset of UV-luminous LBGs, which may explain their lower average obscured fractions compared to the multi-band REBELS sample studied here. On the other hand, both \citet{algera2023} and \citet{mitsuhashi2024_cristal} had to adopt a fixed dust temperature of $T_\mathrm{dust} \approx 42 - 45\,\mathrm{K}$ given a lack of sample-wide multi-band data, which means that a population of low-mass galaxies with warm dust and high obscured SFRs may have been missed.

For the multi-band REBELS sample, we see no clear evidence for the increase in $f_\mathrm{obs}$ with stellar mass expected from lower-redshift relations \citep[e.g.,][]{whitaker2017,casey2026}. Instead, the galaxies appear to separate into two groups: a majority with $f_\mathrm{obs} \gtrsim 0.7$, and four galaxies with $f_\mathrm{obs} \approx 0.4$. The latter are also the four primary targets with the coldest dust ($T_\mathrm{dust} \approx 32 - 35\,\mathrm{K}$), reflecting the steep dependence of $L_\mathrm{IR}$, and hence $f_\mathrm{obs}$, on the fitted dust temperature. Given the limited dynamic range in $M_\star$, selection biases, and the substantial uncertainties on $f_\mathrm{obs}$, we refrain from interpreting this separation as a true bimodality. Taken at face value, however, our results highlight a population of relatively low-mass galaxies ($M_\star < 10^{10}\,M_\odot$) that are nevertheless heavily obscured, in agreement with earlier works on individual galaxies at $z\gtrsim7$ \citep{bakx2025,rajulal2026}.

When factoring in the literature LBG sample, which extends to slightly higher stellar masses ($10^9 \lesssim M_\star/M_\odot \lesssim10^{10.5}$), we infer broadly similar results, including the apparent separation into two groups in the obscured fraction. However, we note that the number of high-redshift galaxies with multi-band follow-up remains limited, while selection effects likely bias our sample to infrared-luminous systems. Overall, we thus conclude that heavily obscured galaxies persist out to $z\sim8$, including at modest stellar masses, while deep multi-band follow-up of less infrared-bright systems across a wide mass range is needed to establish the typical obscured fractions of the high-redshift LBG population.

\subsubsection{IRX-$\beta_\mathrm{UV}$ relation}

\begin{figure}
    \centering
    \includegraphics[width=0.5\textwidth]{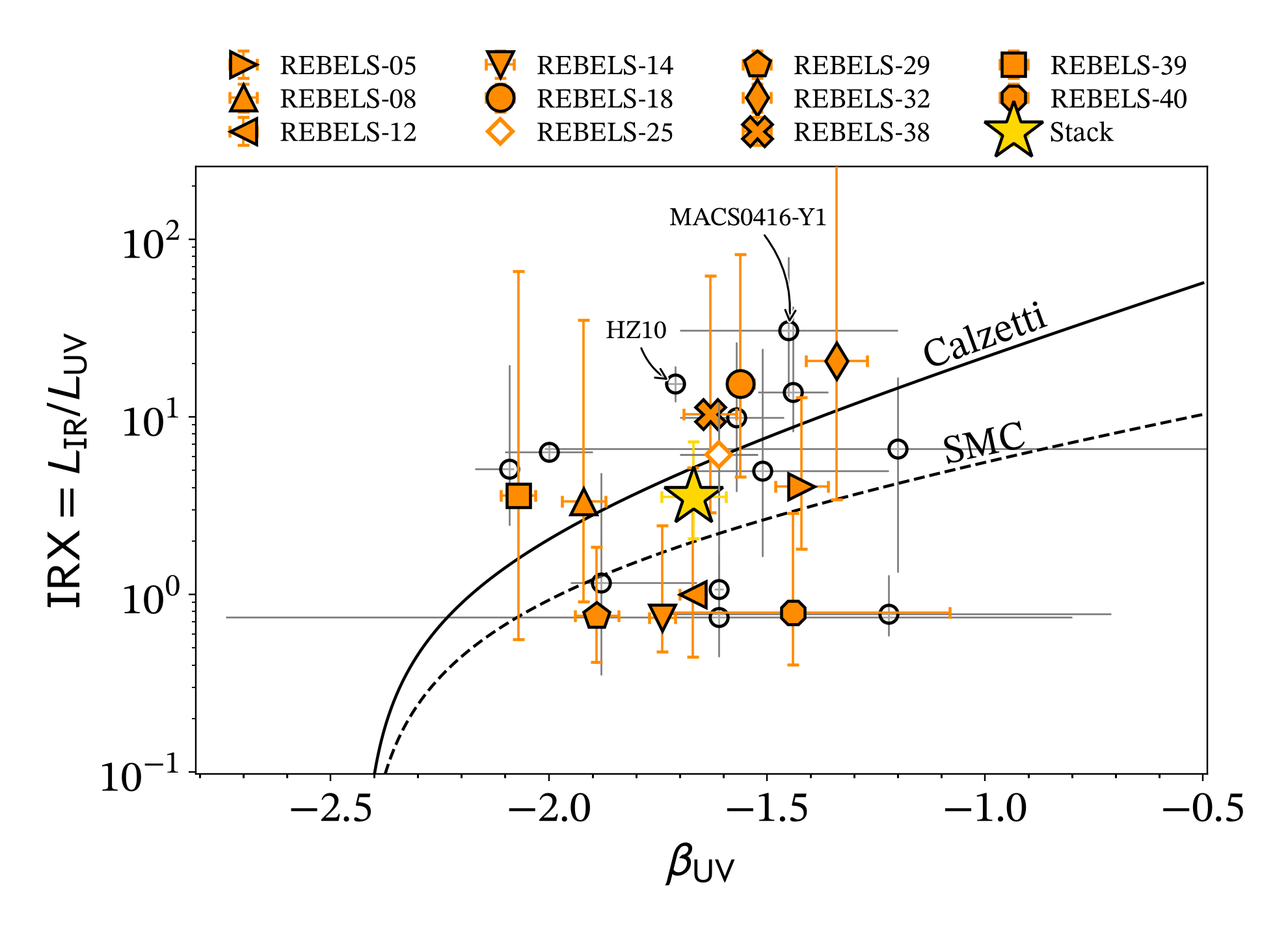}
    \caption{\textbf{The $\mathrm{IRX}-\beta_\mathrm{UV}$ relation of high-redshift LBGs with multi-band ALMA data.} While the REBELS sources (orange) show considerable scatter, the stack (yellow star) lies between the Calzetti and SMC relations, in agreement with previous analyses based on single-band infrared luminosities \citep{bowler2024,fisher2025b}. Open black markers represent literature $z\gtrsim5$ LBGs with multi-band ALMA data; two examples with well-constrained dust SEDs are highlighted \citep{bakx2025,algera2025_hz10}.}
    \label{fig:IRX_beta}
\end{figure}

An alternative means of quantifying the obscuration of our targets, and relating it to dust attenuation at shorter wavelengths, is through the $\mathrm{IRX}-\beta_\mathrm{UV}$ relation. Here $\mathrm{IRX} = L_\mathrm{IR} / L_\mathrm{UV}$ is known as the infrared excess and $\beta_\mathrm{UV}$ represents the UV continuum slope. Both quantities are proxies for dust obscuration, and the exact relation between the two depends on the attenuation curve \citep[e.g.,][]{meurer1999,calzetti2000}, stellar ages, and the star-dust geometry \citep[e.g.,][]{faisst2017,popping2017_irx,vijayan2024}.

We obtain the UV continuum slopes from \citet{fisher2025b}, except for REBELS-40 where we adopt the measurement from \citet{bowler2024} based on ground-based observations. Given their faintness in the rest-UV, no $\beta_\mathrm{UV}$ measurements are available for the two dust-detected serendipitous neighbors, and we therefore omit them from the following analysis.

We obtain the infrared excess directly from the obscured fractions determined in the previous section, as $\mathrm{IRX} = \kappa_\mathrm{UV} / \kappa_\mathrm{IR} \times f_\mathrm{obs} / (1 - f_\mathrm{obs})$, where $\kappa_\mathrm{UV/IR}$ convert the measured UV and IR luminosities to the corresponding SFRs. Under the adopted conversions (Section \ref{sec:introduction}), the ratio $\kappa_\mathrm{UV} / \kappa_\mathrm{IR} \approx 1.21$ \citep[see also][]{mclure2018}.

We show the $\mathrm{IRX}-\beta_\mathrm{UV}$ relation of our targets, as well as the subset of $z\gtrsim5$ LBGs with multi-band dust and $\beta_\mathrm{UV}$ measurements in the literature, in Figure \ref{fig:IRX_beta}. We anchor the relation to $\beta_\mathrm{UV,0} = -2.43$, which is the average value inferred for the REBELS-IFU sample by \citet{fisher2025b}. 

The multi-band REBELS sample is consistent with both the standard Calzetti and SMC attenuation curves, with the dual-band stack falling in between the two relations at a weighted-mean $\beta_\mathrm{UV} = -1.67 \pm 0.07$. This is in agreement with earlier analyses of the REBELS sample by \citet{bowler2024} and \citet{fisher2025b}, which reached similar conclusions based on single-band infrared luminosities, and hence required assuming a dust temperature. Our multi-band measurements now show that the typical dust temperature of our targets is indeed similar to that previously assumed, lending support to these earlier results.

Individual galaxies scatter around the two relations, ranging from slightly above the Calzetti curve (e.g., REBELS-18 and REBELS-32) to slightly below the SMC curve (e.g., REBELS-14 and REBELS-40). However, three of these four galaxies have only weakly constrained dust temperatures (Section~\ref{sec:results_MBB_fits}), such that their infrared luminosities, and thus IRX, remain uncertain. Together with the uncertainties in $\beta_\mathrm{UV}$ -- in particular for the ground-based measurement of REBELS-40 -- and in the intrinsic $\beta_\mathrm{UV,0}$, this scatter among our sample therefore does not necessarily imply genuine variations in the attenuation curve.

The literature LBGs follow a similar distribution, although a few galaxies with well-constrained infrared luminosities, such as MACS0416-Y1, the Big Three Dragons and HZ10, fall above the Calzetti relation \citep{villanueva2024,bakx2025,rajulal2026}. Such offsets are generally attributed to the star-dust geometry, in which a patchy dust distribution allows part of the UV light to escape without being strongly reddened, while heavily obscured regions dominate the infrared luminosity \citep[e.g.,][]{faisst2017,popping2017_irx,behrens2018,liang2019,sommovigo2020,vijayan2024}.

Overall, we conclude that the bulk of dusty high-redshift LBGs are broadly consistent with the standard Calzetti and SMC attenuation curves. This contrasts with recent studies that have inferred relatively flat attenuation curves for some high-redshift galaxies \citep[e.g.,][]{markov2025,shivaei2025}, possibly reflecting an excess of large grains produced by core-collapse supernovae, whose extinction properties may differ from those commonly assumed for local galaxies \citep[e.g.,][]{mckinney2025,otaki2026b}.

For our targets, however, a supernova origin of the bulk of the dust appears less likely. As discussed in Section~\ref{sec:discussion_dust_production}, their relatively high dust masses suggest that additional processes beyond supernova production are required to build up their dust reservoirs. Moreover, the processing of grains in the ISM of such relatively evolved systems -- for example through shattering -- produces small grains that steepen the extinction curve \citep[e.g.,][]{hirashita2009,asano2014,narayanan2025}. 

The flat attenuation curves reported at high redshift are instead generally inferred for (stacks of) low-mass galaxies, which mostly appear dust-poor \citep[e.g.,][]{ciesla2025,markov2025,rodighiero2026}. Their dust reservoirs may therefore predominantly be due to supernova production, potentially resulting in different grain populations from those in our targets. We caution, however, that the $\mathrm{IRX}-\beta_\mathrm{UV}$ diagram does not uniquely constrain the underlying extinction curve: the observed attenuation depends not only on the grain properties that determine the extinction curve, but also on the star-dust geometry. Furthermore, variations in $\beta_\mathrm{UV,0}$ among galaxies would similarly shift their positions in the diagram. 

Overall, our results therefore suggest that the $\mathrm{IRX} - \beta_\mathrm{UV}$ relations of dusty LBGs at $z\sim7$ are broadly consistent with standard attenuation curves, potentially contrasting with the flatter relations inferred for dust-poor, low-mass galaxies based on \textit{JWST}/NIRSpec spectroscopy.

\subsubsection{Possible connection to high-redshift quiescent galaxies}
\label{sec:QG_connection}

\begin{figure*}
    \centering
    \includegraphics[width=0.95\textwidth]{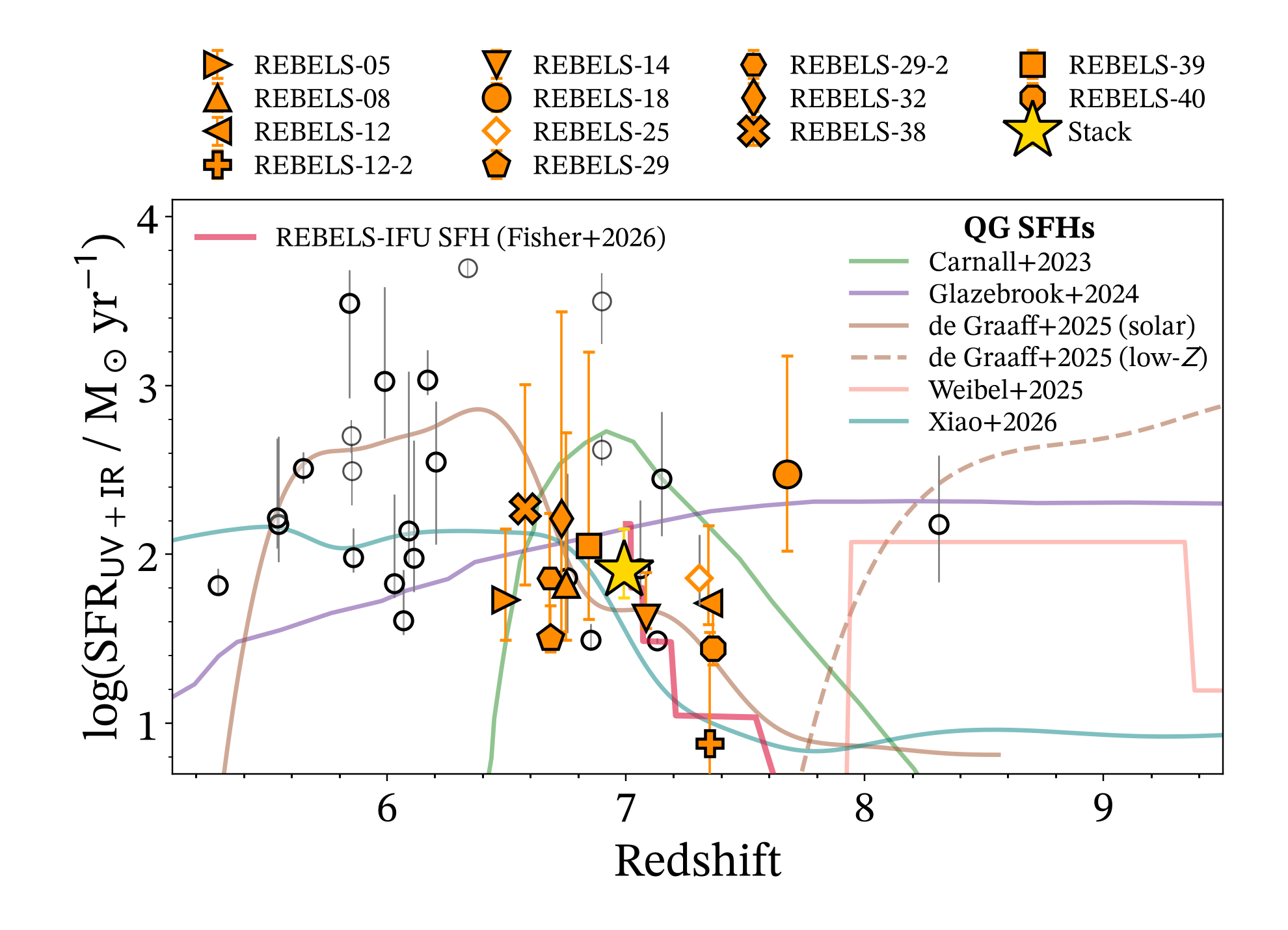}
    \vspace*{-0.4cm}
    \caption{\textbf{REBELS galaxies have star formation rates consistent with the reconstructed growth tracks of some high-redshift quiescent galaxies.} We plot the total (UV+IR) star formation rates of our targets (orange) and literature galaxies with multi-band far-infrared data (grey/black circles) as a function of redshift. The stack (yellow star) has a typical star formation rate of $\mathrm{SFR} = 81_{-25}^{+61}\,M_\odot\,\mathrm{yr}^{-1}$ while some of the REBELS galaxies with warm dust reach considerably higher values (e.g., $\mathrm{SFR} \sim 300\,M_\odot\,\mathrm{yr}^{-1}$ for REBELS-18 at $z=7.67$). The growth tracks of various individual quiescent galaxies \citep{carnall2023,glazebrook2024,degraaff2025,weibel2025} and a stack of $z\approx4-5$ QGs \citep{xiao2026} are overlaid. We also overplot the rising SFH inferred for the REBELS-IFU targets from \citet{fisher2025b} based on NIRSpec IFU data. REBELS galaxies have SFRs at $z\approx6.5 - 8$ similar to those of the reconstructed QG star formation histories, suggesting that (a subset of) them are plausible progenitors of this quiescent population.}
    \label{fig:SFR_vs_redshift}
\end{figure*}

We determine the total star formation rates of our targets by summing their UV and infrared contributions. As recently discussed by \citet{fisher2025b}, the likely rising star formation histories (SFHs) of our targets imply that recently formed, massive stars contribute disproportionately to both their UV emission and dust heating. As a result, these tracers best probe star formation on typical timescales of $\sim20-30\,\mathrm{Myr}$, rather than the canonically assumed $\sim100\,\mathrm{Myr}$ \citep[see also][]{mcclymont2025_burstiness}.

We show the total SFRs of our sample and the multi-band literature galaxies as a function of redshift in Figure \ref{fig:SFR_vs_redshift}. For REBELS-12-2 and REBELS-29-2, we show only the IR-based SFR given the lack of a detection in the rest-UV. The SFRs of our sample range from $\mathrm{SFR}_\mathrm{UV+IR} \sim 10 - 300\,M_\odot\,\mathrm{yr}^{-1}$, and the stack of the 10 primary targets yields an average $\mathrm{SFR}_\mathrm{UV+IR} = 81_{-25}^{+61}\,M_\odot\,\mathrm{yr}^{-1}$. Four sources -- REBELS-18, REBELS-32, REBELS-38 and REBELS-39 -- have median total SFRs which exceed $100\,M_\odot\,\mathrm{yr}^{-1}$. 

REBELS-18, in particular, has a high inferred total star formation rate of $\mathrm{SFR}_\mathrm{UV+IR} = 300_{-190}^{+1180}\,M_\odot\,\mathrm{yr}^{-1}$, which is especially remarkable given that it is our highest-redshift target, at $z=7.67$. While such poorly constrained upper bounds are a direct consequence of the limited frequency coverage, independent physical arguments could help to exclude the most extreme solutions. Extreme value statistics, for example, predict the most massive or luminous galaxy expected within a given survey volume in $\Lambda$CDM \citep{lovell2023,lovell2026,enriquez-vargas2026}. Incorporating such predictions as physically motivated priors on the SFRs, and hence dust temperatures, of the most extreme galaxies presents an interesting avenue for future work.

The large star formation rates of our targets raise the possibility that the most IR-luminous ones among them could be progenitors of the high-redshift quiescent galaxies (QGs) that \textit{JWST} has been discovering in growing numbers \citep[e.g.,][]{carnall2023,glazebrook2024,nanayakkara2024,baker2025,baker2025b,zhang2026_qgs}. While a detailed exploration of this possible connection is beyond the scope of the current work, we overlay possible QG growth tracks on Figure \ref{fig:SFR_vs_redshift} for a visual comparison. In particular, we adopt the fiducial SFH of a $z=4.66$ QG from \citet{carnall2023}; the parametric SFH from \citet{glazebrook2024} of a $z=3.205$ QG found to have formed most of its stars at $z > 10$; the fiducial (lower-metallicity) and alternative (higher-metallicity) growth tracks of a $z=4.90$ QG from \citet{degraaff2025}; and the SFH of the $z=7.29$ QG from \citet{weibel2025}. Moreover, we adopt the average growth history of the most rapidly formed $z\approx4-5$ QGs drawn from the recent compilation by \citet{xiao2026}. We emphasize that the growth histories of individual QGs are notoriously uncertain, as they depend on, for example, the precise wavelength coverage, the adopted functional form of the SFH, the stellar population models, and the metallicity \citep[e.g.,][]{nanayakkara2025,turner2025,mcconachie2026}. These comparisons should therefore be taken as indicative, rather than as evidence for a direct evolutionary link between these populations.

\medskip 

The recent SFRs of our targets at $z\approx6.5-7$ agree with the average growth track of the most rapidly formed $z\sim4-5$ QGs from \citet{xiao2026}, while our targets at $z\gtrsim7.3$ lie above this track, which only reaches comparable SFRs at later times. The higher-metallicity model of \citet{degraaff2025} similarly predicts SFRs at $z\sim7$ comparable to those of the REBELS galaxies, followed by a burst at $z\approx6-6.5$. Moreover, at $z\approx7$, both tracks have assembled a stellar mass of $\log(M_\star/M_\odot)\approx9.5-9.7$, similar to the stellar masses of our targets ($\log(M_\star/M_\odot)=9.2-9.8$).

Notably, the average non-parametric SFH inferred for the REBELS-IFU sample by \citet[][see also \citealt{fisher2025}]{fisher2025b}, anchored to $z=7$, agrees reasonably well with the high-metallicity \citet{degraaff2025} and average \citet{xiao2026} tracks. This rapidly rising SFH was inferred from SED fits to the \textit{JWST}/NIRSpec data on the REBELS-IFU sources alone, and thus does not include the ALMA continuum data. Nevertheless, this SFH matches the stacked UV+IR star formation rate of our sample well, and the semi-independent constraints from rest-optical spectroscopy and far-infrared continuum emission thus paint a consistent picture: the REBELS galaxies not only exhibit recent SFRs comparable to those of the aforementioned QG growth tracks at $z\approx7$, but also appear to have followed a similar growth history up to this epoch.

Some of the more extreme QG star formation histories, on the other hand, peak earlier or at higher SFRs. The $z=7.29$ QG from \citet{weibel2025} and the lower-metallicity model of \citet{degraaff2025} reach SFRs comparable to or exceeding those of our targets, but at earlier times ($z\gtrsim8$). The SFH from \citet{carnall2023}, in turn, peaks at an SFR several times higher than that of the bulk of the REBELS galaxies, but comparable within the uncertainties to those of the warmest and most luminous of our targets. The SFH of the $z=3.2$ QG from \citet{glazebrook2024} reaches SFRs comparable to those of our most star-forming targets, including REBELS-18, at $z\approx7-8$. However, as its SFH suggests it has already been forming stars at this rate since $z\gtrsim10$, it has assembled a stellar mass of $M_\star \sim 10^{11}\,M_\odot$ by $z\approx7$, well in excess of the stellar masses of the REBELS galaxies. 

Taken together, IR-luminous LBGs at $z\approx7$ are plausible progenitors of at least part of the QG population at $z\approx4-5$, with the warmest and most luminous of our targets possibly evolving into some of its more extreme members. The progenitors of the earliest-forming QGs, in contrast, must have assembled the bulk of their stellar mass well before $z\approx7$, and would thus have been considerably more massive than the REBELS galaxies by this epoch.

\medskip

Since star formation in the most IR-luminous REBELS galaxies is already heavily obscured ($f_\mathrm{obs} \gtrsim 0.9$; Section~\ref{sec:discussion_obscured_SF}), our sample may capture only the subset of QG progenitors that remain sufficiently UV-luminous to be selected. A further, fully dust-obscured starbursting population would be missed by such UV-based selections altogether \citep[e.g.,][]{endsley2023,bing2025,sun2026,zavala2026}.

A quantitative test of this evolutionary connection would require a detailed comparison of the number densities of both populations, which is beyond the scope of this work. REBELS targets were drawn from UV-selected candidates across a relatively wide but not exceptionally large area of $\sim7\,\mathrm{deg}^2$ \citep{bouwens2022}, and the multi-band subset studied here does not constitute a complete sample. Wide-area (sub)millimeter surveys, such as the Cycle 11 ALMA Large Program CHAMPS \citep{faisst2026_champs,martinez2026_champs}, in combination with deep rest-UV and -optical ancillary data, are needed to establish the abundance of likely QG progenitors at $z\gtrsim7$, including those that are fully dust-obscured. 

\medskip

Finally, if there is indeed an evolutionary connection between $z \gtrsim 4$ QGs and luminous star-forming galaxies at $z\approx6-8$ such as those in REBELS, two key questions remain: what eventually quenches these galaxies, and what becomes of their dust? Given that quenching must be rapid to produce the earliest QGs, many mechanisms invoke some form of AGN feedback \citep{hartley2023,onoue2025,wu2025,chandro-gomez2026,valentino2026}. Currently, however, there is no clear evidence for AGN activity in our targets \citep{rowland2026}, although this is mainly due to the lack of high-resolution spectroscopy ($R\gtrsim1000$) that prevents the robust identification of broad Balmer emission.

Regarding their dust, $z\sim3 - 5$ QGs generally show little attenuation, and while far-infrared dust emission has now been detected in a number of them, they typically appear considerably less dust-rich than star-forming galaxies at similar epochs \citep[e.g.,][]{magdis2021,suzuki2022,lee2024,nanayakkara2025}. This implies that, if REBELS-like galaxies are indeed their progenitors, a large fraction of their dust reservoirs must be removed or destroyed within the $\lesssim1\,\mathrm{Gyr}$ separating these epochs. Possible channels include the ejection of dust through outflows or its destruction through thermal sputtering in hot gas. Recent studies have revealed a large diversity of dust-to-stellar mass and dust-to-gas ratios in quiescent galaxies, possibly reflecting a variety of quenching mechanisms and timescales \citep{lesniewska2025,spilker2025_qg,lorenzon2025,donevski2026}. Connecting this diversity to a dusty, star-forming progenitor population will require tracing the dust content of galaxies across the transition from star formation to quiescence at $z\gtrsim4$ with \textit{JWST} and ALMA.

\subsection{Dust production}
\label{sec:discussion_dust_production}

\begin{figure*}
    \centering
    \includegraphics[width=0.95\textwidth]{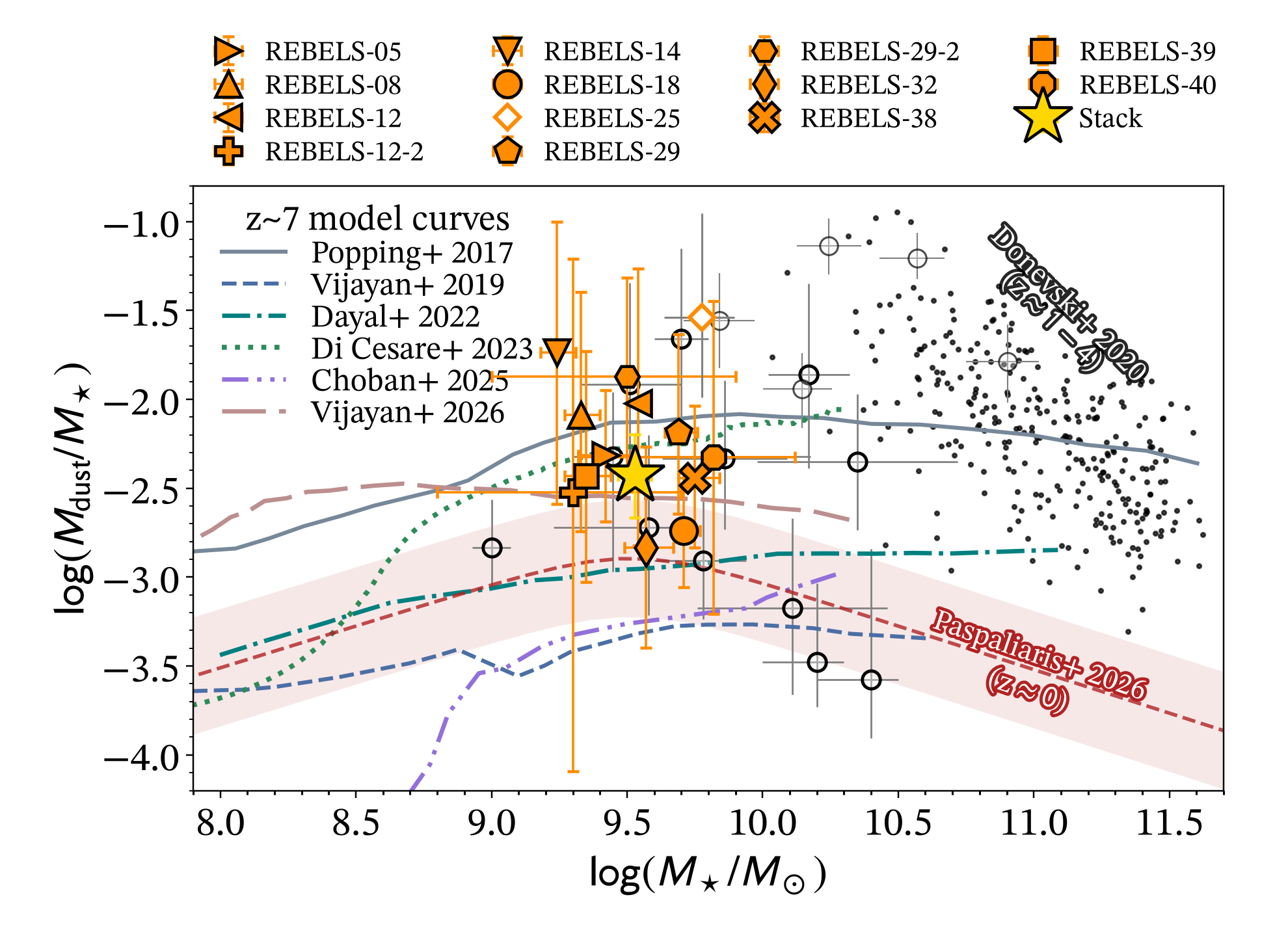}
    \vspace*{-0.4cm}
    \caption{\textbf{Dust-detected, UV-bright Lyman break galaxies at $z\approx7$ show an average dust-to-stellar mass ratio of $\mathbf{M_\mathrm{dust}/M_\star\approx0.4_{-0.2}^{+0.3}\%}$.} The dust-to-stellar mass ratio is plotted against stellar mass for $z \gtrsim 5$ dusty galaxies, $1 \lesssim z \lesssim 4$ DSFGs \citep{donevski2020}, and compared against model tracks from hydrodynamical simulations \citep{dicesare2023,choban2025,vijayan2026} and semi-analytical models \citep{popping2017,vijayan2019,dayal2022}. For comparison, we also plot the local relation from \citet[][red dashed line and shading]{paspaliaris2026}. Taken at face value, dust-to-stellar mass ratios of the REBELS sample are best reproduced by theoretical models in which ISM grain growth is efficient.}
    \label{fig:Mdust_Mstar_vs_Mstar}
\end{figure*}

Finally, we discuss the dust-to-stellar mass ratios ($M_\mathrm{dust}/M_\star$) of our targets in the context of early dust enrichment. This ratio is a powerful proxy for dust build-up \citep[e.g.,][]{donevski2020,dicesare2023,witstok2023_beta,schneider2024,bakx2026}, with most observational studies and theoretical works suggesting that low-mass galaxies are generally dust-poor, with their dust primarily produced by stars, while dust build-up in more massive galaxies is dominated by growth in the ISM \citep[e.g.,][]{remyruyer2014,devis2017,galliano2021,palla2024,osman2025}. However, the role of ISM grain growth at high redshift remains debated \citep[e.g.,][]{lesniewska2019,dayal2022}, with some studies questioning whether it can operate at all under the typical ISM conditions of early galaxies \citep{ferrara2016}.

Whether grain growth can proceed efficiently depends on its timescale, which in turn is set by the local properties of the ISM -- in particular the gas density, temperature and metallicity -- and must be compared with the time available for galaxies to evolve \citep[e.g.,][]{asano2013,graziani2020,choban2025}. The scarcity of dust detections at $z\gtrsim8$ has been interpreted as grain growth not yet having had sufficient time (or sufficient gas-phase metals) to become efficient at these early epochs, such that dust in the earliest galaxies may be predominantly of stellar origin \citep[][]{burgarella2025,burgarella2026,algera2026,bakx2026}. This interpretation, too, remains debated, as the compactness of high-redshift galaxies suggests even modest amounts of supernova dust could render them obscured \citep[e.g.,][]{ferrara2023,ferrara2025_bluemonsters}, although the resulting obscuration also depends on the grain properties and on the (possibly inhomogeneous) distribution of gas and dust in these rapidly assembling systems \citep[e.g.,][]{sommovigo2026}. Consequently, the origin of dust at $z\sim7$ is relevant not only to the dusty galaxy population at this epoch, but also to the possibly dust-poor one at even earlier times.

The multi-band REBELS sample spans a range of dust-to-stellar mass ratios of $-2.9 \lesssim \log(M_\mathrm{dust}/M_\star) \lesssim -1.8$, broadly consistent with $z\gtrsim5$ LBGs in the literature (Figure \ref{fig:Mdust_Mstar_vs_Mstar}). The dual-band stack yields $\log(M_\mathrm{dust}/M_\star) = -2.44 \pm 0.24$, equivalent to $M_\mathrm{dust}/M_\star = 3.6_{-1.5}^{+2.7} \times 10^{-3}$, which we take as typical for the dust-detected REBELS sample at $z\sim7$.

The few known DSFGs at $z\gtrsim5$ show dust-to-stellar mass ratios at the upper end of, or exceeding, the values from REBELS, spanning $-2 \lesssim \log(M_\mathrm{dust}/M_\star) \lesssim -1$. However, these measurements are somewhat uncertain given the possibility of optically thick dust resulting in overestimated dust masses (cf.\ Figure \ref{fig:Tdust_vs_Lir}) and/or stellar masses being underestimated due to strong dust obscuration \citep[e.g.,][]{li2026}. While the latter effect may also play a role in the particularly dust-rich REBELS-25 \citep{algera2026,cescon2026,rowland2026_r25}, the stellar masses of the REBELS sources presented here are less likely to be strongly biased by obscuration \citep{fisher2025,rowland2026}.

Compared to the few $z\gtrsim5$ DSFGs, the $z\approx1-4$ DSFG sample from \citet{donevski2020} shows a wider range of dust-to-stellar mass ratios of $-3 \lesssim \log(M_\mathrm{dust}/M_\star) \lesssim -1$, with an average value of $\log(M_\mathrm{dust}/M_\star) \approx -2.2$ that is slightly higher than, but broadly consistent with, that of the REBELS sample and $z\gtrsim5$ LBGs in the literature.

We also compare our measurements to the recent study of $M_\mathrm{dust}/M_\star$ in local galaxies by \citet[][red dashed line in Figure \ref{fig:Mdust_Mstar_vs_Mstar}]{paspaliaris2026}. Their sample spans a broad range of stellar masses, from $7 \lesssim \log(M_\star/M_\odot) \lesssim 11$. They find a positive correlation between $M_\mathrm{dust}/M_\star$ and stellar mass for low-mass (dwarf) galaxies, which turns over at $M_\star \approx 10^{9.5}\,M_\odot$ -- the average mass of our REBELS sample (Table \ref{tab:dust_properties}). At low masses, this trend likely reflects rapid dust enrichment as galaxies are building up their stellar mass, while at higher masses the turnover may suggest a low gas fraction limiting ISM dust growth or enhanced dust destruction \citep[see also e.g.,][]{delooze2020,galliano2021}. The $z\approx5 - 8$ LBG sample generally falls above the $z\approx0$ trend and its scatter, suggesting more efficient dust build-up in this high-redshift population at fixed stellar mass. 

\medskip

We proceed by comparing our $M_\mathrm{dust}/M_\star$ measurements to a variety of theoretical predictions, including those from hydrodynamical simulations and semi-analytical models (SAMs). In the former category, we compare to the {\sc{dustyGadget}} \citep{graziani2020,dicesare2023}, FIRE \citep{choban2025}, and COLIBRE simulations \citep{trayford2026,vijayan2026}. For the SAMs, we compare to the Santa Cruz \citep{popping2017}, L-GALAXIES \citep{vijayan2019}, and {\sc{delphi}} models \citep{dayal2022}.

Figure \ref{fig:Mdust_Mstar_vs_Mstar} reveals that the {\sc{dustyGadget}} simulations predict dust-to-stellar mass ratios similar to those observed in the REBELS sample. These simulations adopt the supernova dust yields from \citet{bianchi2007}, which imply that only a minor fraction ($\sim2-20\%$) of the dust formed in supernova ejecta survives the reverse shock. As a result, grain growth dominates the build-up of dust mass at $z\approx7$ at the high stellar mass end in {\sc{dustyGadget}}. 

The FIRE simulations, on the other hand, underpredict the measured dust-to-stellar mass ratios of our targets, both in their fiducial model and in a model with enhanced supernova dust yields and shorter grain growth timescales (only the latter is shown in Figure \ref{fig:Mdust_Mstar_vs_Mstar}). As discussed in \citet{choban2025}, this is likely due to the relatively low metallicities the galaxies attain in the simulations as a result of their bursty star formation histories and strong outflows, which limit the metal mass available for grain growth. 

The recent COLIBRE simulations predict dust-to-stellar mass ratios in good agreement with those inferred for our targets. These simulations couple a live dust model to an explicit treatment of the cold ISM, and allow for efficient grain growth in unresolved dense molecular clouds through a density-dependent clumping factor \citep{trayford2026}. In COLIBRE, grain growth is the dominant dust build-up mechanism at stellar masses $M_\star \gtrsim 10^8\,M_\odot$ at $z\approx7$, where $M_\mathrm{dust}/M_\star$ becomes nearly independent of stellar mass at $\log(M_\mathrm{dust}/M_\star) \approx -2.5$ \citep{vijayan2026}, similar to the typical value we measure.

Moving on to the SAMs, we find that the fiducial \citet{popping2017} model matches or even exceeds the dust-to-stellar mass ratios measured for the REBELS sample. In this model too, grain growth dominates the dust build-up of massive galaxies at $z\approx7$, aided by the re-accretion of previously ejected metals and dust. The {\sc{delphi}} SAM \citep{dayal2022}, on the other hand, predicts a lower typical $M_\mathrm{dust}/M_\star \approx 10^{-3}$ at $z=7$. This is despite its relatively high fiducial SN dust yield of $0.5\,M_\odot$ per SN, as efficient dust destruction and ejection substantially reduce the net dust yield, while grain growth remains subdominant at these early times. {\sc{delphi}} matches the least dusty subset of LBGs with multi-band observations, but underpredicts the typical $M_\mathrm{dust}/M_\star$ by $\sim0.5\,\mathrm{dex}$. Lastly, the \citet{vijayan2019} SAM predicts a dust-to-stellar mass ratio that is $\sim1\,\mathrm{dex}$ lower than that measured in the $z\gtrsim5$ LBG population. In their models, grain growth is the dominant channel of dust build-up at this epoch. However, as grain growth is restricted to molecular clouds, and dust is moreover set to be completely destroyed in major mergers, the overall build-up of dust is delayed, resulting in limited dust masses by $z\approx7$.

\medskip

In summary, the dust-to-stellar mass ratios of the multi-band REBELS and $z\gtrsim5$ literature LBGs are best reproduced by models in which grain growth efficiently builds up the dust mass by $z\approx7$ \citep{popping2017,dicesare2023,vijayan2026}, whereas models in which SN production dominates the dust budget \citep{dayal2022}, or in which grain growth is delayed or limited by the available metals \citep{vijayan2019,choban2025}, underpredict them. However, the differences among these theoretical predictions do not solely reflect whether, or how efficiently, grain growth operates. Rather, they reflect the combined effects of the adopted SN dust yields and their survival, the efficiencies of grain growth and destruction, and the metal enrichment histories of the model galaxies as regulated by feedback. Consequently, the agreement of a given model with the observed dust-to-stellar mass ratios does not uniquely identify the dominant channel of dust build-up.

It must also be borne in mind that the galaxies with multi-band dust continuum data constitute a biased subset of especially dust-rich systems. As galaxies of similar stellar mass can differ substantially in dust content depending on their star formation histories and chemical enrichment \citep[e.g.,][]{palla2024,dangelo2026}, we are likely preferentially selecting those that have already developed the ISM conditions required for efficient grain growth to take place. Combining multi- and single-band dust masses, as well as upper limits for continuum-undetected galaxies, with gas-phase metallicities will enable population-wide constraints on dust-to-metal ratios, helping to distinguish between differences in metal enrichment histories and variations in the efficiency of grain growth and destruction. We explore this in more detail in Algera et al.\ (in preparation).

\section{Conclusions}
\label{sec:conclusions}

We have presented a comprehensive overview of the multi-band ALMA observations conducted for a sample of thirteen $z = 6.5 - 7.7$ galaxies from the REBELS survey, comprising ten primary targets and three serendipitously \ciil{}-detected companions that lie within the same pointings (REBELS-12-2, REBELS-29-2 and REBELS-39-2). In total, the data consist of 31 distinct ALMA observations drawn from 8 unique observing projects. All thirteen galaxies have been observed in Bands 6 and 8, with one (REBELS-38) also having additional observations carried out in Bands 3 and 5.

We homogeneously imaged these data and obtained the multi-band flux densities of our targets in a consistent manner through $(u,v)$-plane model fitting. We performed optically thin modified blackbody fitting for this multi-band sample using a novel `inverse temperature' prior, which we demonstrate yields considerably less biased dust properties at limited frequency coverage and signal-to-noise than a standard flat temperature prior. We draw the following conclusions:

\begin{itemize}

    \item \textbf{The typical dust temperature of dust-detected, UV-bright LBGs at $\boldsymbol{z\sim7}$ is $\boldsymbol{T_\mathrm{dust} = 49_{-8}^{+11}\,\mathrm{K}}$.} We thus recommend adopting a typical dust temperature of $50\,\mathrm{K}$ for this population in the absence of multi-band data. That said, the measured dust temperatures of our sample span a broad range of $T_\mathrm{dust} \approx 30 - 70\,\mathrm{K}$ given a fiducial $\beta_\mathrm{IR} = 2.0$. Within the $z > 5$ Lyman break galaxy population alone, we find no significant evidence for redshift evolution of $T_\mathrm{dust}$. 

    \item \textbf{Most star formation in dust-detected LBGs at $z\sim7$ is obscured.} We infer an average obscured fraction of $f_\mathrm{obs} = 0.75 \pm 0.11$ from a dual-band stack of our sample, with a range of $f_\mathrm{obs} \approx 0.4 - 0.95$ for the individual targets. While biased to previously dust-detected galaxies, this shows obscured star formation to frequently be dominant even in galaxies initially selected to be UV-bright.

    \item \textbf{The most IR-luminous galaxies within REBELS are plausible progenitors of some high-redshift quiescent galaxies.} We infer star formation rates of $\mathrm{SFR}_\mathrm{UV+IR} \approx 110 - 300\,M_\odot\,\mathrm{yr}^{-1}$ for four of our targets, which intersect plausible growth histories of a subset of $z > 4$ quiescent galaxies. These SFRs are $f_\mathrm{obs}\approx75 - 95\%$ obscured, suggesting that a population of UV-fainter galaxies with similarly high obscured SFRs may still remain hidden.

    \item \textbf{On average, UV-luminous LBGs lie between the Calzetti and SMC $\mathrm{IRX}-\beta_\mathrm{UV}$ relations.} This suggests standard dust attenuation laws generally apply for this dusty high-redshift population.
    
    \item \textbf{The typical dust-to-stellar mass ratio of dust-detected, UV-bright LBGs at $\boldsymbol{z\sim7}$ is $\boldsymbol{M_\mathrm{dust}/M_\star = 0.4_{-0.2}^{+0.3}\%}$.} Based on a comparison to simulations and semi-analytical models, such dust masses are best reproduced by models in which ISM dust growth is efficient.
    
\end{itemize}

This analysis has more than doubled the current number of $z>6.5$ LBGs detected in ALMA Band 8 continuum emission, from six to 14. Given the presence of several galaxies with seemingly warm or hot dust, subsequent follow-up in higher-frequency ALMA bands is necessary to more accurately constrain their dust temperatures, masses, and IR luminosities. 

The $z\sim7$ galaxy population may mark an important crossroads in early galaxy evolution. Only $\sim100 - 200\,$Myr earlier, at $z\gtrsim8$, the galaxy population appears largely dust-poor \citep[e.g.,][]{burgarella2025,burgarella2026,ciesla2025,algera2026,algera2026b,bakx2026}, suggesting that dust becomes detectable over a short cosmic period. At the same time, the massive quiescent galaxies now found at $z>4$ \citep[e.g.,][]{carnall2023,baker2025,degraaff2025} likely formed their stars in powerful, heavily obscured starbursts at $z>6$, at rates comparable to or even exceeding those we infer for the dusty, UV-bright galaxies studied here. Both a deep, systematic search for dust at $z>8$ and the blind identification of fully obscured galaxies at $z>6$ require a substantial increase in the sensitivity and mapping speed of ALMA, such as envisioned by the ALMA2040 concept \citep[][]{algera2026_alma2040,sun2026_alma2040}. Such observations would allow us to trace the build-up of dust and obscured star formation across the first billion years of cosmic history, from the emergence of the first dusty galaxies to the formation of the first quiescent systems.

\section*{Acknowledgments}

\noindent This paper makes use of the following ALMA data: \\
ADS/JAO.ALMA\#2019.1.01634.L, 
ADS/JAO.ALMA\#2021.1.00318.S, 
ADS/JAO.ALMA\#2021.1.01297.S, 
ADS/JAO.ALMA\#2021.1.01389.S, 
ADS/JAO.ALMA\#2022.1.00446.S, 
ADS/JAO.ALMA\#2022.1.01131.S, 
ADS/JAO.ALMA\#2022.1.01384.S, 
ADS/JAO.ALMA\#2024.1.00406.S. 

ALMA is a partnership of ESO (representing its member states), NSF (USA) and NINS (Japan), together with NRC (Canada), NSTC and ASIAA (Taiwan), and KASI (Republic of Korea), in cooperation with the Republic of Chile. The Joint ALMA Observatory is operated by ESO, AUI/NRAO and NAOJ.

We acknowledge the use of Claude (Anthropic; Opus 5 and 5.5) to assist with code development, data visualization, and text editing. All AI-assisted code was verified by the authors, and the analysis was validated through independent tests. The authors take full responsibility for the content of this paper.

HSBA gratefully acknowledges support from Academia Sinica through grant AS-PD-1141-M01-2. CDC acknowledges support from the European Union (ERC, AGENTS, 101076224). Y.F. acknowledges support from JSPS KAKENHI Grant Numbers JP23K13149 and 26K00736.

\bibliographystyle{mn2e}
\bibliography{main}

\appendix

\section{Tapered imaging}
\label{app:tapered_imaging}

\begin{figure*}
    \centering
    \includegraphics[width=0.95\textwidth]{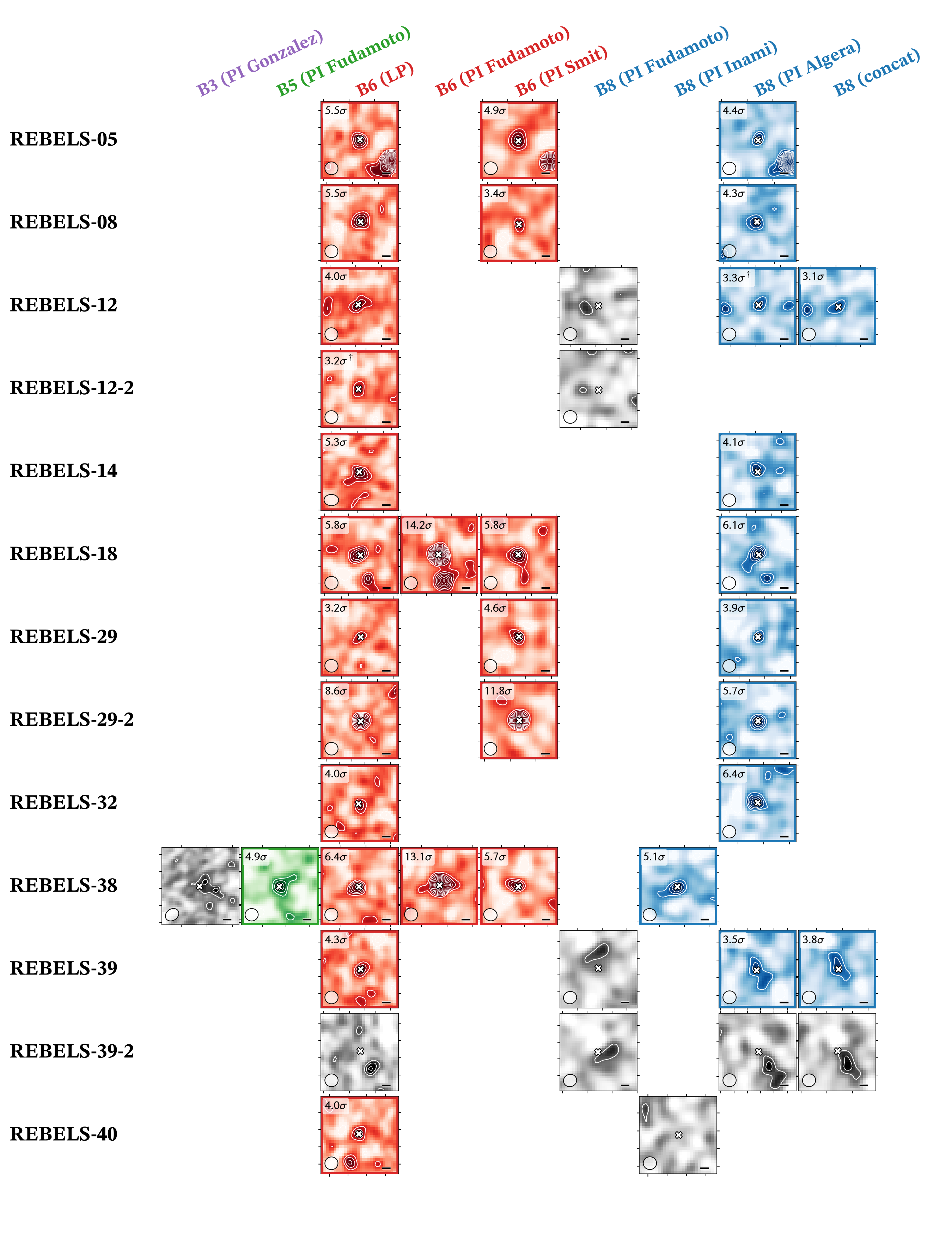}
    \vspace*{-1cm}\caption{\textbf{Compilation of the $1.5''$-tapered, multi-band ALMA observations for the REBELS sample presented in this work.} The cutouts span a fixed area of $8''\times 8''$ and colors and contours are as in Figure \ref{fig:allCutouts}. Galaxies whose S/N is annotated with a dagger are detected only in the tapered maps. All targets except for REBELS-39-2 are detected in at least one band, and 10/13 targets are detected in two or more bands.}
    \label{fig:allCutoutsTapered}
\end{figure*}

\begin{figure*}
    \centering
    \includegraphics[width=0.95\textwidth]{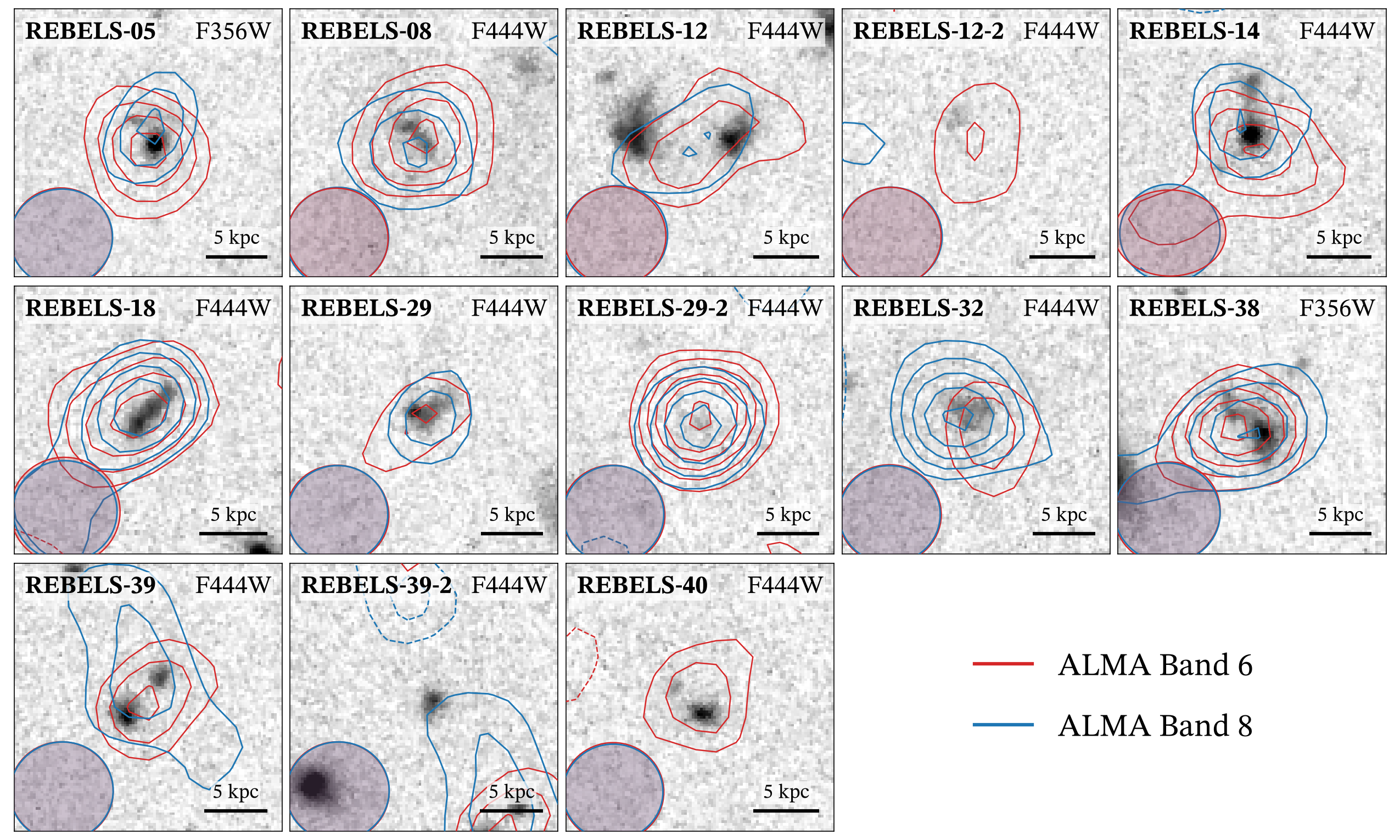}
    \caption{\textbf{$1.5''$-tapered ALMA Band 6 (red) and Band 8 (blue) contours overlaid on NIRCam F356W or F444W imaging.} Data and contours are as in Figure \ref{fig:NIRCamALMAoverlays}.}
    \label{fig:NIRCamALMAoverlaysTapered}
    \vspace*{0.2cm}
\end{figure*}

We show the full set of $1.5''$-tapered dust continuum images in Figure \ref{fig:allCutoutsTapered}. We also show the tapered Band 6 and 8 images overlaid on NIRCam F356W or F444W imaging in Figure \ref{fig:NIRCamALMAoverlaysTapered}. Cutouts with the naturally weighted ALMA data are shown in the main body of the paper (Figures \ref{fig:allCutouts} and \ref{fig:NIRCamALMAoverlays}).

\section{Multi-band centroids and positional offsets}
\label{app:spatialOffsets}

\begin{figure*}
    \centering
    \includegraphics[width=0.95\textwidth]{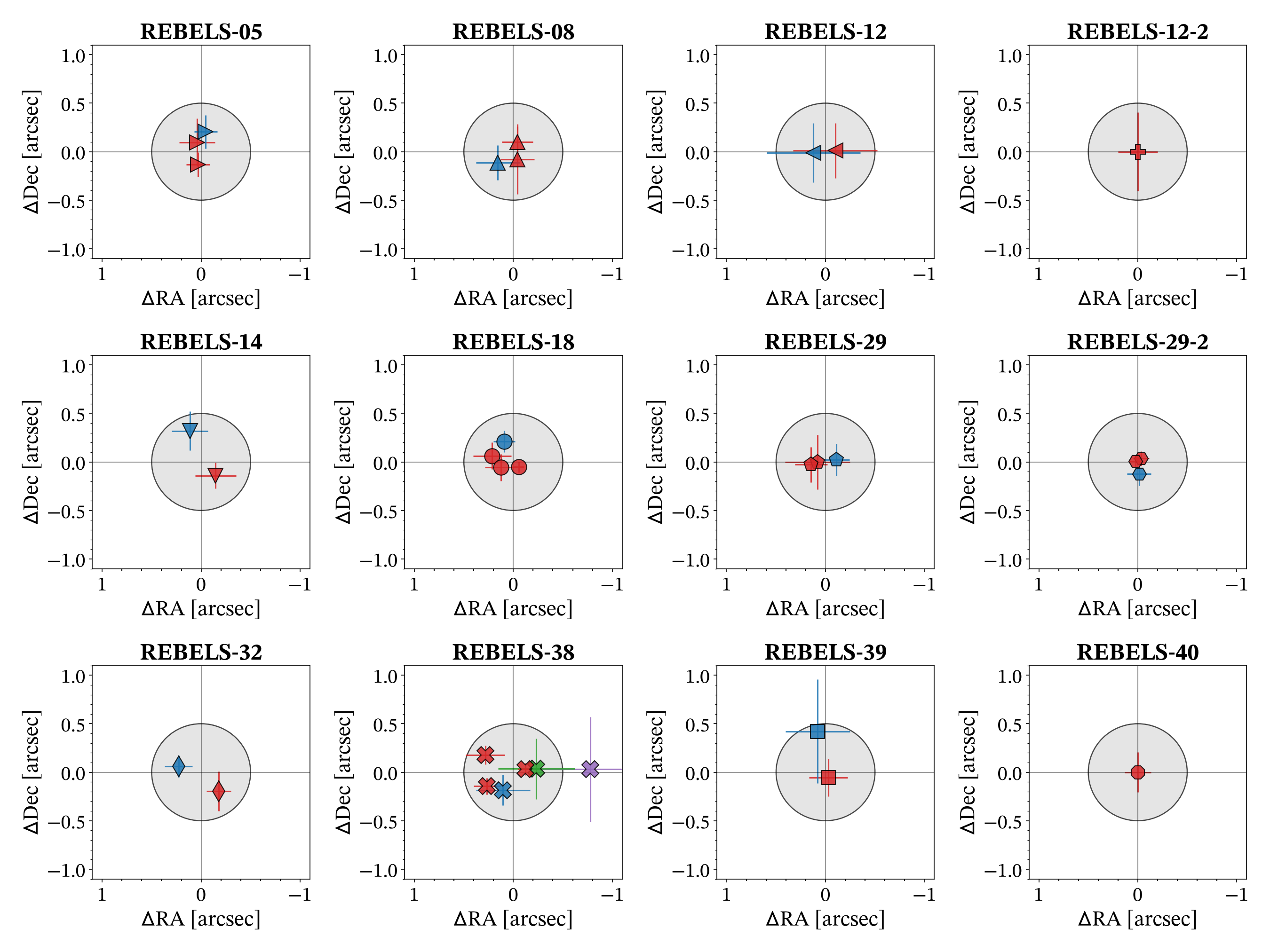}
    \caption{\textbf{The multi-band dust continuum positions of our targets are consistent to within $\boldsymbol{0.5''}$.} We show the offset in RA and Dec with respect to the weighted average coordinates for each of our targets (east is to the left). Points are colored by band (purple, green, red and blue for Bands 3, 5, 6 and 8, respectively), and the grey circle with a radius of $0.5''$ is added to guide the eye. All points except for the faint Band 3 emission in REBELS-38 are consistent with the fiducial position to within $0.5''$. }
    \label{fig:positionalOffsets}
\end{figure*}

As discussed in Section \ref{sec:centroids}, we adopt a fixed, inverse-variance-weighted dust continuum centroid for our targets across all bands, which is necessary to, for example, measure aperture flux densities in a consistent manner across wavelength. We show the offsets between the per-band centroids and the inverse-variance-weighted centroids -- both obtained in $1.5''$-tapered images -- in Figure \ref{fig:positionalOffsets}. In all cases except for the Band 3 observations of REBELS-38, the per-band centroids lie within $0.50''$ of the inverse-variance-weighted centroid, corresponding to one-third of the synthesized beam, with a mean separation of $0.17'' \pm 0.10''$. The Band 3 emission of REBELS-38 is offset by $\sim0.8''$ and implies a flux density well above the extrapolation of its MBB fit, suggesting it is boosted by, or entirely due to, noise. We therefore exclude it when computing the centroid of REBELS-38, and treat the Band 3 data as an upper limit in our analysis.

\section{A systematic comparison of flux density measurements}
\label{app:flux_measurement_comparison}

\begin{figure*}
    \centering
    \includegraphics[width=0.95\textwidth]{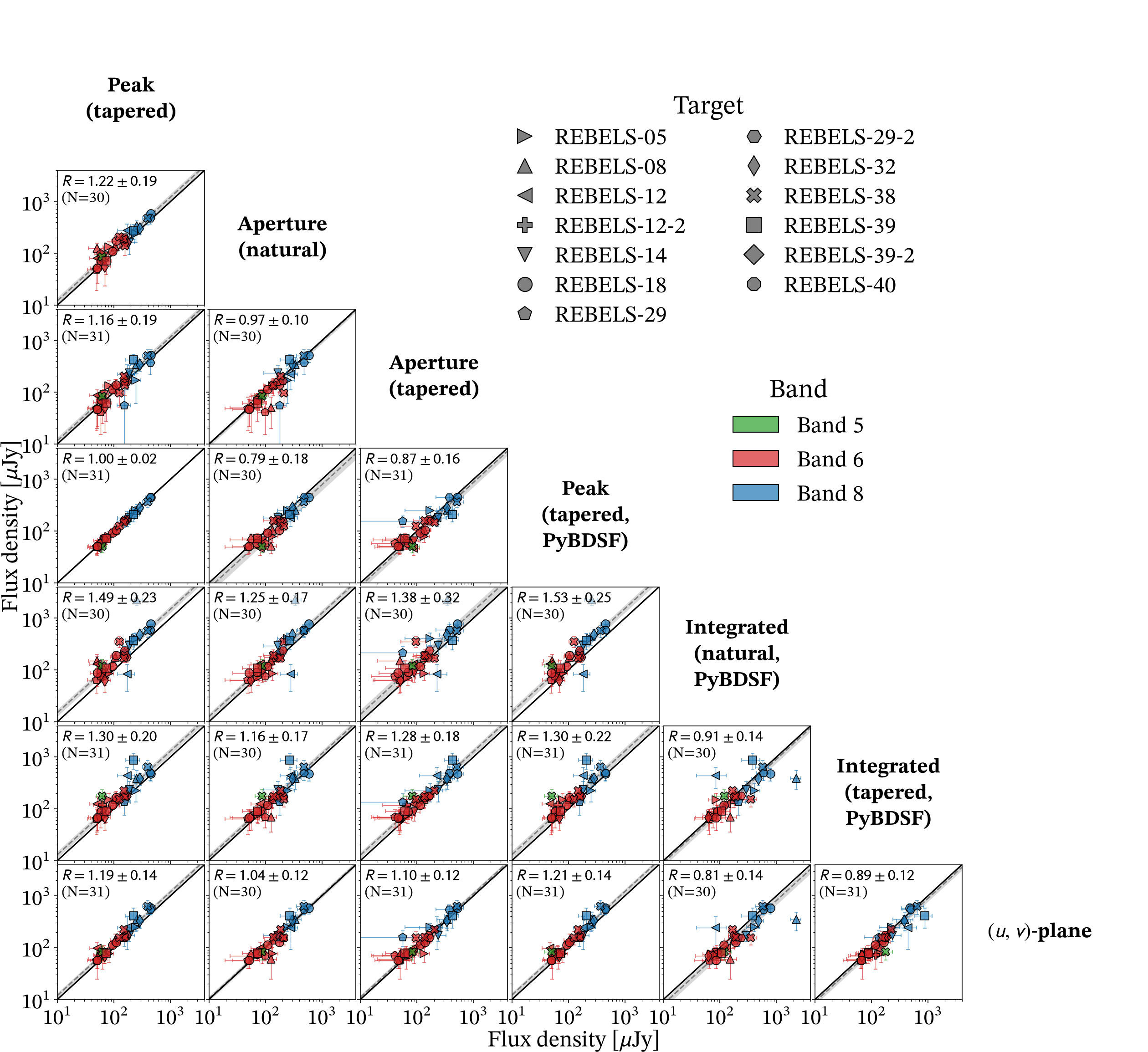}
    \caption{\textbf{Matrix plot comparing the flux densities of our sources measured in various ways.} The column header (row header) corresponds to the flux plotted on the $x$-axis ($y$-axis), such that the top left panel shows the naturally weighted aperture flux ($y$) against the tapered peak flux density ($x$). Galaxies are color-coded by ALMA Band, and upper limits are omitted from the plot, as are galaxies for which no reliable integrated flux could be determined. The number of sources used in the comparison is annotated on each panel, in addition to the median ratio ($R = y/x$) and median absolute deviation (MAD), which are also shown through the dashed line and grey shading. The solid black line represents the one-to-one relation. Our fiducial $(u,v)$-plane fluxes (bottom row) exceed the tapered peak fluxes by a median factor of $R=1.19\pm0.14$, agree with the aperture fluxes ($R=1.04\pm0.12$; naturally weighted, and $R=1.10\pm0.12$; tapered), and fall slightly below integrated fluxes ($R=0.81 \pm 0.14$; naturally weighted, and $R=0.89 \pm 0.12$; tapered).}\label{fig:flux_comparison_matrix}
\end{figure*}

\begin{figure*}
    \centering
    \includegraphics[width=0.95\textwidth]{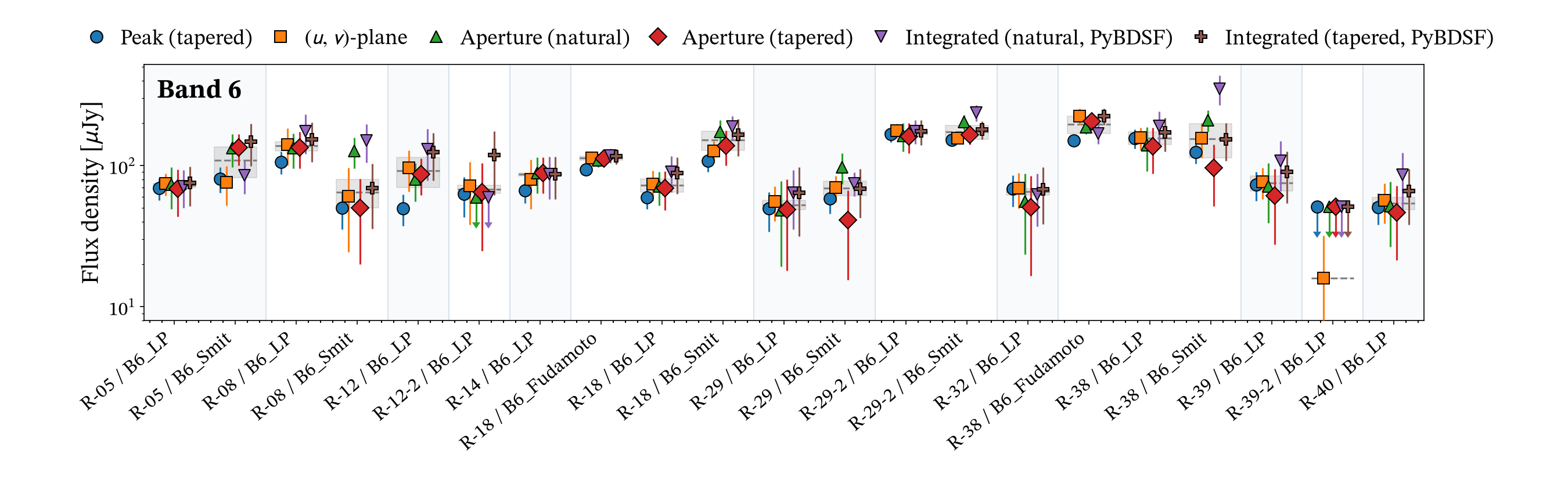}
    \includegraphics[width=0.95\textwidth]{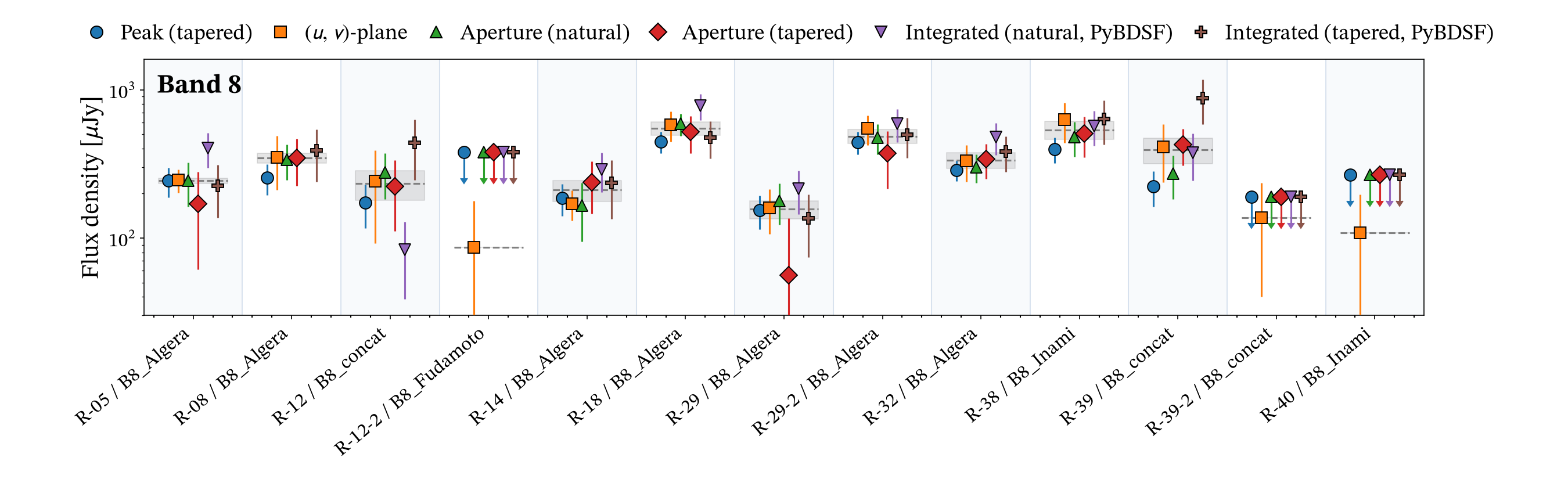}
    \caption{\textbf{ALMA Band 6 (\textit{top}) and Band 8 (\textit{bottom}) flux densities per galaxy, as obtained through various methods.} For each project, we compare six flux densities and show their median and typical spread through the grey dashed line and shading. Note that while we adopt $3\sigma$ upper limits for the image-plane fluxes in case of a non-detection, we use the raw fitted values for the $(u,v)$-plane measurements (orange squares). }\label{fig:flux_comparison_B6_B8}
\end{figure*}

In Section \ref{sec:methods_fluxdensity_measurements}, we adopt $(u,v)$-plane model fitting as our fiducial means of obtaining the flux densities of our targets. However, we explore several additional techniques, as discussed in detail below: peak flux density measurements (Section \ref{app:methods_peak_fluxes}), integrated fluxes from 2D Gaussian fitting (Section \ref{app:methods_gaussian_fitting}), and aperture flux densities (Section \ref{app:methods_aperture_fluxes}). A detailed comparison between the different flux density measurements in the form of a matrix plot is shown in Figure \ref{fig:flux_comparison_matrix}. In Figure \ref{fig:flux_comparison_B6_B8}, we compare these fluxes on a per-galaxy basis for Bands 6 and 8, which make up the bulk of our data.

\subsection{Peak-pixel flux densities}
\label{app:methods_peak_fluxes}

For an unresolved source, the peak flux density (formally an intensity, given its units of $\mathrm{Jy\,beam}^{-1}$) accurately represents its total flux. It is generally obtained from 2D Gaussian fitting, where the peak flux then represents the amplitude of the Gaussian while its uncertainty equals the (local) RMS \citep[e.g.,][]{condon1997}. For a fully unresolved source, whose apparent size thus equals that of the (Gaussian) synthesized beam, this is equivalent to reading off the flux density in the peak pixel of the image, at least in an idealized, noiseless scenario. In practice, we read off the maximum pixel value within a circular region centered on the dust continuum centroid of our target, with a radius of half the beam FWHM (adopting the geometric mean of its major and minor axes, i.e., $\approx0.75''$ for the $1.5''$-tapered maps), which comfortably encloses all positional offsets between the different bands (Section~\ref{sec:centroids}). We refer to this measurement as the `peak-pixel flux density' to distinguish it from the peak flux density obtained from Gaussian fitting (Section \ref{app:methods_gaussian_fitting}). In any case, we find that these two peak-flux methods yield fully consistent results (Figure \ref{fig:flux_comparison_matrix}).

Given the range of angular resolutions of our observations, not all of our targets are expected to be unresolved in each naturally weighted map. We therefore measure peak-pixel flux densities in tapered images, adopting a fixed $(u,v)$-taper of $1.5''$. We note that selecting the brightest pixel within a given radius may bias peak-pixel flux densities high at low S/N. That these nevertheless fall below all integrated and aperture flux densities therefore suggests that our targets are marginally resolved even at $1.5''$ resolution (Figure \ref{fig:flux_comparison_matrix}).

\subsection{Two-dimensional Gaussian fitting}
\label{app:methods_gaussian_fitting}

A common approach to measuring the fluxes of (moderately resolved) galaxies in radio and (sub)millimeter maps is through 2D Gaussian fitting. This is motivated by the fact that, once convolved with the clean beam, galaxies are often well described by one or more Gaussian components. Their flux density is then simply defined as the integrated area under the 2D Gaussian(s), and is therefore often referred to as the `integrated flux'. A downside of this approach, however, is that at limited S/N integrated fluxes are more readily biased high than peak flux densities, as nearby noise peaks can inflate the fitted source size \citep[e.g.,][]{condon1997,algera2020a,leslie2020}. This is most evident in Figure \ref{fig:flux_comparison_matrix}, where the integrated flux density determined from naturally weighted, high-resolution data can in some cases far exceed the other flux density measurements (e.g., the Band 8 data for REBELS-08). We therefore focus on integrated and peak flux densities obtained from the 2D Gaussian fitting carried out with {\sc{PyBDSF}} on the $1.5''$-tapered images (Section \ref{sec:methods_source_detection}), and show the integrated fluxes from the naturally weighted images for comparison only.

\subsection{Aperture flux densities}
\label{app:methods_aperture_fluxes}

With a well-defined source center per target (Section \ref{sec:centroids}), it is possible to consistently derive aperture flux densities across multiple bands. In practice, we adopt simple circular apertures and obtain the flux measurements via {\sc{interferopy}} \citep{boogaard2021}. The aperture radius is set to be $1.2\times$ the geometric mean of the synthesized beam FWHM ($\bar{\theta} = \sqrt{\theta_M \theta_m}$, where $\theta_M$ and $\theta_m$ denote the major and minor axes), or $1.0''$, whichever is larger. This ensures that the aperture exceeds the typical source size in the high-resolution observations, and the beam size in the low-resolution and tapered data, such that $\gtrsim95\%$ of the flux of a marginally resolved source is enclosed. For the $1.5''$-tapered images, this corresponds to an aperture radius of $1.8''$. We show the aperture fluxes obtained from the naturally weighted and $1.5''$-tapered maps in Figures \ref{fig:flux_comparison_matrix} and \ref{fig:flux_comparison_B6_B8}.

We note that aperture flux measurements, in particular in high-resolution observations, can be biased high due to the so-called \citet{jorsater1995} effect \citep[see also e.g.,][]{novak2019,czekala2021,meyer2022}. This arises because the dirty and clean beams are generally not identical in area, while the residual flux in the dirty map is nevertheless summed together with the fluxes in the cleaned components to create the final, clean image. This bias can be circumvented through very deep cleaning (such that little residual flux remains), or through the `residual scaling method', which effectively computes the ratio of the dirty-to-clean-beam area, and uses this to correct the total flux in the residual components \citep[e.g.,][]{novak2019,czekala2021}. Tapering also reduces this effect, as the tapered dirty beam more closely resembles a Gaussian. We therefore do not explicitly correct for the \citet{jorsater1995} effect in this work, but note that aperture flux measurements in the naturally weighted, high-resolution data may be biased high (e.g., the PI Smit Band 6 data for REBELS-08 and REBELS-29; Figures~\ref{fig:flux_comparison_matrix} and \ref{fig:flux_comparison_B6_B8}).

\subsection{Summary of the flux density comparison}
\label{app:methods_flux_comparison}

From our systematic comparison of different flux measurements, we can draw the following conclusions: naturally weighted integrated fluxes generally yield the highest values, and show several strong outliers due to complex source morphologies and low-S/N data (Figure~\ref{fig:flux_comparison_matrix}). Integrated fluxes obtained from $1.5''$-tapered maps are better behaved, and only slightly exceed our fiducial $(u,v)$-plane measurements (median ratio of $0.89 \pm 0.12$). Aperture fluxes from both the naturally weighted ($1.04 \pm 0.12$) and tapered ($1.10 \pm 0.12$) images moreover agree well with the $(u,v)$-plane measurements. However, the former can be affected by the aforementioned \citet{jorsater1995} effect, potentially biasing the flux in high-resolution or multi-configuration data, while the latter may be noisier for individual sources, owing to the reduced sensitivity of tapered images and the large apertures, which may include faint or empty pixels. Finally, peak flux densities are consistently slightly lower than those from the other approaches, by a median factor of $1.19 \pm 0.14$ compared to our fiducial $(u,v)$-plane fits, suggesting that most of our targets are marginally resolved even at $1.5''$ resolution. This thus motivates our choice of $(u,v)$-plane fluxes as fiducial (Section \ref{sec:methods_fluxdensity_measurements}). We explore how the choice of flux density measurement impacts the inferred dust properties in Appendix \ref{app:mbb_properties_across_flux_densities}.

\section{Band 5 and 6 flux density consistency checks}
\label{app:fluxdensity_measurements_B6consistency}

\begin{figure*}
    \centering
    \begin{minipage}[c]{0.65\textwidth}
        \centering
        \includegraphics[width=\textwidth]{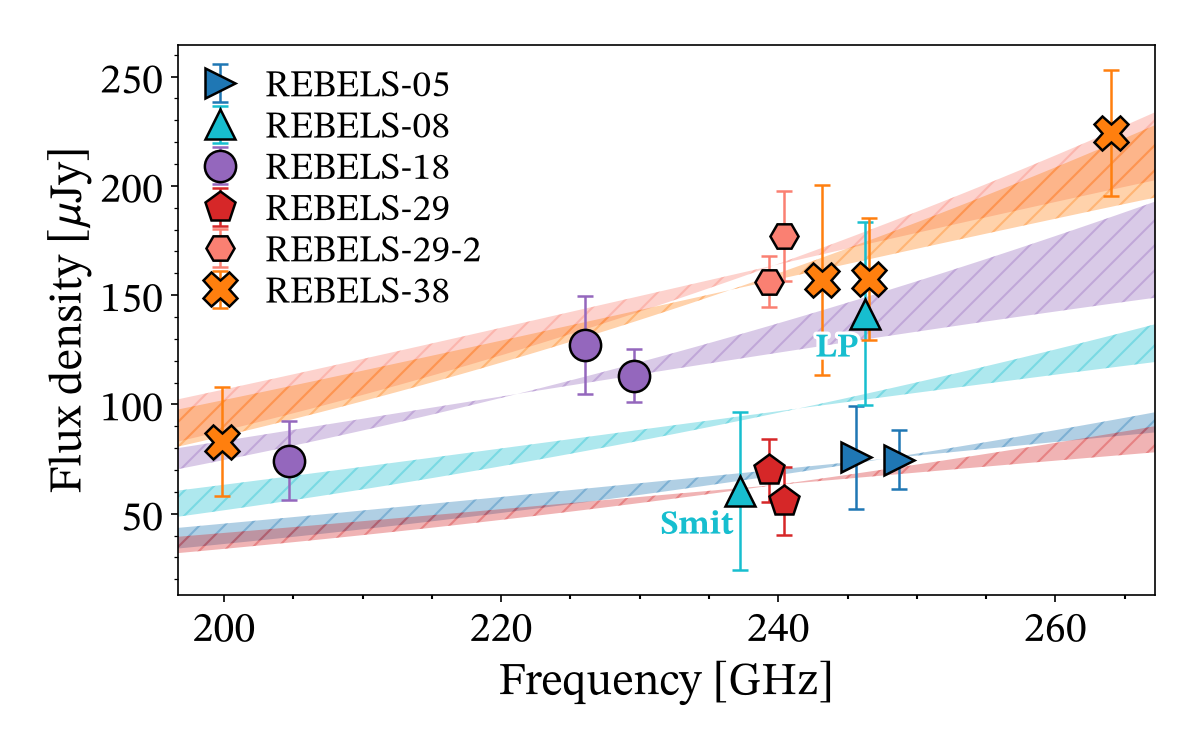}
    \end{minipage}%
    \hfill
    \begin{minipage}[c]{0.30\textwidth}
        \caption{\textbf{The ALMA Band 5 and 6 flux densities of our targets are consistent between different observing projects.} A zoom-in of the Band 5 and 6 $(u,v)$-plane flux densities of the six targets with observations from multiple ALMA projects in these bands. The hatched areas represent MBBs with $T_\mathrm{dust} = 30 - 70\,\mathrm{K}$ and $\beta_\mathrm{IR} = 1.8 - 2.0$, normalized at the average frequency of the observations, illustrating the expected variation in flux density with frequency. The largest difference is found for REBELS-08, whose LP and high-resolution (PI Smit) Band 6 flux densities differ by a factor of $\sim2$, but remain consistent at the $\sim1.5\sigma$ level given their sizable uncertainties.}
        \label{fig:fluxComparisonB6}
    \end{minipage}
\end{figure*}

Our compilation of ALMA observations comprises two or more Band 6 projects for six sources: REBELS-05, REBELS-08, REBELS-29 and REBELS-29-2 (2 separate projects each), and REBELS-18 and REBELS-38 (3 projects). REBELS-38 has moreover been observed in ALMA Band 5, whose central frequency of $\nu_\mathrm{obs} = 199.9\,\mathrm{GHz}$ (Table \ref{tab:observations}) is close to the lower edge of Band 6 ($211\,\mathrm{GHz}$). This set of ALMA projects targeting the narrow frequency range of $\nu_\mathrm{obs} \approx 200 - 265\,\mathrm{GHz}$ can therefore be used to assess the accuracy of our flux density measurements, as well as highlight potential systematic offsets in flux calibration, or possible discrepancies due to the difference in native resolution of the data.

We show our fiducial Band 5 and 6 $(u,v)$-plane flux density measurements of the aforementioned six galaxies in Figure \ref{fig:fluxComparisonB6}. To guide the eye, we overlay MBBs with a range of temperatures ($T_\mathrm{dust} = 30 - 70\,\mathrm{K}$) and emissivity indices ($\beta_\mathrm{IR} = 1.8 - 2.0$), arbitrarily normalized at the noise-weighted average frequency and flux density of the individual Band 5 and 6 measurements per galaxy. This set of $T_\mathrm{dust}$ and $\beta_\mathrm{IR}$ matches the typical range of dust properties inferred for our targets in Section \ref{sec:results_MBB_fits}.

For all six targets, the flux densities are consistent with the expectations from standard optically thin MBBs, without any clear systematic offsets related to angular resolution. The largest difference is found for REBELS-08, whose LP and PI Smit flux densities differ by a factor of $\sim2$, although the large uncertainties on both measurements render them consistent at the $\sim1.5\sigma$ level. This difference persists even when different means of measuring the fluxes are adopted (upper panel in Figure \ref{fig:flux_comparison_B6_B8}), and is thus not unique to the $(u,v)$-plane measurements.

REBELS-08 is only weakly detected in the high-resolution Band 6 imaging, despite a clear $5.5\sigma$ peak detection in the LP data (Figures \ref{fig:allCutouts} and \ref{fig:allCutoutsTapered}). Considering the naturally weighted Band 8 data on REBELS-08 -- where the galaxy is detected at an intermediate resolution -- reveals a complex morphology. REBELS-08 appears to consist of a possible extended dust reservoir (albeit at a modest $\sim2\sigma$ level) and a brighter $\sim4\sigma$ clump towards the southwest. This morphology is not as clearly seen in the high-resolution Band 6 data, although some extended $\sim2\sigma$ emission towards the southwest is visible in the naturally weighted map as well. This complex morphology could therefore possibly bias the flux density measurements of REBELS-08, though deeper, low-resolution Band 6 observations of the galaxy are required for a more definitive conclusion. 

Overall, given that the flux densities are consistent for all six targets, with even the largest discrepancy (REBELS-08) amounting to only $\sim1.5\sigma$, we find no evidence for systematic offsets in our flux density measurements between the different ALMA Band 5 and 6 projects.

\section{Comparison to the REBELS LP Band 6 flux densities}
\label{app:comparison_inami2022}

\begin{figure*}
    \centering
    \includegraphics[width=0.45\textwidth]{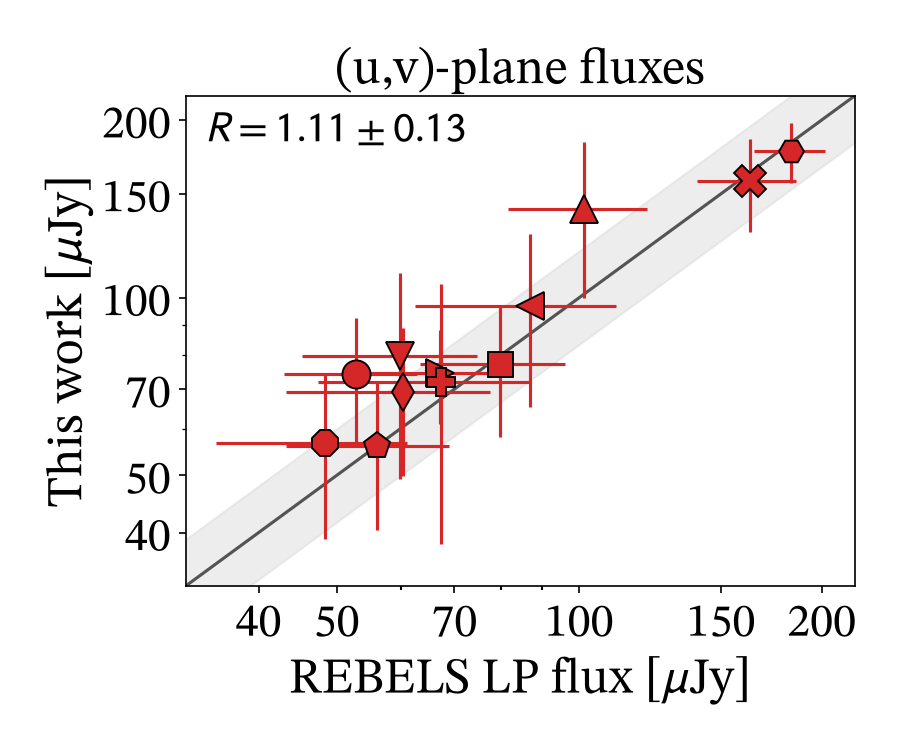}
    \includegraphics[width=0.45\textwidth]{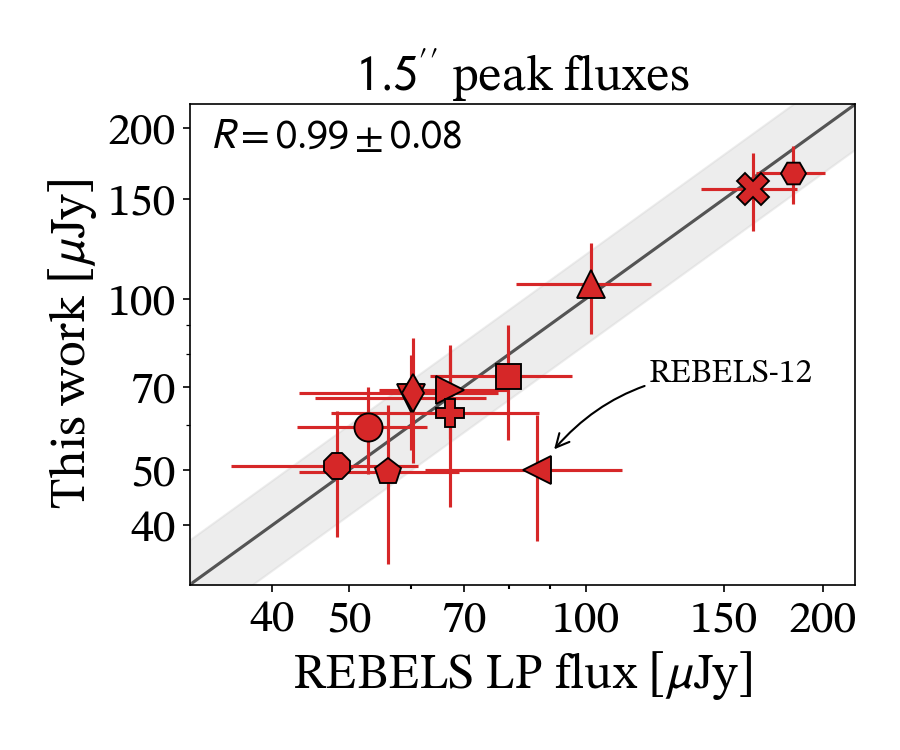}
    \caption{\textbf{Comparison of the fiducial REBELS LP Band 6 $\mathbf{(u,v)}$-plane (\textit{left}) and peak-pixel flux densities (\textit{right}), the latter obtained in $1.5''$-tapered versions of the original LP data, to those measured by \citet{inami2022}.} Markers correspond to the same galaxies as in earlier figures, and the solid line shows the one-to-one relation, with the shading representing a $20\%$ deviation from it. The $(u,v)$-plane flux densities slightly exceed the peak flux densities from \citet{inami2022} by a median factor of $1.11 \pm 0.13$. On the other hand, our $1.5''$ peak fluxes and those in \citet{inami2022} are in good agreement (median ratio of $0.99 \pm 0.08$). The one exception is REBELS-12, which shows a complicated morphology, and for which \citet{inami2022} adopt the summed peak fluxes of its two separate clumps.}
    \vspace*{0.1cm}
    \label{fig:flux_comparison_inami2022}
\end{figure*}

We compare the Band 6 flux densities of the REBELS LP data from \citet{inami2022} to the fiducial $(u,v)$-plane ones measured in this work in the left panel of Figure \ref{fig:flux_comparison_inami2022}. In their work, \citet{inami2022} adopted peak flux densities for all sources (except REBELS-25, which is not analyzed here), and we therefore also compare our peak flux density measurements -- obtained in $1.5''$-tapered maps -- to their values measured in the naturally weighted images (right panel).

We find that our $(u,v)$-plane fluxes are slightly larger than the \citet{inami2022} ones, by a median factor of $1.11 \pm 0.13$, as expected from the comparison of $(u,v)$-plane and peak measurements in Appendix \ref{app:flux_measurement_comparison}. The two largest offsets are seen for REBELS-08 and REBELS-14, which appear slightly spatially extended beyond the $\sim1.3 - 1.5''$ native REBELS beam (Figure \ref{fig:NIRCamALMAoverlays}). However, both are detected at only moderate S/N, and their fluxes remain consistent with those from \citet{inami2022} within the uncertainties.

The peak flux densities obtained from $1.5''$-tapered maps are furthermore in excellent agreement with those measured by \citet{inami2022}, with a median ratio of $S_\mathrm{B6,this\,work} / S_\mathrm{B6,I22} = 0.99 \pm 0.08$, as expected given that the native LP beam is already comparable to our adopted taper. Only for REBELS-12 do we see a significant offset, with the inferred flux density from \citet{inami2022} being $\sim1.7\times$ larger than our measurement. This is because they fitted two Gaussians to REBELS-12, which appears to consist of two clumps in the naturally weighted imaging, and subsequently summed the two peak flux densities to obtain the total Band 6 flux. Indeed, the $(u,v)$-plane fit yields a similar flux density for REBELS-12 as that inferred from the summed peak fluxes in \citet{inami2022}.

\section{Comparison to single-band dust temperature estimates}
\label{app:singleband_predictions}

Prior to the availability of multi-band continuum data, the dust temperatures of REBELS galaxies were estimated from their single-band continua using physically motivated models. \citet{ferrara2022} inferred dust properties from the UV continuum slope and the UV and IR flux densities, while \citet{sommovigo2022} leveraged the \cii{} luminosity and underlying continuum. We briefly compare these estimates to our MBB-based dust temperatures here.

We first compare to \citet{sommovigo2022}, who provide dust temperature estimates for all ten primary REBELS targets analyzed in this work. For these galaxies, they estimate a median temperature of $T_\mathrm{dust}(\mathrm{S22}) = 45 \pm 5\,\mathrm{K}$, in excellent agreement with our sample average and dual-band stack. Their estimates, however, span a narrower range than our measurements, and there is no clear correlation between the \cii{}-based and multi-band dust temperatures for individual galaxies. This may reflect assumptions in the single-band modeling that are now better constrained by \textit{JWST}, such as the metallicity, dust-to-gas ratio, and stellar masses \citep[cf.][]{fisher2025,rowland2026,algera2026}.

\citet{ferrara2022} provide dust temperature estimates for eight of our ten primary targets; for REBELS-08 and REBELS-38 their analytical model does not find a solution, which they attribute to spatial offsets between the UV and IR emission. Among these eight galaxies, they infer a median dust temperature of $T_\mathrm{dust}(\mathrm{F22}) = 56 \pm 8\,\mathrm{K}$, which is likewise consistent with that inferred for our sample. Notably, \citet{ferrara2022} identified REBELS-18 as the warmest galaxy among the REBELS sample, with a dust temperature in excellent agreement with our MBB-fitted value. Their estimates for REBELS-29 and REBELS-40 ($\approx58\,\mathrm{K}$), on the other hand, are considerably warmer than our measured $T_\mathrm{dust}\approx32\,\mathrm{K}$ (although this value is prior-dominated for REBELS-40; Appendix~\ref{app:dust_temperature_posteriors}), such that, as for \citet{sommovigo2022}, there is no significant correlation across the full set of individual galaxies.

Because of the good agreement between the sample-averaged dust temperatures from the single-band estimates and multi-band measurements, the corresponding average dust masses, IR luminosities, and obscured fractions are broadly consistent as well. This suggests that such single-band approaches can recover the typical dust properties of high-redshift LBGs, while multi-band data remain necessary for robust constraints on individual galaxies.

\section{Dust temperature prior in MBB-fitting at low signal-to-noise}
\label{app:mock_mbb_fitting}

In our MBB-fitting framework, we adopt an `inverse-temperature' prior on the dust temperature (Section \ref{sec:methods_mbb_fitting}). We first discuss the physical motivation for this approach, followed by MBB fits to mock sources with both this inverse-temperature prior and a `standard' prior that is uniform in $T_{\mathrm{dust},z}$, which demonstrates that the former better recovers the intrinsic dust properties of galaxies for low-S/N, dual-band data.

\subsection{Motivation for an inverse-temperature prior}

\begin{figure*}
    \centering
    \includegraphics[width=0.95\linewidth]{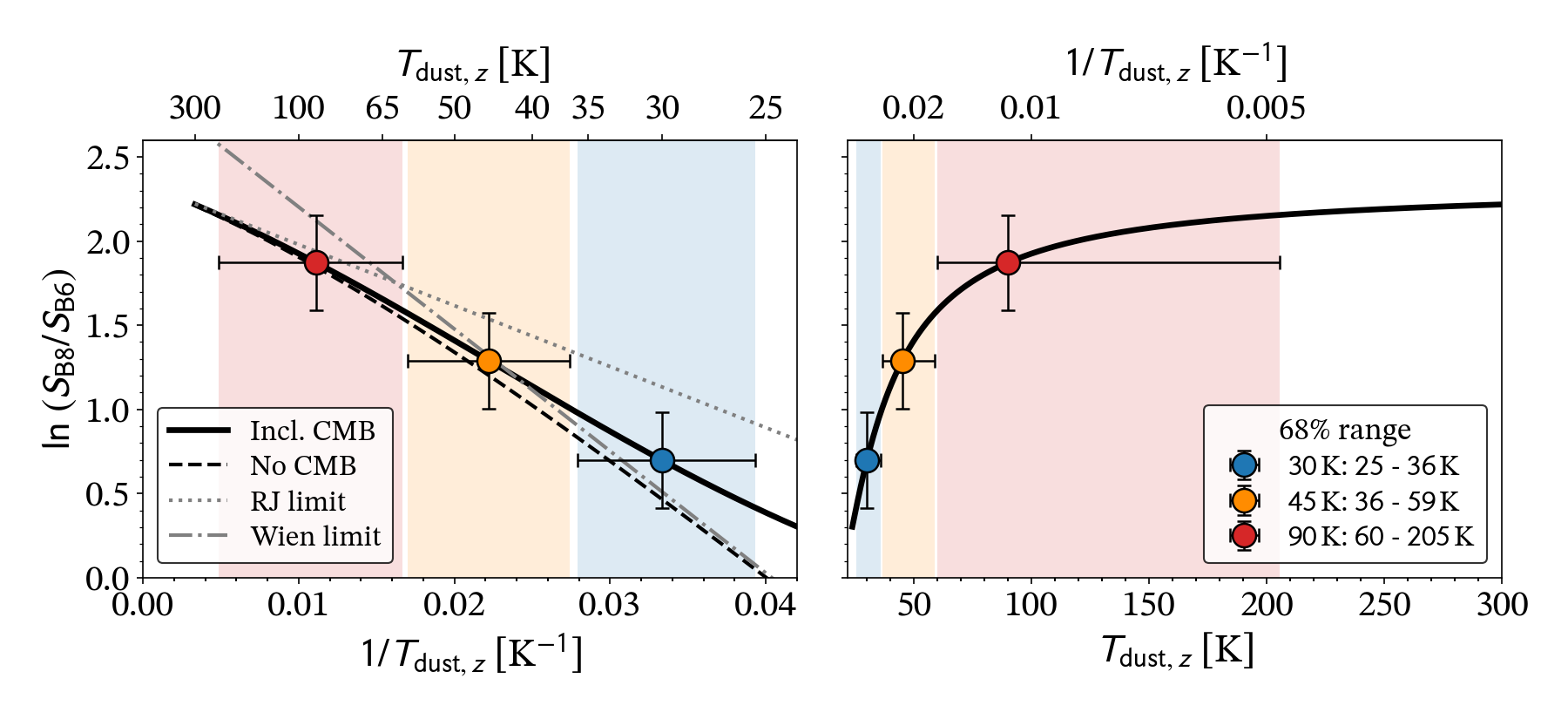}
    \caption{\textbf{The quantity $1 / T_{\mathrm{dust},z}$ can be constrained with near-fixed precision for any temperature through a dual-band MBB fit.} \textit{Left:} the logarithm of the Band 8 to Band 6 flux ratio, $\ln(r) = \ln(S_\mathrm{B8}/S_\mathrm{B6})$ as a function of inverse dust temperature. As discussed in the text, the uncertainty on $\ln(r)$ depends only on the S/N of the dual-band observations. The solid black line shows that $\ln(r)$ scales approximately linearly with $1 / T_{\mathrm{dust},z}$, which implies that the uncertainty on the inverse dust temperature is also set by the S/N, and is nearly independent of the dust temperature itself. The three datapoints, corresponding to the Band 8 / Band 6 flux ratio at $30$, $45$, and $90\,\mathrm{K}$ with an $\mathrm{S/N}=5$ in both bands, illustrate this: for a fixed uncertainty on $\ln(r)$ (vertical errorbars), the uncertainty on $1 / T_{\mathrm{dust},z}$ is similar in all three cases (indicated through the horizontal errorbars and colored shading). The dotted and dash-dotted lines show the analytical scalings for hot dust -- when Bands 6 and 8 probe the Rayleigh-Jeans tail -- and cold dust (Wien limit) in the absence of the CMB, respectively (Equations \ref{eq:fluxRatioRayleighJeansLimit} and \ref{eq:fluxRatioWienLimit}). The dashed black line shows the full relation between $\ln(r)$ and $1 / T_{\mathrm{dust},z}$ when the CMB term is neglected. \textit{Right:} unlike the inverse dust temperature, $T_{\mathrm{dust},z}$ cannot be accurately constrained from a dual-band MBB fit in the case of hot dust (red marker); instead, the uncertainty on $T_{\mathrm{dust},z}$ depends strongly on the value of the temperature itself.}
    \vspace*{0.1cm}
    \label{fig:Tdust_prior}
\end{figure*}

Consider the scenario where flux densities at two rest-frame frequencies are available, $S_{\nu_1}$ and $S_{\nu_2}$, with $\nu_1 > \nu_2$. In our case, these frequencies correspond to those probed by Band 8 and Band 6, respectively. The goal is to infer the dust mass and temperature from these flux measurements through an optically thin MBB fit, keeping the dust emissivity index fixed. In this scenario, the dust temperature ($T_{\mathrm{dust},z}$, denoted as just $T$ in the remainder of this section) is constrained by the flux ratio $r = S_{\nu_1} / S_{\nu_2}$, which is independent of the dust mass, as the latter only sets the normalization (Equation \ref{eq:mbb_fit}). For simplicity, we neglect the CMB term, $B_\nu(T_{\mathrm{CMB},z})$, in the following analytical approximation, but include it below in a numerical check, which yields consistent results. 

From standard error propagation assuming Gaussian independent errors on the flux densities, it follows that the error on the flux ratio can be written as 

\begin{align}
    \sigma_r = r \times \sqrt{\mathrm{S/N}_1^{-2} + \mathrm{S/N}_2^{-2}} \ ,
    \label{eq:flux_ratio_snr}
\end{align}

\noindent where $\mathrm{S/N}_i = S_{\nu_i} / \sigma_{\nu_i}$, with $\sigma_{\nu_i}$ being the uncertainty on $S_{\nu_i}$. The fractional error $\sigma_r / r$ thus depends only on the S/N of the data, and not on the flux ratio itself. For a given S/N, which is set by the source brightness and the sensitivity of the observations, the error on $\ln(r)$, denoted $\sigma_{\ln r} = \sigma_r / r$, is therefore a constant value. For an optically thin MBB, $\ln(r)$ can be written as (Equation \ref{eq:mbb_fit})

\begin{align}
    \ln(r) = \left(3 + \beta_\mathrm{IR}\right) \ln\left(\frac{\nu_1}{\nu_2}\right) + \ln\left(e^{x_2} - 1\right) - \ln\left(e^{x_1} - 1\right) \ ,
\end{align}

\noindent where we defined $x_i = h\nu_i / k_\mathrm{B} T$. For hot dust, $x_i \ll 1$, and under a Taylor expansion this expression simplifies to

\begin{align}
    \ln(r) \simeq \left(2 + \beta_\mathrm{IR}\right)\ln\left(\frac{\nu_1}{\nu_2}\right) - \frac{h}{2k_\mathrm{B}T}\left(\nu_1 - \nu_2\right) \ ,
\label{eq:fluxRatioRayleighJeansLimit}
\end{align}

\noindent which is linear in $1 / T$. For cold dust ($x_i \gg 1$), a similar expansion yields

\begin{align}
    \ln(r) \simeq \left(3 + \beta_\mathrm{IR}\right)\ln\left(\frac{\nu_1}{\nu_2}\right) - \frac{h}{k_\mathrm{B}T}\left(\nu_1 - \nu_2\right) \ ,
\label{eq:fluxRatioWienLimit}
\end{align}

\noindent which again is linear in $1 / T$, with a slope twice as large. Between these two limits, $\ln(r)$ therefore varies nearly linearly with $1 / T$, with a slope that changes by at most a factor of two (and by only a factor of $\sim1.4$ across $T \approx 25 - 300\,\mathrm{K}$ at $z=7$ when including the CMB; see below). Consequently, a fixed uncertainty on $\ln(r)$ corresponds to an approximately fixed uncertainty on $1 / T$, irrespective of the temperature of the source. The uncertainty on the temperature $T$ itself instead increases rapidly for warmer dust.

We illustrate this in Figure \ref{fig:Tdust_prior}, which shows $\ln(r)$ for an optically thin MBB at $z = 7$ with $\beta_\mathrm{IR} = 2.0$, sampled at rest-frame wavelengths of $88$ and $158\,\mu\mathrm{m}$, and now including the effect of the CMB. The full expression remains close to linear in $1 / T$ (solid black line). Assuming an S/N of 5 in both bands, a source at $T_{\mathrm{dust},z} = 30$, $45$ or $90\,\mathrm{K}$ is constrained to a nearly identical precision in $1 / T$ (colored markers). In contrast, the uncertainty on $T$ increases rapidly for hot dust; for a source with $T_{\mathrm{dust},z} = 90\,\mathrm{K}$, the $68\%$ range spans $60 - 205\,\mathrm{K}$ (red marker and shading).

Since the data constrain $1 / T$ with a precision that is nearly independent of the temperature, a prior that is uniform in $1 / T$ assigns equal weight to all temperatures that can be distinguished by the observations. This is approximately a \citet{jeffreys1946} prior, which in terms of the temperature has a density $p(T) \propto T^{-2}$. A prior that is uniform in $T$ instead places most of its weight at high temperatures, where the data are least constraining, thereby biasing the fitted dust temperatures high. We demonstrate this empirically in the next subsection.

\subsection{Dual-band modified blackbody fits of mock sources at low signal-to-noise}

\begin{figure*}
    \centering
    \includegraphics[width=0.85\linewidth]{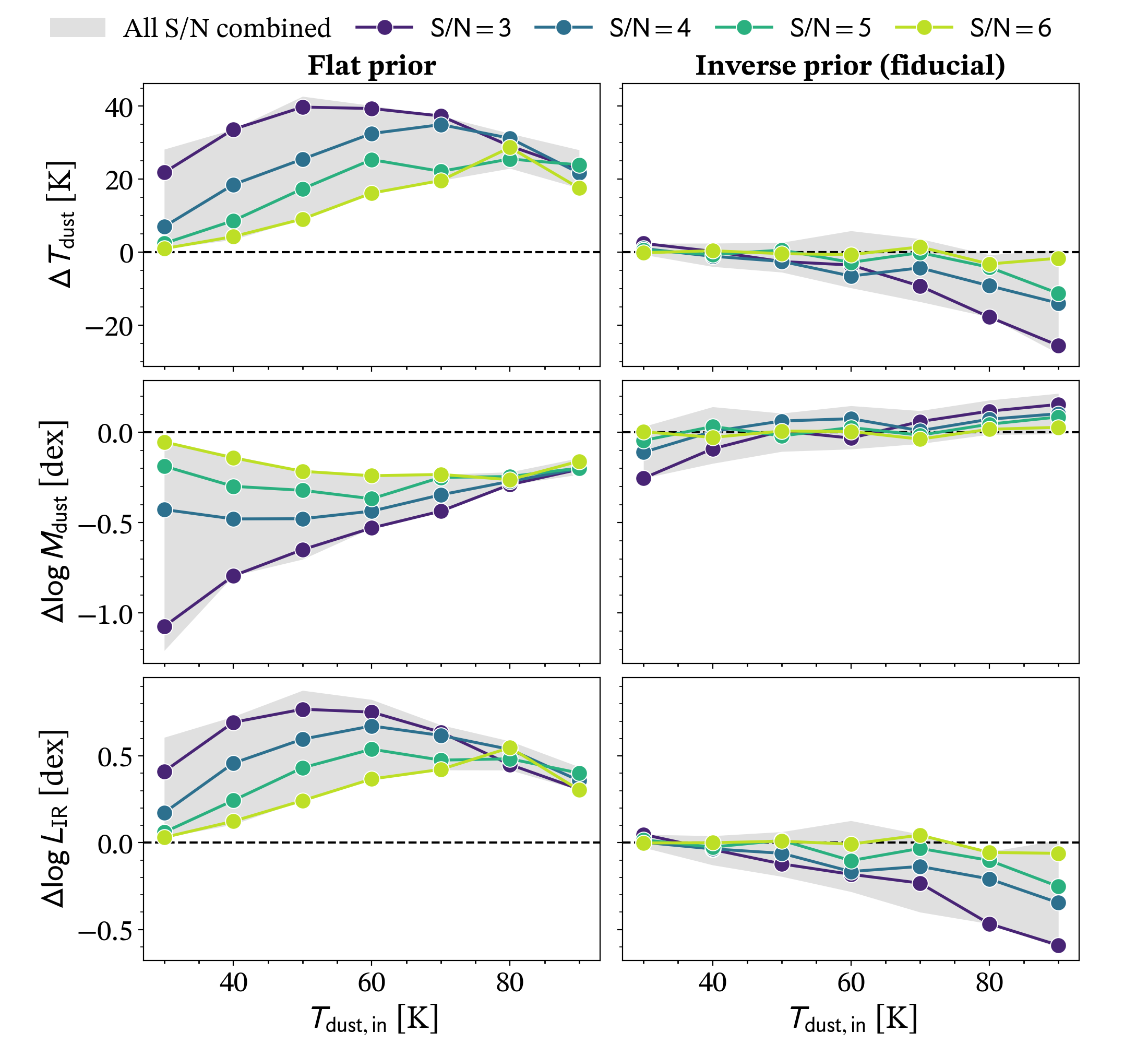}
    \caption{\textbf{The fiducial `inverse' dust temperature prior (\textit{right}) recovers the input dust properties of dual-band mock MBB fits with considerably smaller bias than a flat prior (\textit{left}).} The columns show the results of mock MBB fits adopting a flat prior on $T_{\mathrm{dust},z}$ with a smooth taper above $150\,\mathrm{K}$ (\textit{left}), and the fiducial inverse prior, $p(T_{\mathrm{dust},z}) \propto T_{\mathrm{dust},z}^{-2}$ bounded at $300\,\mathrm{K}$ (\textit{right}). The rows show the offset between the recovered and input dust temperature (\textit{top}; $\Delta T_\mathrm{dust} = T_\mathrm{dust,fit} - T_\mathrm{dust,in}$), dust mass (\textit{middle}) and infrared luminosity (\textit{bottom}) as a function of the input dust temperature. Colored lines correspond to an equal S/N of 3, 4, 5 and 6 in both Bands 6 and 8, while the grey shading shows the full range spanned by all 16 combinations of S/N in the two bands. With a flat prior, the dust temperature and infrared luminosity are systematically overestimated and the dust mass is underestimated, by up to $\approx40\,\mathrm{K}$, $0.8\,\mathrm{dex}$ and $1.1\,\mathrm{dex}$, respectively, with biases persisting at $\mathrm{S/N} = 6$ for $T_\mathrm{dust,in} \gtrsim 50\,\mathrm{K}$. With the inverse prior, all three quantities are recovered with little bias for $\mathrm{S/N} \geq 4$; at $\mathrm{S/N} = 3$, the dust temperature and infrared luminosity are underestimated only for $T_\mathrm{dust,in} \gtrsim 70\,\mathrm{K}$, by up to $\approx25\,\mathrm{K}$ and $0.6\,\mathrm{dex}$, respectively.}
    \label{fig:mock_mbb_fitting}
\end{figure*}

To assess the accuracy with which we can infer the dust properties of $z\approx7$ galaxies with limited frequency coverage and modest signal-to-noise, as well as under the assumption of different dust temperature priors, we perform mock simulations. These build upon the tests carried out in \citet{algera2024b}, and the fits to mock MBBs used to assess the effects of multi-temperature dust in \citet{sommovigo_algera2025}. 

Briefly, we assume a two-band setup (Bands 6 and 8) and that the dust can be described as an MBB with one of seven fixed dust temperatures, $T_\mathrm{dust} = 30, 40, \ldots, 90\,\mathrm{K}$. The dust mass is defined implicitly by the MBB normalization, which is fixed to $100\,\mu\mathrm{Jy}$ at $220\,\mathrm{GHz}$. We adopt a fixed redshift $z=7$, a dust emissivity index $\beta_\mathrm{IR} = 2.0$, and a grid of $\mathrm{S/N} = \left\{3, 4, 5, 6\right\}$ in both bands to explore the effect of fitting MBBs to noisy data. For each temperature and combination of dual-band S/N (i.e., $7 \times 4 \times 4 = 112$ combinations), we draw the fluxes in the two bands and add a Gaussian offset based on the flux uncertainty. We repeat this process 200 times for each combination (i.e., with 200 random error draws), and record the fitted dust temperature, mass, and infrared luminosity for each run. We perform the mock fits twice: once for the fiducial `inverse' dust temperature prior adopted in Section \ref{sec:methods_mbb_fitting}, and once for the prior used in \citet{algera2024,algera2024b}, which is flat between the CMB temperature and $150\,\mathrm{K}$, and smoothly decreases beyond this upper limit as a Gaussian with standard deviation $\sigma = 30\,\mathrm{K}$. As already discussed in \citet{algera2024} and verified by mock tests in \citet{algera2024b}, this flat prior is prone to overestimating $T_\mathrm{dust}$ at high temperatures and low signal-to-noise. We note that, given the higher S/N and multi-band coverage available in those previous works, the dust temperatures reported there are unlikely to have been overestimated.

We show the results of the mock fits with the `inverse' and flat priors in Figure \ref{fig:mock_mbb_fitting}. The flat prior significantly overestimates dust temperatures by up to $+40\,\mathrm{K}$ at low $\mathrm{S/N}=3$ for input dust temperatures of $T_\mathrm{dust,in}\approx50-60\,\mathrm{K}$. Even at higher S/N, this bias persists for warm dust, and consequently dust masses and infrared luminosities can be significantly underestimated and overestimated, respectively, by up to an order of magnitude. The inverse prior, on the other hand, recovers input dust temperatures to within $10\,\mathrm{K}$ even at low S/N=3 for $T_\mathrm{dust} \lesssim 70\,\mathrm{K}$, and up to $T_\mathrm{dust} = 80\,\mathrm{K}$ for $\mathrm{S/N} \geq 4$, with offsets of at most $\approx 13\,\mathrm{K}$ at $90\,\mathrm{K}$ (and up to $\approx25\,\mathrm{K}$ at $\mathrm{S/N}=3$). As a result, dust masses are recovered to within $\lesssim0.25\,\mathrm{dex}$ across all input temperatures and S/N, and infrared luminosities are, at most, underestimated by $\sim0.6\,\mathrm{dex}$ for low S/N and high temperatures. For more typical dust temperatures of $T_\mathrm{dust} \approx 30 - 60\,\mathrm{K}$, all dust properties can be accurately recovered even at a low $\mathrm{S/N} = 3$. In summary, the inverse dust temperature prior $p(T) \propto T^{-2}$ thus significantly improves the accuracy with which galaxy dust properties can be inferred from MBB fits at low S/N compared to a standard flat prior.

\section{Comparison of MBB dust properties across flux density measurements}
\label{app:mbb_properties_across_flux_densities}

\begin{figure*}
    \centering
    \includegraphics[width=0.95\textwidth]{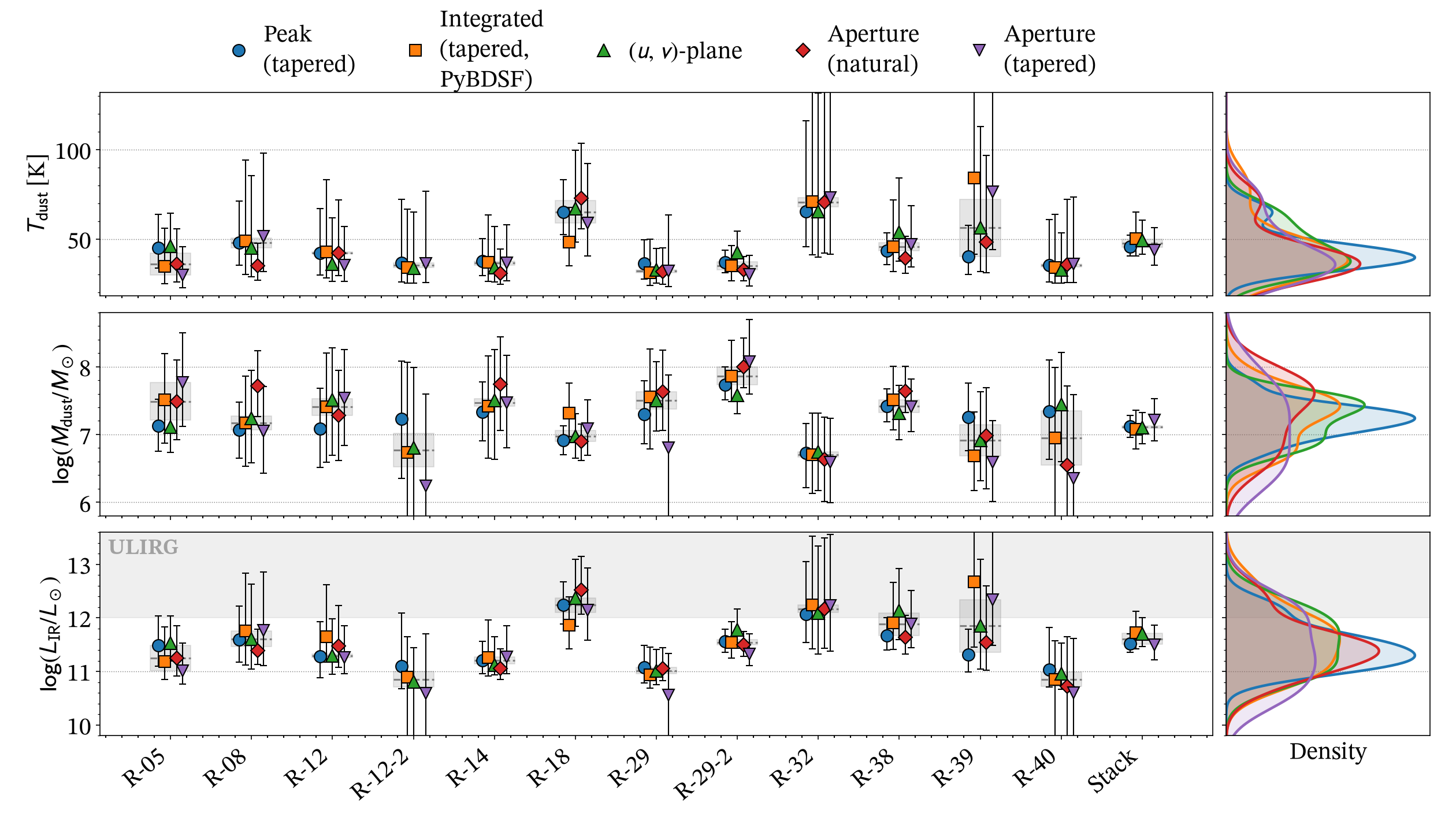}
    \caption{\textbf{Fitted MBB parameters as a function of flux measurement method, for each of our 12 dust-detected targets plus the dual-band stack.} The different rows show the dust temperature (top), dust mass (middle), and infrared luminosity (bottom) per galaxy, with a kernel density estimate (KDE) shown on the right. While the inferred MBB properties are generally consistent within the uncertainties for a given source, the measurement approach can introduce additional scatter.}
    \label{fig:MBB_fits_across_flux_measurements}
\end{figure*}

For the various flux density measurement methods outlined in Appendix \ref{app:flux_measurement_comparison}, we perform optically thin MBB fitting as in Section \ref{sec:methods_mbb_fitting}. For each galaxy -- as well as for the stack -- we show the fitted dust temperature, dust mass, and infrared luminosity in Figure \ref{fig:MBB_fits_across_flux_measurements} to reveal the spread in inferred dust properties based on how the flux densities are measured.

Across the twelve galaxies, the median absolute deviation between methods has a median value of $2.3\,$K in $T_\mathrm{dust}$, $0.12\,$dex in $\log(M_\mathrm{dust})$ and $0.14\,$dex in $\log(L_\mathrm{IR})$, well below the statistical uncertainties on the fiducial measurements. The dust temperatures of individual sources can nevertheless vary considerably between methods, most notably for REBELS-39 ($40-84\,$K), although given its large uncertainties, all of these values are consistent with its fiducial dust temperature to within $1\sigma$.

Relative to the fiducial $(u,v)$-plane measurements, the alternative methods yield similar dust temperatures and masses, with median offsets of $-3.4$ to $+1.2\,$K and $-0.09$ to $+0.12\,\mathrm{dex}$, respectively. The infrared luminosities from the peak and integrated flux densities likewise agree with the fiducial values (median offsets $\leq0.02\,$dex), while those based on aperture flux densities are systematically lower by $\sim0.2\,$dex, albeit still well within the typical uncertainties. The number of galaxies with a median $L_\mathrm{IR} > 10^{12}\,L_\odot$ correspondingly varies between two and three depending on the adopted method. For the stack, the different methods yield $T_\mathrm{dust} = 44 - 50\,$K, $\log(M_\mathrm{dust}/M_\odot) = 7.1 - 7.2$ and $\log(L_\mathrm{IR}/L_\odot) = 11.5 - 11.7$, consistent with those inferred from the fiducial $(u,v)$-plane measurements in all cases. Overall, the average dust properties of our targets are thus robust to the adopted flux measurement approach.

\section{Dust temperature posteriors}
\label{app:dust_temperature_posteriors}

\begin{figure*}
    \centering\includegraphics[width=0.95\textwidth]{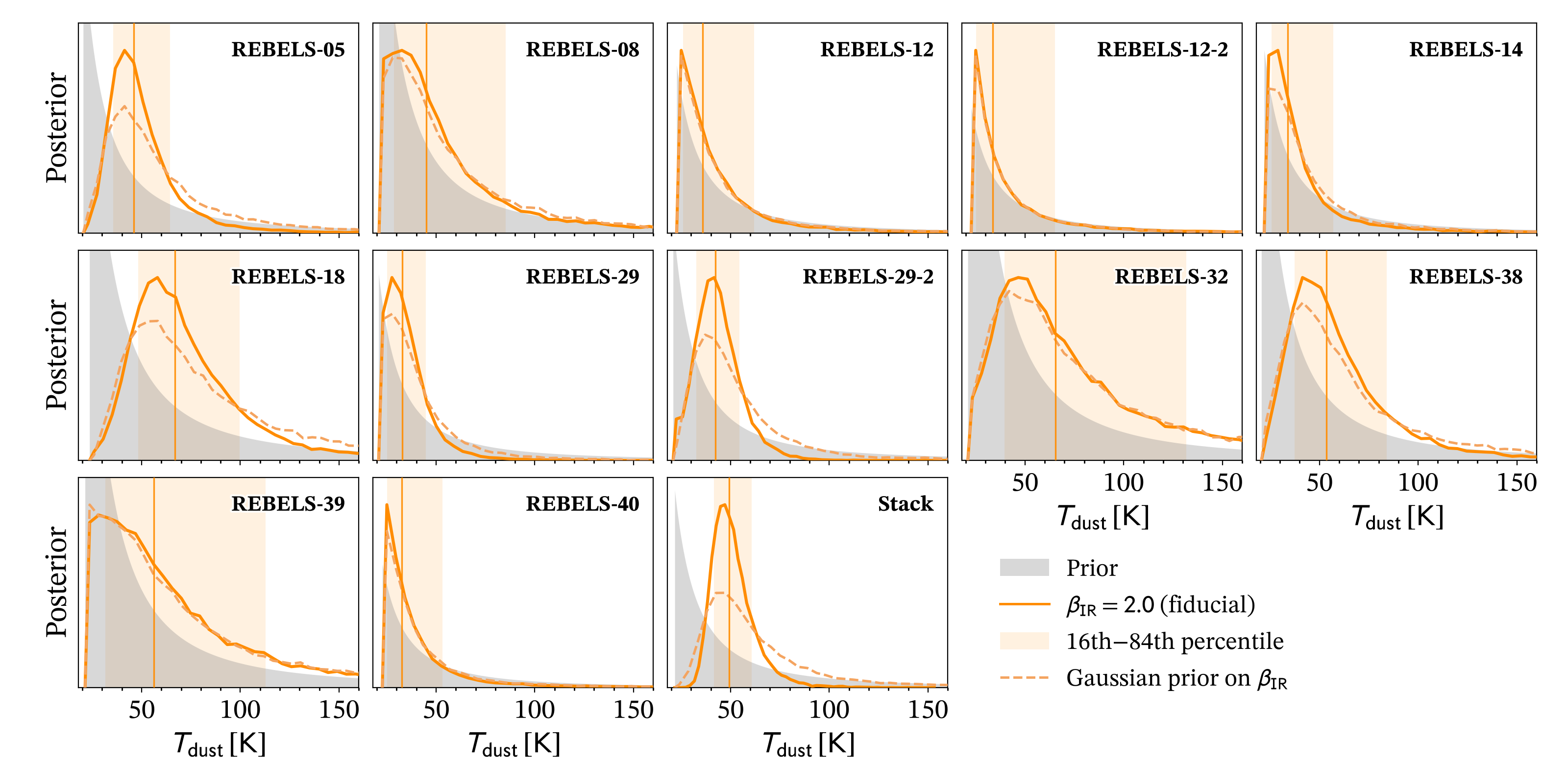}
    \caption{\textbf{Dust temperature posteriors for our 12 continuum-detected targets and the dual-band stack (final panel).} The posteriors for the fiducial $\beta_\mathrm{IR} = 2.0$ are shown through the solid lines, and the median and $16-84^\mathrm{th}$ percentiles are shown through the vertical line and shading. Dashed lines represent the posteriors obtained from the fit with a Gaussian prior on $\beta_\mathrm{IR}$. The inverse-temperature prior is shown through the grey shading.}
    \label{fig:Tdust_posteriors}
\end{figure*}

We show the dust temperature posteriors for the MBB fits using the inverse-temperature prior (Section \ref{sec:methods_mbb_fitting}) in Figure \ref{fig:Tdust_posteriors}. The posteriors for both the fiducial $\beta_\mathrm{IR} = 2.0$ (solid line) and the Gaussian prior on $\beta_\mathrm{IR}$ (dashed) are plotted; in what follows, we focus on the former.

As can be seen from the figure, the dust temperature posterior for a subset of sources reflects the prior. To quantify this, we compare the $16-84^\mathrm{th}$ percentile width of the posterior with that of the prior, both measured in $1 / T_{\mathrm{dust},z}$ in which the prior is uniform. For reference, the prior has a median of $T_{\mathrm{dust},z} \approx 40\,\mathrm{K}$ and a $16-84^\mathrm{th}$ percentile range of approximately $25 - 100\,\mathrm{K}$, with a lower bound set by the CMB temperature, and thus depending slightly on redshift. We define the width ratio $W$ as the (posterior width) / (prior width), such that a larger value implies a more prior-dominated fit. Among the 12 individual targets, it ranges from $W = 0.39$ for REBELS-29-2 (best constrained) to $W=0.90$ for REBELS-12-2 (least), with a median of $W = 0.66$.

We identify six sources with $W > 0.7$, which we consider prior-dominated: REBELS-08, REBELS-12, REBELS-12-2, REBELS-14, REBELS-39, and REBELS-40. Four of these have relatively cold dust temperatures (median $T_\mathrm{dust} < 40\,\mathrm{K}$), while REBELS-08 and REBELS-39 are warmer (median $T_\mathrm{dust} \approx 45\,\mathrm{K}$ and $56\,\mathrm{K}$, respectively). As the inverse-temperature prior favors low temperatures, the cold dust temperatures of the former four are likely in part driven by the prior, although their posterior medians lie below that of the prior itself ($\approx40\,\mathrm{K}$), indicating that the data also prefer relatively cold dust. Nevertheless, the individual dust properties of these six sources should be interpreted with caution.

The dust temperature of the stack, on the other hand, is well constrained ($W = 0.24$), and is consistent with the sample median, suggesting that the prior does not significantly bias the dust temperatures of our sample as a whole. For comparison, adopting the flat temperature prior in Appendix \ref{app:mock_mbb_fitting} yields a stacked dust temperature consistent with our fiducial value, while the median dust temperature of the individual targets increases by $\approx 10 - 15\,\mathrm{K}$. The inverse-temperature prior thus reconciles the individual dust temperatures with that of the well-constrained stack.

\section{Modified blackbody fits with a Gaussian prior on $\beta_\mathrm{IR}$}
\label{app:mbb_fits_gaussian_prior}

\begin{figure*}
    \centering\includegraphics[width=1.0\textwidth]{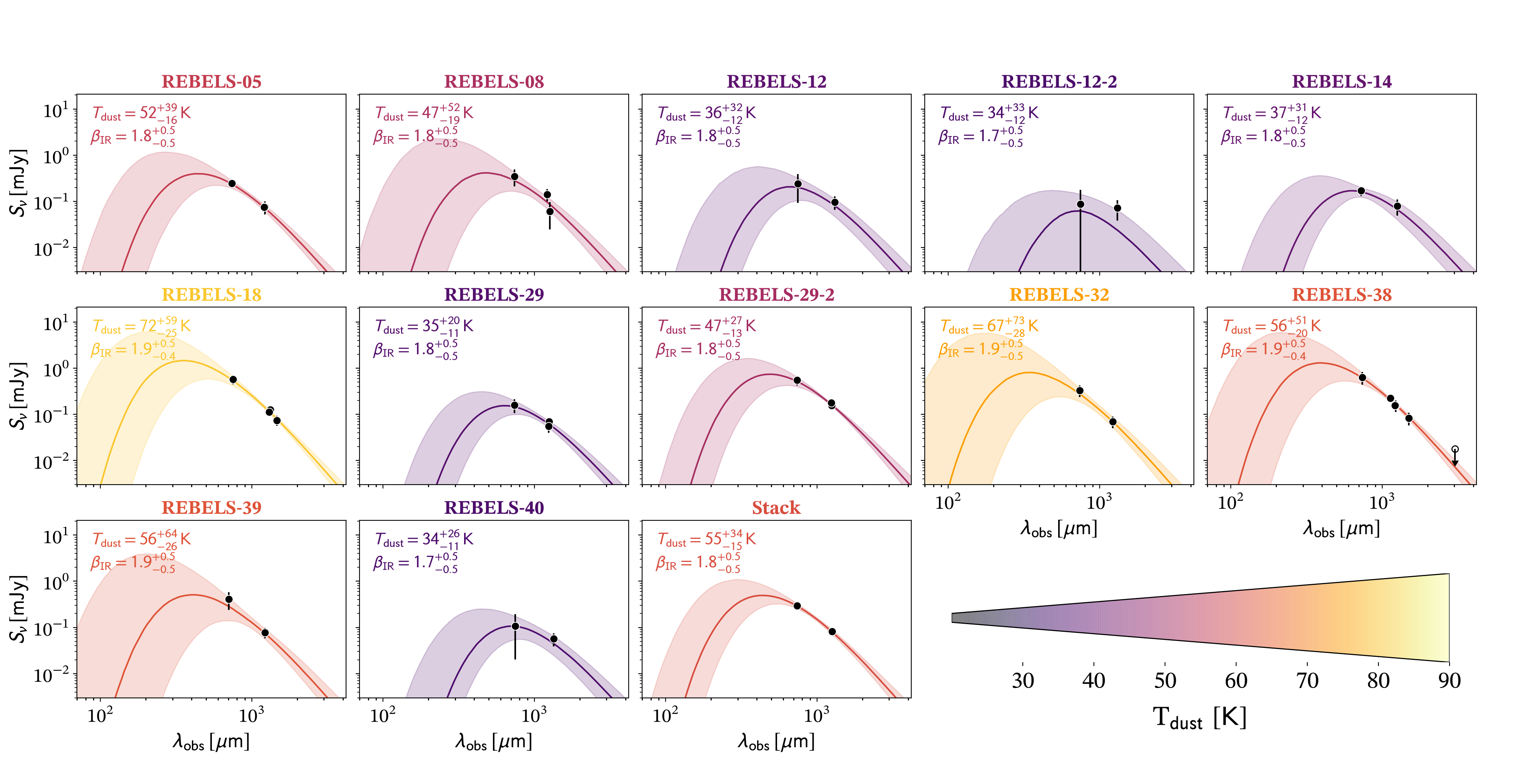}
    \caption{\textbf{Modified blackbody fits of our high-redshift sample, adopting a Gaussian prior on $\beta_\mathrm{IR}$ with mean $1.8$ and standard deviation $0.5$.} As in Figure~\ref{fig:MBBfitting}, fiducial $(u,v)$-plane flux measurements are adopted for all 12 targets and the stack.}
    \label{fig:MBBfitting_gaussianPrior}
\end{figure*}

In Section \ref{sec:results_MBB_fits}, we fit the dust SEDs of our 12 continuum-detected targets and a dual-band stack with a fixed dust emissivity index $\beta_\mathrm{IR} = 2.0$. For completeness, we show fits with a Gaussian prior on $\beta_\mathrm{IR}$ in Figure \ref{fig:MBBfitting_gaussianPrior}, using our fiducial $(u,v)$-plane flux density measurements. This prior adopts a mean of $1.8$ and standard deviation of $0.5$ (Section \ref{sec:methods_mbb_fitting}). Given that the dust emissivity index cannot be constrained from the current data, its posterior largely reflects the prior, and the median dust emissivity indices range from $\beta_\mathrm{IR} = 1.7 - 1.9$ with a typical uncertainty of $0.5$. The dust temperatures are consistent with those obtained from the $\beta_\mathrm{IR} = 2.0$ fit, albeit with larger uncertainties as expected. For the stack, we now infer a dust temperature of $T_\mathrm{dust} = 55_{-15}^{+34}\,\mathrm{K}$, compared to the fiducial value of $T_\mathrm{dust} = 49_{-8}^{+11}\,\mathrm{K}$ when a fixed $\beta_\mathrm{IR} = 2.0$ is adopted.

\section{Tabulated flux densities}
\label{app:fluxdensity_table}

In Table \ref{tab:flux_measurements} we provide the fiducial multi-band $(u,v)$-plane flux density measurements for our 13 targets. All but REBELS-39-2 are detected in at least one ALMA band. For REBELS-12 and REBELS-39, the MBB fits adopt the concatenated Band 8 data (labeled `B8-concat' in the table). However, we also provide the individual measurements from the PI Fudamoto and PI Algera Band 8 projects for completeness.

\begin{sidewaystable*}
\centering
\caption{Fiducial $(u,v)$-plane flux densities of our targets, corrected for primary beam attenuation. Columns are labeled by ALMA band and PI; `B8-concat' denotes the concatenated Band 8 data from PIs Fudamoto and Algera (Section~\ref{sec:methods_concat}).}
\label{tab:flux_measurements}
\resizebox{\textwidth}{!}{%
\begin{tabular}{lcccccccccccccccccc}
\toprule
Project & \multicolumn{2}{c}{B3-Gonzalez} & \multicolumn{2}{c}{B5-Fudamoto} & \multicolumn{2}{c}{B6-LP} & \multicolumn{2}{c}{B6-Smit} & \multicolumn{2}{c}{B6-Fudamoto} & \multicolumn{2}{c}{B8-Inami} & \multicolumn{2}{c}{B8-Fudamoto} & \multicolumn{2}{c}{B8-Algera} & \multicolumn{2}{c}{B8-concat} \\
 & Freq & Flux & Freq & Flux & Freq & Flux & Freq & Flux & Freq & Flux & Freq & Flux & Freq & Flux & Freq & Flux & Freq & Flux \\
Galaxy & [GHz] & [$\mu$Jy] & [GHz] & [$\mu$Jy] & [GHz] & [$\mu$Jy] & [GHz] & [$\mu$Jy] & [GHz] & [$\mu$Jy] & [GHz] & [$\mu$Jy] & [GHz] & [$\mu$Jy] & [GHz] & [$\mu$Jy] & [GHz] & [$\mu$Jy] \\
\midrule
REBELS-05 & - &  & - &  & 248.7 & $74.6 \pm 13.5$ & 245.6 & $75.9 \pm 23.6$ & - &  & - &  & - &  & 405.0 & $245.4 \pm 44.2$ & - &  \\
REBELS-08 & - &  & - &  & 246.2 & $141.5 \pm 41.7$ & 237.2 & $60.5 \pm 36.1$ & - &  & - &  & - &  & 405.0 & $349.0 \pm 138.0$ & - &  \\
REBELS-12 & - &  & - &  & 228.1 & $96.6 \pm 31.4$ & - &  & - &  & - &  & 401.4 & $<354.3$ & 401.5 & $306.7 \pm 156.6$ & 401.5 & $241.3 \pm 149.2$ \\
REBELS-12-2 & - &  & - &  & 228.1 & $72.0 \pm 33.7$ & - &  & - &  & - &  & 401.4 & $<380.1$ & - &  & - &  \\
REBELS-14 & - &  & - &  & 238.6 & $79.7 \pm 30.3$ & - &  & - &  & - &  & - &  & 412.8 & $170.4 \pm 39.5$ & - &  \\
REBELS-18 & - &  & - &  & 204.7 & $74.3 \pm 18.1$ & 226.1 & $127.1 \pm 22.4$ & 229.6 & $113.2 \pm 12.3$ & - &  & - &  & 398.2 & $578.0 \pm 133.5$ & - &  \\
REBELS-29 & - &  & - &  & 240.4 & $56.0 \pm 15.6$ & 239.3 & $69.9 \pm 14.4$ & - &  & - &  & - &  & 405.0 & $159.1 \pm 53.1$ & - &  \\
REBELS-29-2 & - &  & - &  & 240.4 & $177.1 \pm 20.8$ & 239.3 & $156.2 \pm 11.5$ & - &  & - &  & - &  & 405.0 & $547.9 \pm 123.8$ & - &  \\
REBELS-32 & - &  & - &  & 244.5 & $69.3 \pm 19.4$ & - &  & - &  & - &  & - &  & 405.0 & $330.2 \pm 90.8$ & - &  \\
REBELS-38 & 99.1 & $<12.3$ & 199.9 & $83.0 \pm 25.1$ & 246.6 & $157.4 \pm 28.0$ & 243.1 & $156.9 \pm 43.4$ & 264.0 & $224.1 \pm 28.9$ & 405.0 & $629.4 \pm 190.5$ & - &  & - &  & - &  \\
REBELS-39 & - &  & - &  & 244.5 & $77.3 \pm 19.2$ & - &  & - &  & - &  & 425.6 & $<326.0$ & 425.6 & $382.5 \pm 187.2$ & 425.6 & $410.8 \pm 173.1$ \\
REBELS-39-2 & - &  & - &  & 244.5 & $<51.2$ & - &  & - &  & - &  & 425.6 & $<346.5$ & 425.6 & $<212.8$ & 425.6 & $<189.0$ \\
REBELS-40 & - &  & - &  & 223.3 & $56.8 \pm 17.8$ & - &  & - &  & 400.5 & $<267.5$ & - &  & - &  & - &  \\
\bottomrule
\end{tabular}}
\begin{tablenotes}
        \footnotesize
        \item Upper limits are $3\sigma$. However, the modified blackbody fits use the fitted $(u,v)$-plane fluxes even for sources not detected at $>3\sigma$ in the image plane (Section~\ref{sec:methods_fluxdensity_measurements}), with the exception of the Band 3 data of REBELS-38, which is treated as an upper limit (Section \ref{sec:centroids}). REBELS-39-2 is undetected in all bands and is not fitted.
\end{tablenotes}
\end{sidewaystable*}

\end{document}